\documentclass{emulateapj}
\usepackage{graphicx}
\usepackage{epstopdf}
\usepackage{epsfig}
\usepackage{natbib}
\usepackage[dvips]{color}
\begin{document}
\newcommand{\z}{\emph{z} $\sim$ }
\newcommand{\kms}{km s$^{-1}$}
\makeatletter
\newcommand{\Rmnum}[1]{\expandafter\@slowromancap\romannumeral #1@}
\makeatother

\title{The Properties and Prevalence of Galactic Outflows at \z1 in the Extended Groth Strip\altaffilmark{1}}
\author{\sc Katherine A. Kornei and Alice E. Shapley\altaffilmark{2}}
\affil{Department of Physics and Astronomy, University of California, Los Angeles, CA 90025, USA}
\author{\sc Crystal L. Martin}
\affil{Physics Department, University of California, Santa Barbara, CA 93106, USA}
\author{\sc Alison L. Coil\altaffilmark{3}}
\affil{Center for Astrophysics and Space Sciences, Department of Physics, University of California, San Diego, CA 92093, USA}
\author{\sc Jennifer M. Lotz}
\affil{Space Telescope Science Institute, Baltimore, MD 21218, USA}
\author{\sc David Schiminovich}
\affil{Department of Astronomy, Columbia University, New York, NY 10027, USA}
\author{\sc Kevin Bundy}
\affil{Kavli Institute for the Physics and Mathematics of the Universe \\ Todai Institutes for Advanced Study \\ University of Tokyo, Kashiwa, Japan 277-8583 (Kavli IPMU, WPI)}
%\author{and}
\author{\sc Kai G. Noeske}
\affil{Space Telescope Science Institute, Baltimore, MD 21218, USA}

\altaffiltext{1}{Based, in part, on data obtained at the W.M. Keck Observatory, which is operated as a scientific partnership among the California Institute of Technology, the University of California, and NASA, and was made possible by the generous financial support of the W.M. Keck Foundation.} 
\altaffiltext{2}{Packard Fellow.}
\altaffiltext{3}{Alfred P. Sloan Fellow.}

\begin{abstract}
We investigate galactic-scale outflowing winds in 72 star-forming galaxies at \z1 in the Extended Groth Strip. Galaxies were selected from the DEEP2 survey and follow-up LRIS spectroscopy was obtained covering Si~II, C~IV, Fe~II, Mg~II, and Mg~I lines in the rest-frame ultraviolet. Using  \emph{GALEX}, \emph{HST}, and \emph{Spitzer} imaging available for the Extended Groth Strip, we examine galaxies on a per-object basis in order to better understand both the prevalence of galactic outflows at \z1 and the star-forming and structural properties of objects experiencing outflows. Gas velocities, measured from the centroids of Fe~II interstellar absorption lines, are found to span the interval [--217, +155] \kms. We find that $\sim$ 40\% (10\%) of the sample exhibits blueshifted Fe~II lines at the 1$\sigma$ (3$\sigma$) level. We also measure maximal outflow velocities using the profiles of the Fe~II and Mg~II lines; we find that Mg~II frequently traces higher velocity gas than Fe~II. Using quantitative morphological parameters derived from the \emph{HST} imaging, we find that mergers are not a prerequisite for driving outflows. More face-on galaxies also show stronger winds than highly inclined systems, consistent with the canonical picture of winds emanating perpendicular to galactic disks. In light of clumpy galaxy morphologies, we develop a new physically-motivated technique for estimating areas corresponding to star formation. We use these area measurements in tandem with \emph{GALEX}-derived star-formation rates to calculate star-formation rate surface densities. At least 70\% of the sample exceeds a star-formation rate surface density of 0.1 M$_{\odot}$ yr$^{-1}$ kpc$^{-2}$, the threshold necessary for driving an outflow in local starbursts. At the same time, the outflow detection fraction of only 40\% in Fe~II absorption provides further evidence for an outflow geometry that is not spherically symmetric. We see a $\sim$ 3$\sigma$ trend between outflow velocity and star-formation rate surface density, but no significant trend between outflow velocity and star-formation rate. Higher resolution data are needed in order to test the scaling relations between outflow velocity and both star-formation rate and star-formation rate surface density predicted by theory. \end{abstract}

\section{Introduction}
Far from being closed boxes, galaxies are known to affect their environment by expelling gas and metals into the intergalactic medium (IGM) via ``superwinds" \citep[e.g.,][]{heckman1990,steidel1996,franx1997,martin1999,pettini2000,pettini2001,shapley2003,martin2005,veilleux2005,rupke2005,tremonti2007,weiner2009,steidel2010,coil2011}. These outflows may contribute to the limiting of black hole and spheroid growth \citep[possibly resulting in the correlation between black hole and bulge mass; e.g.,][]{ferrarese2000,robertson2006}, the enrichment of the IGM \citep{oppenheimer2006}, and the regulation of star formation through the ejection of cold gas \citep{scannapieco2005,croton2006}. At high redshifts, winds may have played a critical role in reionization by clearing paths for ionizing radiation to escape from galaxies \citep{dove2000,steidel2001,heckman2001,gnedin2008}. 

\begin{figure*}
\begin{center}$
\begin{array}{c} %match_info_histograms.sm
\includegraphics[width=6in]{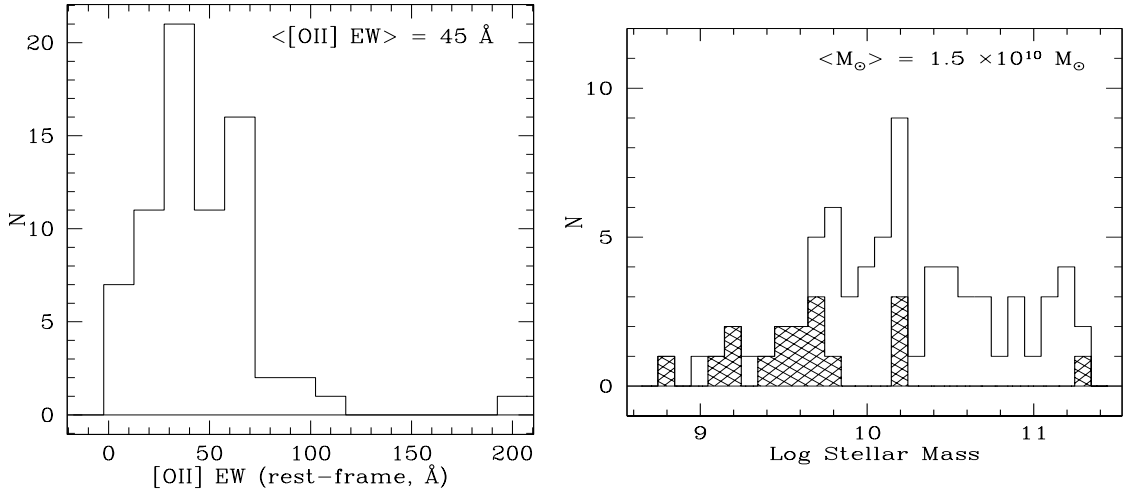}
\end{array}$
\end{center}
\caption{Distributions of rest-frame [O~II] equivalent widths (\emph{left}) and stellar masses (\emph{right}) for our sample of 72 EGS objects at 0.7 $<$ $z$ $<$ 1.3. The hatched histogram in the stellar mass panel denotes galaxies without $K$-band detections (17 objects). Stellar masses were estimated from SED fitting using \emph{BRI} (and \emph{K}, when available) photometry \citep{bundy2006}, assuming a Chabrier \citeyearpar{chabrier2003} IMF.} %Use DEEP2.EGS.z02.err.dat from Alison on 7/15/11 to plot background [OII] values?
\label{OII_mass_histograms}
\end{figure*}

The kinematics of interstellar absorption lines provide one of the key probes of outflowing gas in galaxies. Observations of blueshifted interstellar absorption lines at a variety of rest wavelengths and ionization states have been observed in both local and high-redshift samples \citep[e.g.,][]{heckman2000,martin2005,pettini2002,shapley2003,steidel2010,coil2011,kulas2011,law2012}. The presence of outflows with blueshifted velocities relative to stars and H~II regions appears to be associated with objects undergoing starbursts: UV-selected star-forming galaxies at \emph{z} = 2--3 with large star-formation rates (SFRs) \citep{pettini2002,shapley2003,steidel2010}, local ULIRGs \citep{heckman2000,rupke2002,martin2005,rupke2005}, and local dwarf starbursts \citep{lequeux1995,heckman1997,heckman2001b,schwartz2004}. Studies of X-ray and H$\alpha$ emission in local samples have focused on correlating outflows with spatially-resolved properties such as star-formation rate surface density ($\Sigma_{\rm SFR}$), morphology, and galaxy inclination \citep[e.g.,][]{lehnert1996,heckman2002,strickland2004}. Only recently, however, have absorption-line studies of winds at higher redshifts begun similar investigations \citep{sato2009,weiner2009,rubin2010a,steidel2010,talia2012,law2012}. Studies of spatially-resolved properties necessitate high-resolution imaging and careful measurement of the galactic regions in which stars are forming. These studies are particularly relevant around \z1; examining the processes regulating star formation is critical to understanding why the global star-formation rate density declines between \z1 and \z0 \citep{reddy2008}. 

The study of outflows in \z1 objects to date has relied primarily on composite spectra and visual classification of galaxy morphology \citep[e.g.,][]{weiner2009,rubin2010a}. Here, we use a sample of galaxies drawn from the \z1 DEEP2 redshift survey \citep{newman2012} to examine the relationship between outflows and host galaxy properties. For these objects, we obtain observations with the Low Resolution Imaging Spectrometer \citep[LRIS;][]{oke1995,steidel2004} on Keck \Rmnum{1}. These LRIS spectra cover a bluer wavelength range than the DEIMOS discovery spectra from the DEEP2 survey and are inclusive of a variety of rest-frame UV features from which wind kinematics can be measured (Si~II $\lambda$1526, C~IV $\lambda$1549, Fe~II $\lambda$1608, Al~II $\lambda$1670, Fe~II $\lambda$2344, Fe~II $\lambda \lambda$2374/2382, Fe~II $\lambda \lambda$2587/2600, Mg~II $\lambda \lambda$2796/2803, and Mg~I $\lambda$2852). In this paper, we focus specifically on objects in the Extended Groth Strip (EGS) field, where an extensive multiwavelength dataset enables SFR, $\Sigma_{\rm SFR}$, morphology, inclination, and dust attenuation to be estimated and in turn correlated with outflow properties on a per-object basis. We aim to determine which, if any, of these properties most strongly drives outflows at \z1. Furthermore, with our unique data set we can statistically probe the geometry of galactic winds. 

In Section \ref{sec: data}, we present the imaging and spectroscopic data used in our analysis, including extensive datasets of EGS observations. We discuss SFR, a new technique for estimating galaxy areas, and $\Sigma_{\rm SFR}$ in Section \ref{sec: sfrsd}. Absorption-line modeling is reviewed in Section \ref{sec: modeling}, while Section \ref{sec: results} is devoted to results. In Section \ref{sec: discussion}, we present a discussion of the results. Our conclusions are summarized in Section \ref{sec: conclusions}. Throughout the paper, we assume a standard $\Lambda$CDM cosmology with \emph{H}$_{\rm 0}$ = 70 km s$^{-1}$ Mpc$^{-1}$, $\Omega_{\rm M}$ = 0.3, and $\Omega_{\Lambda}$ = 0.7. All wavelengths are measured in vacuum. At \emph{z} = 0.7 (1.3), an angular size of 1$''$ corresponds to 7.1 (8.4) kpc. 

\section{Sample and Observations} \label{sec: data}

\subsection{DEEP2 Survey}
The objects presented here were drawn from the DEEP2 survey conducted from 2002--2005 using the DEep Imaging Multi-Object Spectrograph (DEIMOS) on Keck \Rmnum{2} \citep{newman2012}. With four fields totaling three square degrees, the DEEP2 survey obtained high-quality redshifts for $\sim$ 30,000 galaxies at 0.7 $\le$ $z$ $\le$ 1.5 brighter than $R_{\rm AB}$ = 24.1 to study clustering and evolution in galactic properties out to \z1. 

The DEIMOS spectra were obtained with a 1200 lines mm$^{-1}$ grating and a 1.$''$0 slit and have a resolution of \emph{R} $\approx$ 5000 \citep{newman2012}. Wavelength coverage extends from $\sim$ 6500--9100 \AA\, inclusive of the [O~II] $\lambda \lambda$3727/3729 doublet, which is typically resolved. Galaxy properties, including luminosities, colors, stellar masses ($M_*$), and [O~II] equivalent widths, have been measured \citep[Figure \ref{OII_mass_histograms};][]{bundy2006,willmer2006}. Stellar masses were calculated from SED modeling with \emph{BRIK} photometry, assuming \citet{bruzual2003} spectral templates and a \citet{chabrier2003} initial mass function (IMF). For our study, modeling was done with \emph{BRI} photometry alone if objects lacked \emph{K}-band detections (24\% of the sample). 

Photometric preselection based on \emph{BRI} colors was applied in three out of the four DEEP2 fields in order to isolate galaxies at $z$ $\ge$ 0.7. In the fourth field, the EGS, no color cuts were applied and galaxies at lower redshifts were accordingly targeted for spectroscopy. The EGS dataset is unique not only for its inclusion of more local galaxies, but also for its extensive multi-wavelength coverage from the All-Wavelength Extended Groth Strip International Survey \citep[AEGIS;][]{davis2007}. The objects presented in this paper are all in the EGS and consequently have, in addition to the original DEIMOS spectra, a variety of complementary data spanning from the X-ray to the radio regimes (Section \ref{sec: multiwavelength}). 

\subsection{LRIS Observations}
While the DEIMOS spectra are generally dominated by nebular emission features ([O~II], [Ne~III], H$\gamma$, H$\delta$, H$\beta$, [O~III]), the majority of low- and high-ionization interstellar absorption features tracing outflows are in the rest-frame UV and are observed at shorter wavelengths than the blue edge of the typical DEIMOS spectra in the DEEP2 survey ($\sim$ 6500 \AA\ in the observed frame). In order to probe these outflow features (e.g., Fe~II $\lambda$2344, Fe~II $\lambda \lambda$2374/2382, Fe~II $\lambda \lambda$2587/2600, Mg~II $\lambda \lambda$2796/2803), we obtained complementary spectroscopic data for 212 objects using the LRIS spectrograph on Keck \Rmnum{1}. These observations, covering all four DEEP2 survey fields, are described in further detail in \citep{martin2012}.

%LRIS selection info: /Users/aes/deep2\_lris/lris/masks/sample\_summary.egs.dat. 31/72 obj. in primary redshift range; also selected on basis of [O\Rmnum{2}] at lower redshift, pairs (10\% of sample), and green valley.

\begin{figure}
\centering
\includegraphics[width=3.5in]{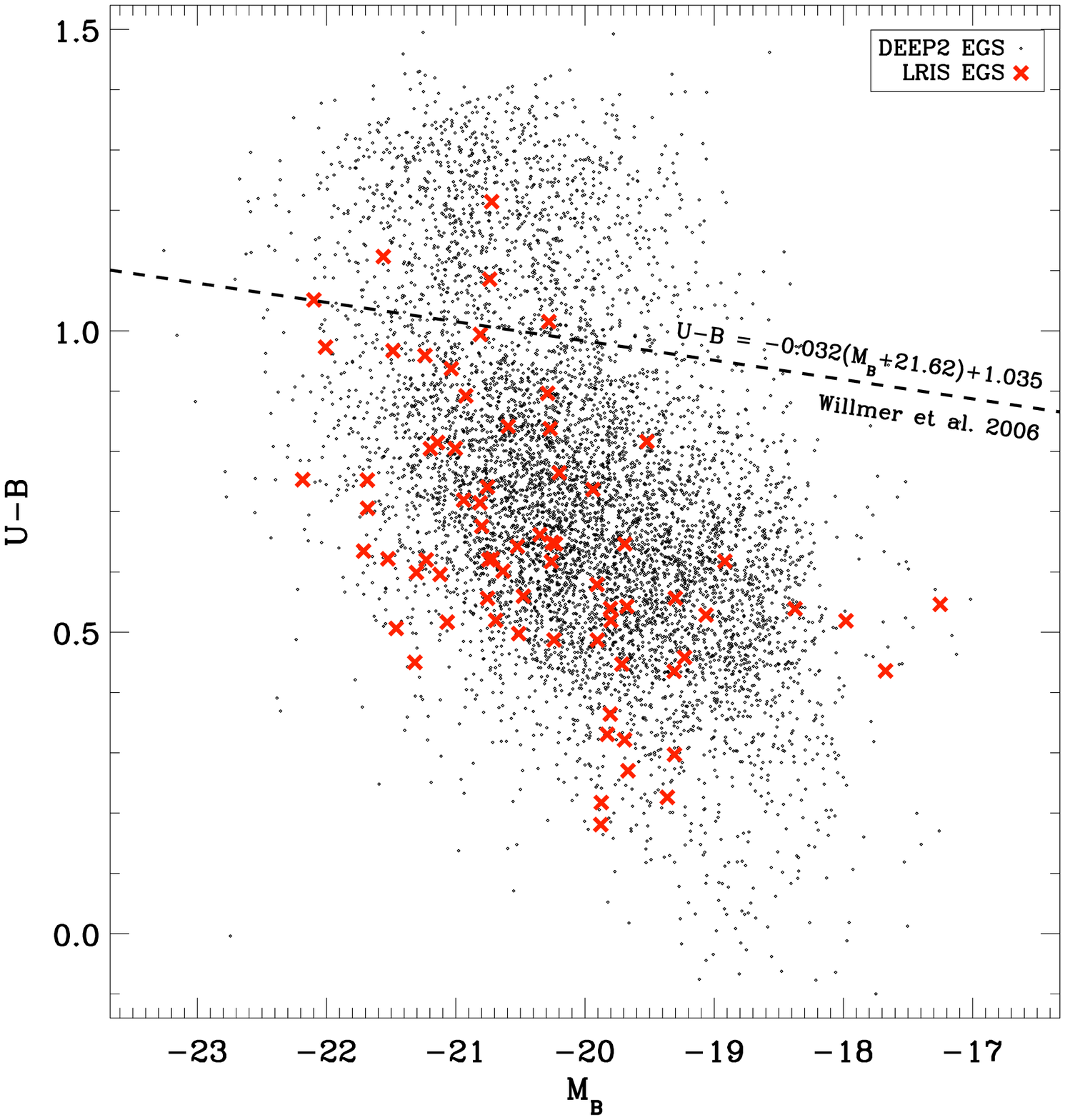} 
\caption{Rest-frame color-magnitude diagram with photometry corrected for Galactic extinction, assuming $H_{\rm 0}$ = 100 km s$^{-1}$ Mpc$^{-1}$ \citep{willmer2006}. Small dots show objects at 0.70 $<$ $z$ $<$ 1.35 in the EGS field of the DEEP2 survey and red crosses mark the 72 EGS objects with LRIS follow-up discussed in this paper. The red sequence and blue cloud are delineated by the dashed line \citep{willmer2006}; we emphasize that the majority of the objects presented in this work are brighter, blue cloud galaxies.}
\label{CMD}
\end{figure}

\begin{figure}
\centering
\includegraphics[width=3.5in]{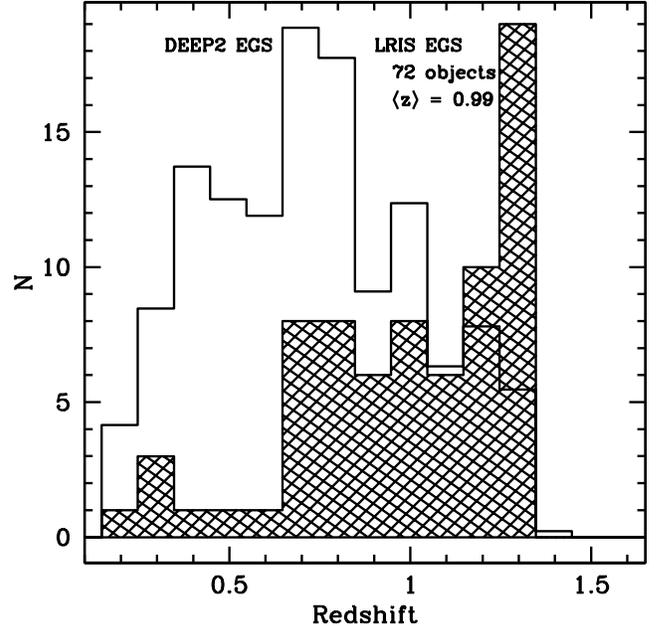} 
\caption{The redshift distribution of the LRIS EGS sample (hatched histogram), compared with a normalized histogram of redshifts of EGS objects in the parent DEEP2 survey. Objects at \emph{z} $>$ 0.7 were prioritized for LRIS follow-up, given the target wavelength coverage of 1500--2800 \AA\ in the rest-frame. The significant fraction of objects at larger redshifts reflects our heightened prioritization of galaxies at 1.19 $\le$ $z$ $\le$ 1.35. Observations of galaxies in this redshift interval yield simultaneous coverage of the C~IV $\lambda$1549 and Mg~I $\lambda$2852 features for the typical LRIS set-up, enabling a comparison of the kinematics of neutral, low-, and high-ionization species (Shapley et al., in prep.).}
\label{redshift}
\end{figure}

In this paper, we discuss 72 objects in the EGS, which are drawn from our larger sample of 212 spectroscopically-confirmed DEEP2 objects with LRIS follow-up \citep{martin2012}. In Figure \ref{CMD}, we show the color-magnitude diagram for both the present sample of 72 objects and also the parent sample of $\sim$ 7,000 spectroscopically-confirmed DEEP2 EGS objects at 0.70 $<$ \emph{z} $<$ 1.35. Objects with \emph{B} $<$ 24.5 were targeted for LRIS observations, resulting in a sample dominated by brighter, bluer galaxies. The redshift distribution of our EGS sample is presented in Figure \ref{redshift}, where $\langle z \rangle$ = 0.99 $\pm$ 0.29. 
%77 obj at 1.19 < z < 1.35
%20 obj are GV
%52 obj have OII > 10
%41 obj have OII > 20
%of 12 obj with OII < 20, 7 have OII > 10: 12015320, 12019542, 12019674, 22029066, 22036912, 22036975, 32017258

The LRIS data, collected in October 2007, June 2008, September 2008, and June 2009, were obtained using 1$.''$2 slits on nine multi-object slitmasks targeting 20--28 objects each. The dichroic capability of LRIS was employed with the grism on the blue side and the grating on the red side. We used two set-up configurations, both with the Atmospheric Dispersion Corrector: the d680 dichroic with the 400 line mm$^{-1}$ grism and the 800 line mm$^{-1}$ grating (145 objects; 47 in the EGS) and the d560 dichroic with the 600 line mm$^{-1}$ grism and the 600 line mm$^{-1}$ grating (67 objects; 25 in the EGS). The resolutions of the 800, 600 and 400 line mm$^{-1}$ grisms/gratings are $R$ = 2000, 1100, and 700, respectively. Features bluer than Mg~II $\lambda \lambda$2796/2803 generally fell on the blue side while Mg~II $\lambda \lambda$2796/2803 and longer wavelength lines (e.g., [O~II]) were recorded on the red side. 

Continuum signal-to-noise (S/N) ratios ranged from $\sim$ 1--20 per pixel over the rest wavelength interval 2400--2500 \AA. In Figure \ref{composite_blue_EGS_smooth}, we show several individual LRIS spectra with a range of measured S/N ratios. Integration times varied from 3--9 hours per slitmask, where objects observed with the d560 dichroic had typically shorter exposures (3--5 hours) than objects observed with the d680 dichroic (5--9 hours). The reduction procedure -- flat-fielding, cosmic ray rejection, background subtraction, extraction, wavelength and flux calibration, and transformation to the vacuum wavelength frame -- was completed using {\tt IRAF} scripts. The spectra were continuum-normalized and composite spectra were assembled from stacks of mean-combined rest-frame spectra. For the composite spectra, we smoothed the objects obtained with the 600 line mm$^{-1}$ grism or grating in order to account for the difference in resolution between the objects observed with the 600 line mm$^{-1}$ and 400 line mm$^{-1}$ setups. 

\subsection{AEGIS Multiwavelength Data} \label{sec: multiwavelength}

The 72 EGS objects presented in this paper have extensive multi-wavelength coverage from the AEGIS dataset. These data products cover observations over nine decades in wavelength, including \emph{Chandra} X-ray, \emph{GALEX} FUV and NUV imaging, \emph{HST} ACS F606W ($V$) and F814W ($I$) imaging, optical CFHT and MMT observations, Palomar/WIRC \emph{J} and $K_s$ imaging, \emph{Spitzer} IRAC and MIPS pointings, and VLA radio observations. This wealth of data permits many analyses, including the estimation of SFR, specific star-formation rate (sSFR = SFR/$M_*$), $\Sigma_{\rm SFR}$, morphology, inclination, and dust attenuation. Here, we discuss the diffraction-limited \emph{HST} Advanced Camera for Surveys \citep[ACS;][]{ford2003} imaging that forms the backbone of our spatially-resolved analyses. In the sample of 72 EGS objects presented here, 56 galaxies have $HST$ coverage. 

Reduced $HST$ images (Cycle 13, Program 10134, PI Marc Davis) were drizzled from four exposures to produce final 8000 $\times$ 8000 pixel images with a sampling of 0.$''$03 pixel$^{-1}$ and a point spread function (PSF) FWHM of $\sim$ 0.$''$1. These deep ($V$ = 28.75 AB [5$\sigma$]; $I$ = 28.10 AB [5$\sigma$]) images were used for several estimates of galaxy extent and light distribution, including the Petrosian radius, the Gini coefficient, and $M_{20}$. The Gini coefficient originated from the field of economics to describe the distribution of wealth in a society; the term is used in astronomy to parameterize how a galaxy's light is distributed. Gini coefficients close to 1 indicate that a galaxy's flux is concentrated in a few bright pixels while values closer to 0 mean that the flux is more evenly distributed over many pixels. $M_{20}$ refers to the normalized second-order moment of the brightest 20\% of a galaxy's flux; larger $M_{20}$ values are associated with merging systems while smaller $M_{20}$ values describe more compact galaxies. We direct the reader to \citet{lotz2004} for a more complete description of the Gini coefficient and $M_{20}$. 

\section{Star-Formation Rates, Galaxy Areas, and Star-Formation Rate Surface Densities} \label{sec: sfrsd}

The multi-wavelength observations of the AEGIS dataset enable detailed measurements of the star-forming properties of the sample. Based on \emph{HST} imaging, we note that the majority of objects appear to have clumpy morphologies (Figure \ref{HST1}). The UV-bright clumps may be star-forming regions embedded in lower surface brightness disks \citep{forsterschreiber2006}. We consequently derive a new estimate of galaxy size inclusive of only the brightest regions likely associated with star formation and use these areas in tandem with SFR estimates in order to calculate $\Sigma_{\rm SFR}$. We also use standard Petrosian size measurements to calculate $\Sigma_{\rm SFR}$ as well. 

\subsection{Star-Formation Rates}

\begin{figure}
\centering
\includegraphics[width=3.5in]{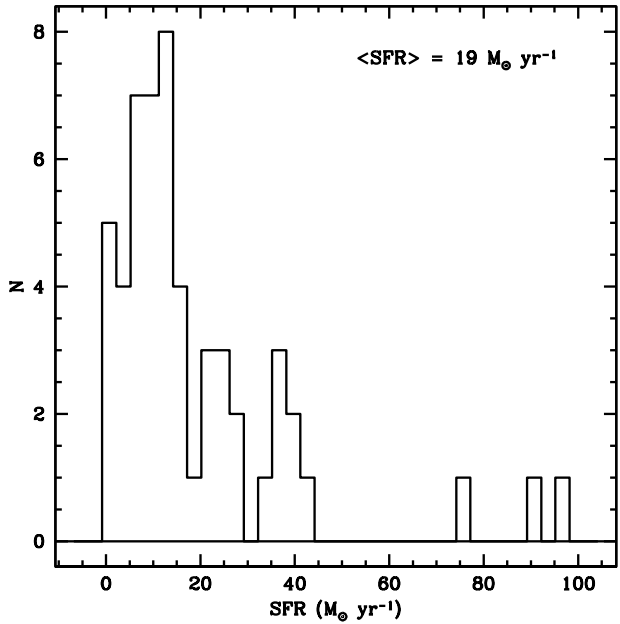} 
\caption{Histogram of SFRs, assuming a Chabrier \citeyearpar{chabrier2003} IMF. The sample is characterized by a mean value of $\langle$SFR$\rangle$ = 19 $M_{\odot}$ yr$^{-1}$.}
\label{match_info_histograms_SFR_EGS}
\end{figure}

We used \emph{Galaxy Evolution Explorer} \citep[\emph{GALEX};][]{martin_galex2005} and \emph{Spitzer Space Telescope} imaging from the AEGIS dataset to estimate the SFRs of the sample. Given the significant attenuation experienced by UV radiation, SFRs inferred from \emph{GALEX} measurements must be corrected for dust absorption. Data from \emph{GALEX}'s FUV and NUV detectors, along with \emph{B}-band observations (for the higher-redshift objects in the sample), were used to calculate a spectral slope, $\beta$, where the flux level over the rest-frame interval 1250--2500 \AA\ is parametrized as $f_{\lambda}$ $\propto$ $\lambda^{\beta}$. Following \citet{meurer1999}, \citet{seibert2005} derived a relationship between $\beta$ and the UV attenuation, A$_{\rm UV}$, based on a sample of several hundred nearby galaxies with both \emph{GALEX} and \emph{Infrared Astronomical Satellite} \citep{neugebauer1984} imaging. For the sample presented here, \citet{schiminovich2007} used the $\beta$/A$_{\rm UV}$ relationship from \citet{seibert2005} to correct \emph{GALEX} luminosities for attenuation. The median A$_{\rm UV}$ value in the sample is 1.8, which corresponds to a factor of $\sim$ 5 correction. UV SFRs corrected for dust -- SFR$_{\rm UV,corr}$ -- were then calculated for a Salpeter \citeyearpar{salpeter1955} IMF over 0.1--100 $M_{\odot}$, according to \citet{salim2007}. We converted all values to the Chabrier \citeyearpar{chabrier2003} IMF by dividing the Salpeter \citeyearpar{salpeter1955} values by 1.8. 54/72 objects ($\sim$ 75\%) have SFR$_{\rm UV,corr}$ measurements, where the 18 objects lacking $\beta$ measurements were observed by \emph{GALEX} but were either not significantly detected or suffered from confusion with neighboring objects\footnote{\emph{GALEX} has a PSF of $\sim$ 5$''$--6$''$ for FUV and NUV observations \citep{martin_galex2005}.}. The SFR$_{\rm UV,corr}$ measurements of the sample are characterized by a median value of 12 M$_{\odot}$ yr$^{-1}$ and extrema of [1, 97] M$_{\odot}$ yr$^{-1}$ (Figure \ref{match_info_histograms_SFR_EGS}). 

\begin{figure}
\centering
\includegraphics[width=3.5in]{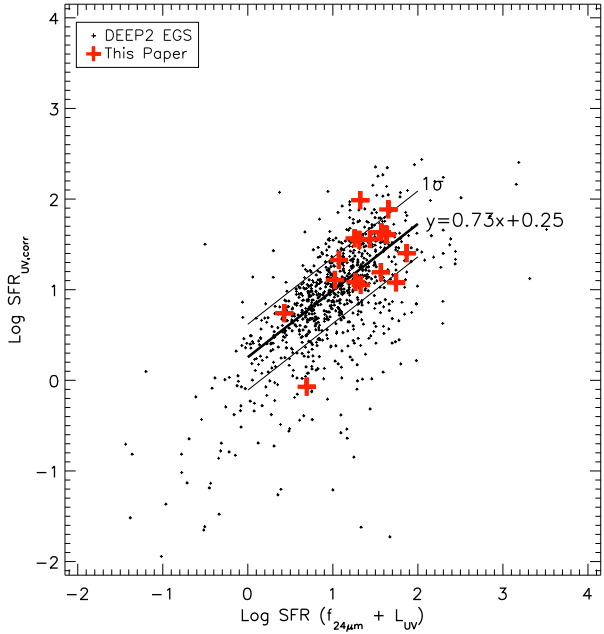} 
\caption{Comparison of two SFR indicators: a UV-corrected SFR (SFR$_{\rm UV,corr}$, where the dust correction is inferred from UV colors) and a SFR inferred from the sum of 24$\mu$m data and uncorrected UV observations: SFR ($f_{24\mu m}$ + L$_{\rm UV}$). The units of SFR are $M_{\odot}$ yr$^{-1}$. The bold crosses indicate the 17 EGS objects in this paper with robust 24$\mu$m data  and the smaller crosses mark objects in the larger DEEP2 EGS parent sample.  Data in the interval 1 $<$ SFR ($f_{24\mu m}$ + L$_{\rm UV}$) $<$ 100, the approximate range of SFRs measured in this paper, have been used to calculate the best-fit line (y = 0.73x + 0.25). The good agreement between SFR$_{\rm UV,corr}$ and SFR ($f_{24\mu m}$ + L$_{\rm UV}$), $\sigma$ $\sim$ 0.4 dex, supports adopting SFR$_{\rm UV,corr}$ given the relatively small number of objects with 24$\mu$m data.}
\label{SFR_galex_0609_SFRs}
\end{figure}

These SFR$_{\rm UV,corr}$ values are consistent with SFRs inferred from longer wavelength measurements. Robust \emph{Spitzer} 24$\mu$m detections ($f_{\rm 24\mu m}$ $\ge$ 60 $\mu$Jy) were used to calculate a total infrared luminosity, L$_{\rm IR}$, based on templates relating L$_{\rm IR}$ and $f_{\rm 24\mu m}$ \citep{chary2001}. While these templates are representative only of quiescent and starbursting galaxies -- not Active Galactic Nuclei (AGN) -- we note that AGN are likely a minimal contaminant in the sample given that a cross-correlation with \emph{Chandra} X-ray catalogs yielded only 1 match for our sample above the limiting fluxes of 5.3 $\times$ 10$^{-17}$ (3.8 $\times$ 10$^{-16}$) erg s$^{-1}$ cm$^{-2}$ in the soft (hard) bands \citep{laird2009}. Here, soft corresponds to 0.5--2 keV and hard corresponds to 2--10 keV. At the mean redshift of our sample, $\langle z \rangle$ = 0.99, these flux limits correspond to AGN luminosities of 2.7 $\times$ 10$^{41}$ (1.9 $\times$ 10$^{42}$) erg s$^{-1}$ in the soft (hard) bands. Based on bolometric corrections from \citet{hopkins2007}, the limits for both \emph{Chandra} bands correspond to $\sim$ 10$^{43}$ erg s$^{-1}$ in bolometric luminosity\footnote{As we do not include the one X-ray detected object (12015320) in our analyses, we believe that the winds we observe are driven by star formation as opposed to AGN activity. Based on rest-frame UV spectra, the objects in our sample do not exhibit signatures of AGN activity although we acknowledge possible contributions from obscured AGN. \citet{krug2010} note that starburst and Seyfert 2 systems show similar outflow kinematics.}. A SFR was then estimated from the addition of SFRs derived from L$_{\rm IR}$ \citep{kennicutt1998} and L$_{\rm UV}$ \citep{schiminovich2007}, where the latter term is a UV luminosity uncorrected for dust. While only 17/72 objects presented here ($\sim$ 25\%) have both UV detections and $f_{\rm 24\mu m}$ $\ge$ 60 $\mu$Jy, a larger sample of objects with these measurements shows a correlation between the two SFRs with a scatter of $\sim$ 0.4 dex (Figure \ref{SFR_galex_0609_SFRs}). In light of this consistency and the relative scarcity of objects with $f_{\rm 24\mu m}$ detections, we use SFR$_{\rm UV,corr}$ (``SFR") in all of the following analyses. 

It is useful to compare the SFRs of the present sample with the SFRs of other recent outflow studies at \z1. \citet{weiner2009} presented an outflow survey of DEEP2 objects at \z1.4 with 25$^{\rm th}$ and 75$^{\rm th}$ percentile values of SFR of 14 and 28 M$_{\odot}$ yr$^{-1}$, respectively, derived from UV measurements assuming a Kroupa \citeyearpar{kroupa2001} IMF. Work by \citet{rubin2010a} utilized objects with smaller SFRs: 25$^{\rm th}$ and 75$^{\rm th}$ percentile values of SFR of 4 and 9 M$_{\odot}$ yr$^{-1}$, respectively. The \citet{rubin2010a} SFRs were also derived assuming a Kroupa \citeyearpar{kroupa2001} IMF using UV measurements. The objects in this paper have 25$^{\rm th}$ and 75$^{\rm th}$ percentile values of 6 and 25 M$_{\odot}$ yr$^{-1}$, respectively, assuming a Chabrier \citeyearpar{chabrier2003} IMF. For direct comparison with the \citet{weiner2009} and \citet{rubin2010a} samples, these values correspond to 8 and 31 M$_{\odot}$ yr$^{-1}$, respectively, for a Kroupa \citeyearpar{kroupa2001} IMF.

In Figure \ref{Mstar_SFR}, we plot SFR versus stellar mass and sSFR versus stellar mass. We find a pronounced correlation between SFR and stellar mass. As this relationship is consistent with the correlation observed in the larger DEEP2 sample, the subsample of objects discussed here do not appear to have star-formation histories grossly inconsistent with other datasets.

\subsection{Calculating a ``Clump Area"} \label{sec: clump}

It has been suggested that there exists a threshold star-formation rate surface density of $\Sigma_{\rm SFR}$ = 0.1 M$_{\odot}$ yr$^{-1}$ kpc$^{-2}$ for driving an outflow in local starbursts \citep{heckman2002}. As estimating $\Sigma_{\rm SFR}$ requires areal information, it is important to characterize properly the extent over which star formation is occuring. While values in the literature often refer to $r$ = half-light radius or Petrosian radius \citep{rubin2010a,steidel2010}, the clumpy morphologies of high-redshift galaxies observed in the rest-frame UV necessitate a different treatment. \citet{rubin2010a} noted that the simple approximation of $\pi r^2$ for the area of a galaxy may result in an overestimate; these authors proposed that the inclusion of only star-forming knots above a specified luminosity threshold may be the most physically motivated method for calculating a galaxy area associated with outflowing material. 

\begin{figure*}
\centering
\includegraphics[width=7in]{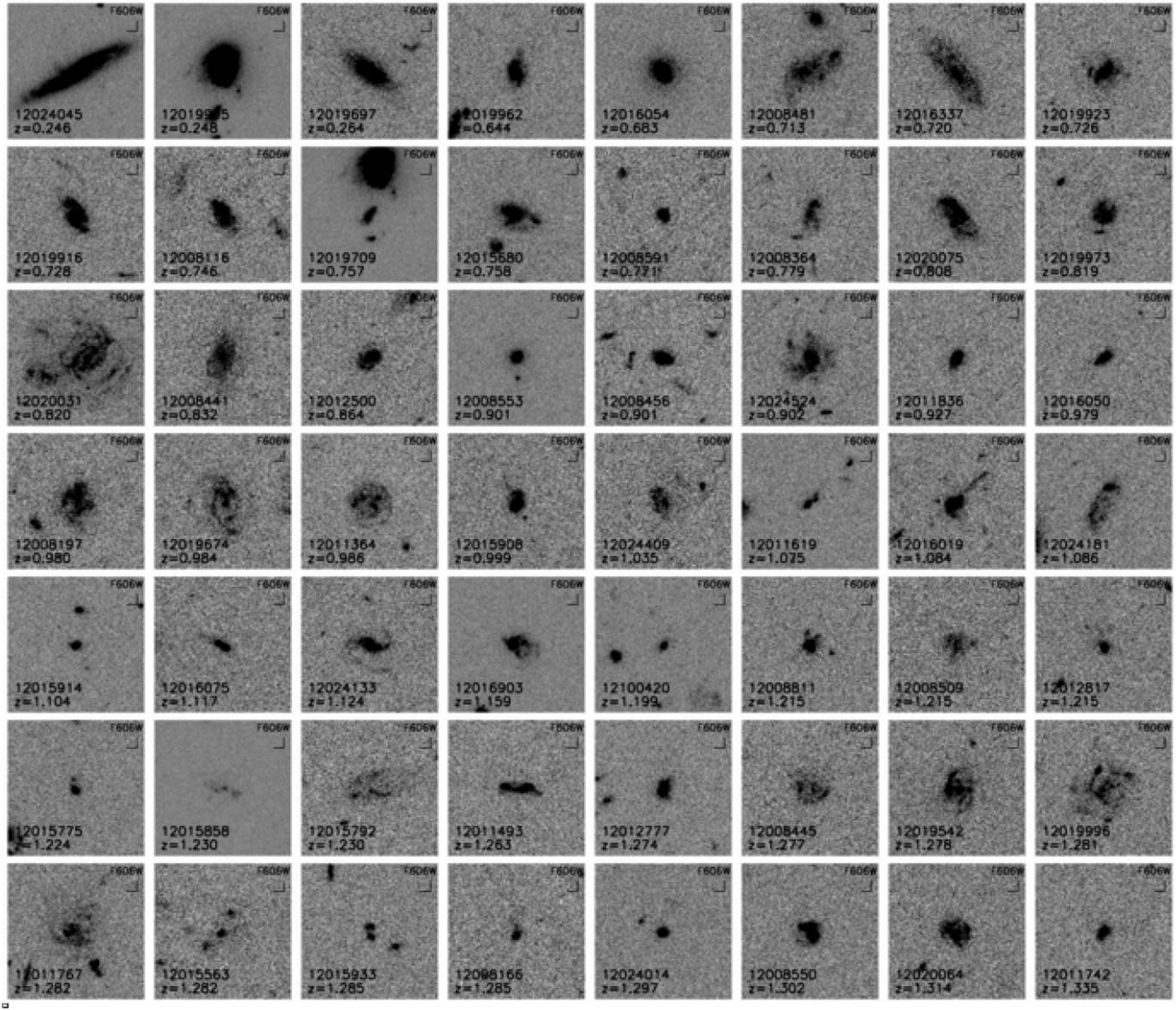} 
\caption{$V$-band thumbnails of the 56 objects with $HST$ imaging, ordered by ascending redshift. Each postage stamp is 6$''$ on a side and oriented with North up and East to the left. Note the clumpy morphologies and discrete star-forming clumps of many objects. At the mean redshift of the sample, $\langle$\emph{z}$\rangle$ = 0.99, the \emph{V}-band traces $\sim$ 3000 \AA\ in the rest-frame.}
\label{HST1}
\end{figure*}

As the majority of objects in our EGS sample appear to have irregular knots of star formation instead of being clearly disk-dominated (Figure \ref{HST1}), we used diffraction-limited $HST$ $V$-band imaging to investigate appropriate areal measurements. $V$-band imaging traces rest-frame $\sim$ 2970 \AA\ at \emph{z} = 0.99, the mean redshift of the sample. We began by estimating the total non-contiguous area of each galaxy above a certain surface brightness limit. The motivation behind this technique stemmed from flagging pixels that corresponded to some measurable physical quantity (in this case, surface brightness or, equivalently, $\Sigma_{\rm SFR}$). Schematically, the conversion from counts per pixel in the images to $\Sigma_{\rm SFR}$ can be illustrated as follows:

\begin{equation} \small{\label{eqn: heckman_criterion} {\frac{\rm \textbf{counts}}{\rm \textbf{pixel}}} \hspace{3mm} \overrightarrow{\rm via~zpt} \hspace{3mm} \frac{\textbf{f}_{\nu}}{\rm \textbf{pixel}} \hspace{3mm} \overrightarrow{\rm via~\emph{z}} \hspace{3mm}  \frac{\rm \textbf{L}_{\nu}}{\rm \textbf{kpc}^2} \hspace{3mm} \overrightarrow{\rm via~ K98} \hspace{3mm} \frac{\rm \textbf{SFR}}{\rm \textbf{kpc}^2}} \end{equation}
\nocite{kennicutt1998}

\begin{figure*}
\centering
\includegraphics[width=5in]{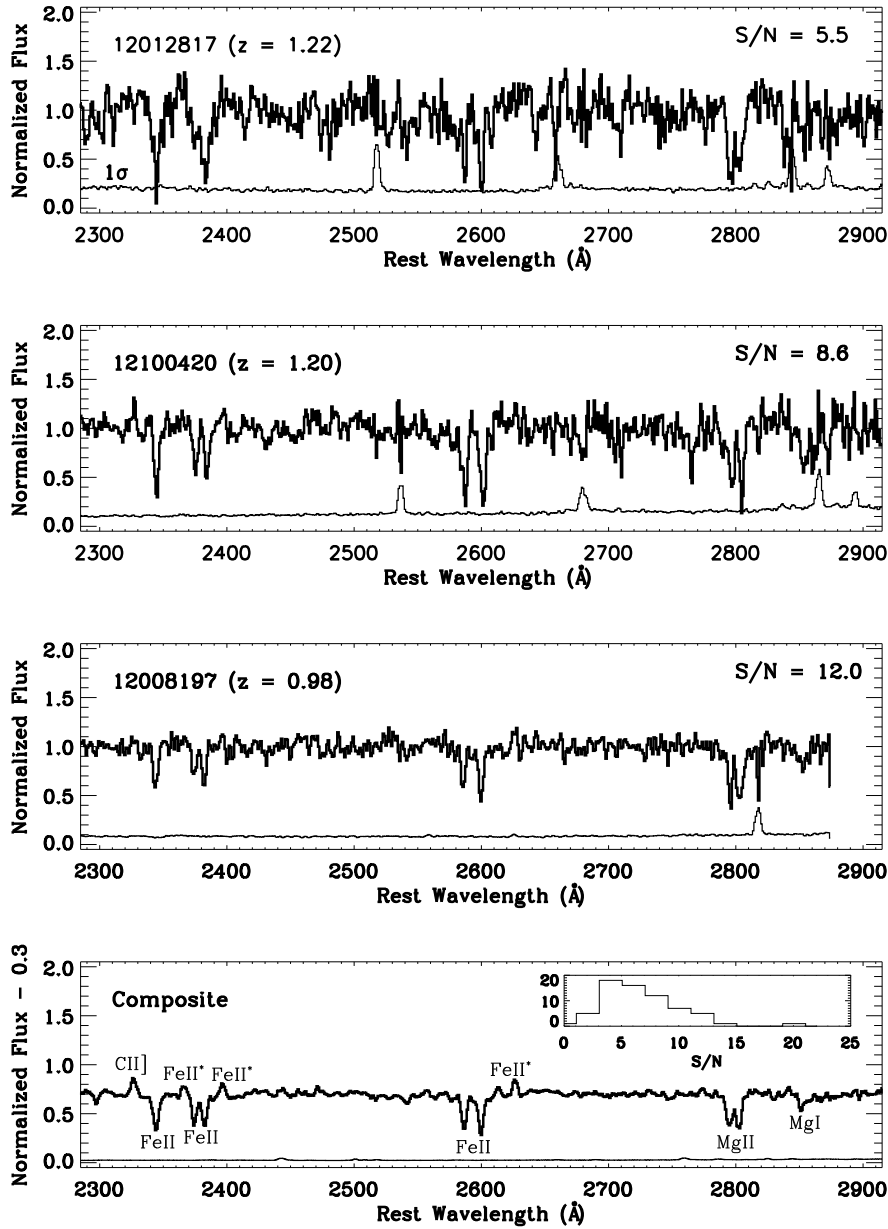} 
\caption{\emph{Top three panels}: Individual LRIS spectra exhibiting a range in continuum S/N. The 1$\sigma$ error spectrum is plotted for each object on the same scale. In each panel, the S/N per pixel measured over 2400--2500 \AA\ is indicated in the upper right-hand corner. We note that Fe~II and Mg~II absorption lines are detected in objects with a variety of S/N values. \emph{Bottom:} A composite spectrum assembled from all EGS spectra (S/N = 39.1 pixel$^{-1}$), with emission and absorption lines of interest labeled. The inset shows a histogram of continuum S/N ratios, where $\langle$S/N$\rangle$ $\sim$ 6.7 pixel$^{-1}$.}
\label{composite_blue_EGS_smooth}
\end{figure*}

\noindent where zpt is the $HST$ $V$-band zeropoint (26.5 AB) and K98 refers to the conversion between rest-frame UV luminosity over 1500--2800 \AA\ ($L_{\rm UV}$) and SFR assuming a Salpeter \citeyearpar{salpeter1955} IMF over 0.1--100 $M_{\odot}$ \citep[SFR = 1.4 $\times$ 10$^{-28}$ $L_{\rm UV}$;][]{kennicutt1998}. With images now in units of $\Sigma_{\rm SFR}$, a simple $\Sigma_{\rm SFR}$ threshold can be implemented. In light of local work by \citet{heckman2002}\footnote{As the \citet{heckman2002} threshold is only approximate, and was calculated assuming a Salpeter \citeyearpar{salpeter1955} IMF, we adopt the same IMF in converting from luminosity to SFR.}, we adopted the criterion $\Sigma_{\rm SFR}$ = 0.1 M$_{\odot}$ yr$^{-1}$ kpc$^{-2}$. While imposing this threshold produces areas that visually trace luminous galaxy clumps, we note two limitations to this methodology: 1) no correction for dust attenuation is applied, as estimates of the UV slope $\beta$ are available for only a subset of the sample and 2) we do not account for effects of galaxy inclination \emph{i}, given the uncertainties of estimating \emph{i} for clumpy objects with angular sizes $\sim$ 0.$''$5. Increased dust and higher \emph{i} will both act to reduce the measured luminosity.

While the unknowns of dust attention and inclination translate into uncertainties in the derived galaxy area (and consequently in $\Sigma_{\rm SFR}$), there is a more fundamental limitation to adopting the aforementioned methodology: the available $V$-band imaging ceases to trace below rest-frame 2800 \AA\ -- the edge of the window over which the \citet{kennicutt1998} conversion is valid -- for galaxies at lower redshifts ($z$ $<$ 1.10; 32 objects). For these comparably closer objects -- comprising $\sim$ 60\% of the sample -- the \citet{kennicutt1998} relation between $L_{\rm UV}$ and SFR cannot be applied. 

It was accordingly necessary to develop a new technique for estimating areas of the entire sample, irrespective of redshift. To this end, we focused on characterizing the areas of the higher redshift galaxies in the sample ($z$ $>$ 1.10; 24 objects) for which the $V$-band imaging \emph{does} trace rest-frame 1500--2800 \AA. For each of these objects, we noted the area calculated using the methodology described above and the threshold 0.1 M$_{\odot}$ yr$^{-1}$ kpc$^{-2}$. We then sought to parametrize these physically-motivated areas as containing a simple percentage (N) of the total flux within the Petrosian radius, R$_{\rm P}$. This technique enabled us to estimate an average value for N and then apply this value uniformly to both the higher and lower redshift galaxies.

We found that, with the exception of several diffuse disks and extremely compact spheroids, the majority of objects required N = [40--80]. The median of the sample was N = 74; we adopted this value as representative of the fractional flux within R$_{\rm P}$ that traced regions of intense star formation. For each galaxy with \emph{HST} imaging, we calculated the area derived by adding flux-ordered pixels, brightest first, until the enclosed flux was 74\% of the total flux within R$_{\rm P}$\footnote{Two objects were excluded from this analysis (12008364 and 12015792), given that their surface brightness levels were too low to measure R$_{\rm P}$.}. This method permits a single, systematic area calculation (A$_{74}$) for each galaxy independent of redshift while still being based on the physical grounding of a surface brightness threshold. We make the implicit assumption that N does not vary significantly over the range of look-back times probed by our sample.

\begin{figure*}
%\begin{center}$
%\begin{array}{c} 
%\includegraphics[width=1.3in]{12020031_clump.eps} %clip_area_do.pro 
%\includegraphics[width=1.3in]{12008441_clump.eps} 
%\includegraphics[width=1.3in]{12012500_clump.eps} 
%\includegraphics[width=1.3in]{12008553_clump.eps} \\
%\includegraphics[width=1.3in]{12008456_clump.eps} 
%\includegraphics[width=1.3in]{12024524_clump_clumparea.eps} 
%\includegraphics[width=1.3in]{12011836_clump.eps} 
%\includegraphics[width=1.3in]{12016050_clump.eps} \\
%\includegraphics[width=1.3in]{12008197_clump.eps} 
%\includegraphics[width=1.3in]{12019674_clump_clumparea.eps} 
%\includegraphics[width=1.3in]{12011364_clump_clumparea.eps} 
%\includegraphics[width=1.3in]{12015908_clump.eps} \\
%\includegraphics[trim = 0.5in 0.15in 0in 0in,clip,width=3in]{area_ratio.eps}  %area_ratio.pro
%\end{array}$
%\end{center}
\centering
\includegraphics[height=7in]{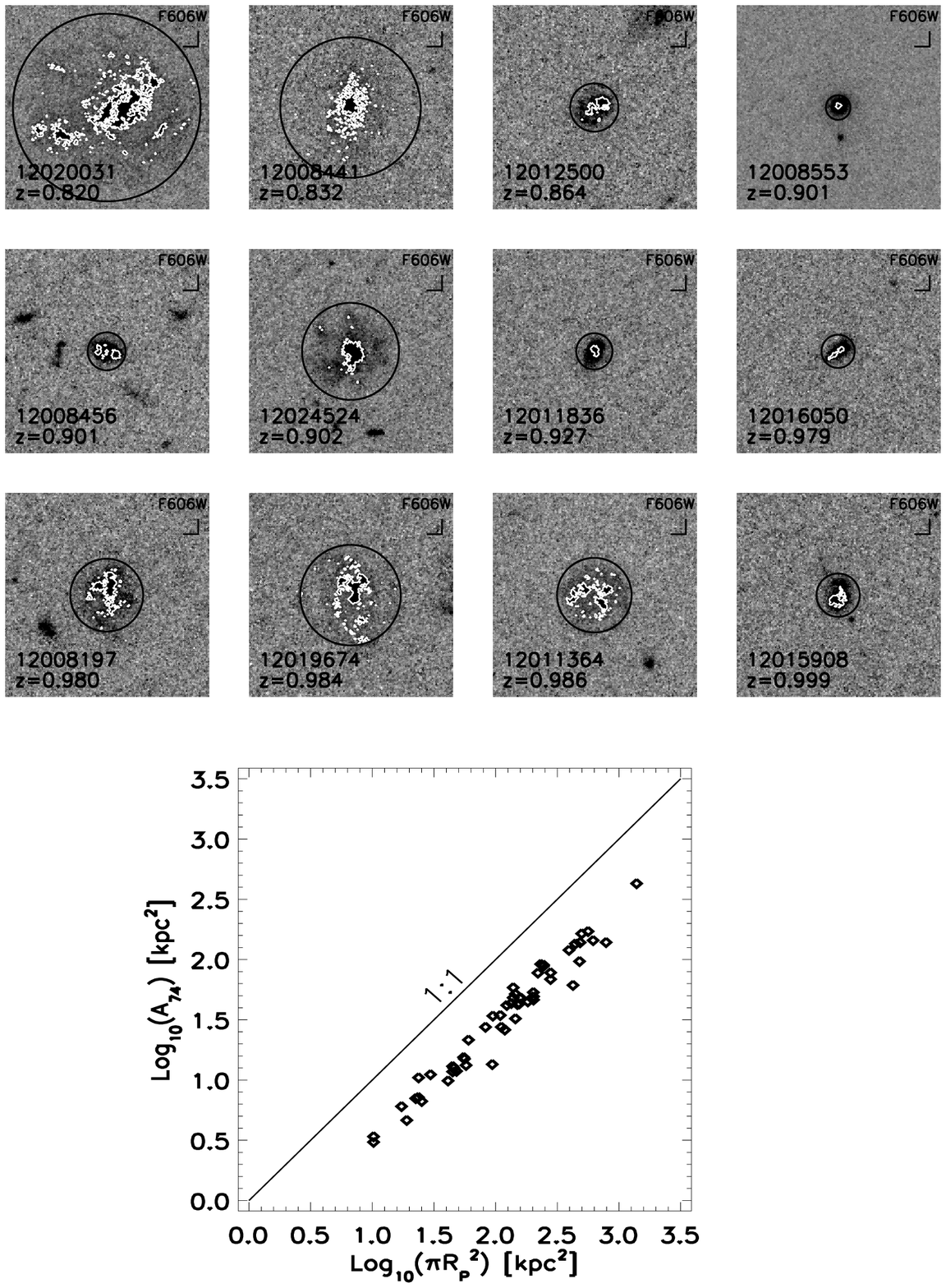} %/scripts/f9_inprogress.fig to edit using "Xfig"
\caption{$V$-band $HST$ thumbnails, 6$''$ on a side, are shown for a subset of 12 objects to illustrate the A$_{74}$ clump areas derived in Section \ref{sec: sfrsd}. These 12 objects were chosen as representing a diversity of clump areas, from compact, contiguous areas to more diffuse, segmented areas. Petrosian radii are plotted as black circles while the clump areas are shown as white contours. The ratio between the areas -- $\pi$R$_{\rm P}^2$/A$_{74}$ -- is indicated in the lower right of each thumbnail. The lower plot compares the $\pi$R$_{\rm P}^2$ and A$_{74}$ areas for the sample, where $\langle$$\pi$R$_{\rm P}^2$/A$_{74}$$\rangle$ $\sim$ 3.7. We note that the strong correlation between the Petrosian area and the clump area explains the similar results obtained with the two $\Sigma_{\rm SFR}$ measurements (Section \ref{sec: results}).}
\label{Contour1}
\end{figure*}

We find that areas calculated using this new technique are, on average, a factor of 3.7 smaller than the areas inferred from the $V$-band Petrosian radii: A$_{74}$ $\sim$ $\pi$R$_{\rm P}^2$/3.7. As seen in Figure \ref{Contour1}, A$_{74}$ appears to closely trace luminous clumpy regions\footnote{The \emph{HST} resolution, while excellent, is finite (FWHM $\sim$ 0.$''$1), so these areas represent upper limits.} while the Petrosian area consistently overestimates the area likely associated with star formation. We consequently favor the adoption of A$_{74}$ over $\pi$R$_{\rm P}^2$, although we use both area estimates below in tandem with derived SFRs in order to calculate $\Sigma_{\rm SFR}$. Given the strong correlation between A$_{74}$ and $\pi$R$_{\rm P}^2$, we note that our primary results remain unchanged regardless of which area estimate is adopted (Section \ref{sec: results}). While the initial motivation for calculating A$_{74}$ was to isolate an area measurement distinct from, and uncorrelated with, $\pi$R$_{\rm P}^2$, our finding that these two area estimates are in fact strongly linked is evidence that the structural morphologies of the galaxies in the sample are rather similar. If the light distributions of the galaxies were widely disparate, we would not expect that galaxies with larger Petrosian areas would also have larger A$_{74}$ areas while galaxies with smaller $\pi$R$_{\rm P}^2$ would likewise have smaller clump areas. We verified that the correlation between $\pi$R$_{\rm P}^2$ and clump area holds for a variety of N values. 

\subsection{Star-Formation Rate Surface Densities} \label{sec: sfrsd1}

With two estimates of galaxy area (A$_{74}$ and $\pi$R$_{\rm P}^2$) and one estimate of SFR, we calculated the star-formation rate surface densities of the sample. We define two $\Sigma_{\rm SFR}$ quantities, depending on which area measurement we adopt:

\begin{equation} \Sigma_{\rm SFR}\rm (A_{74}) =  SFR/A_{74} \end{equation}
\begin{equation} \Sigma_{\rm SFR}\rm (R_{\rm P}) =  SFR/\pi R_{\rm P}^2 \end{equation}

\begin{figure}
\centering
\includegraphics[width=3.5in]{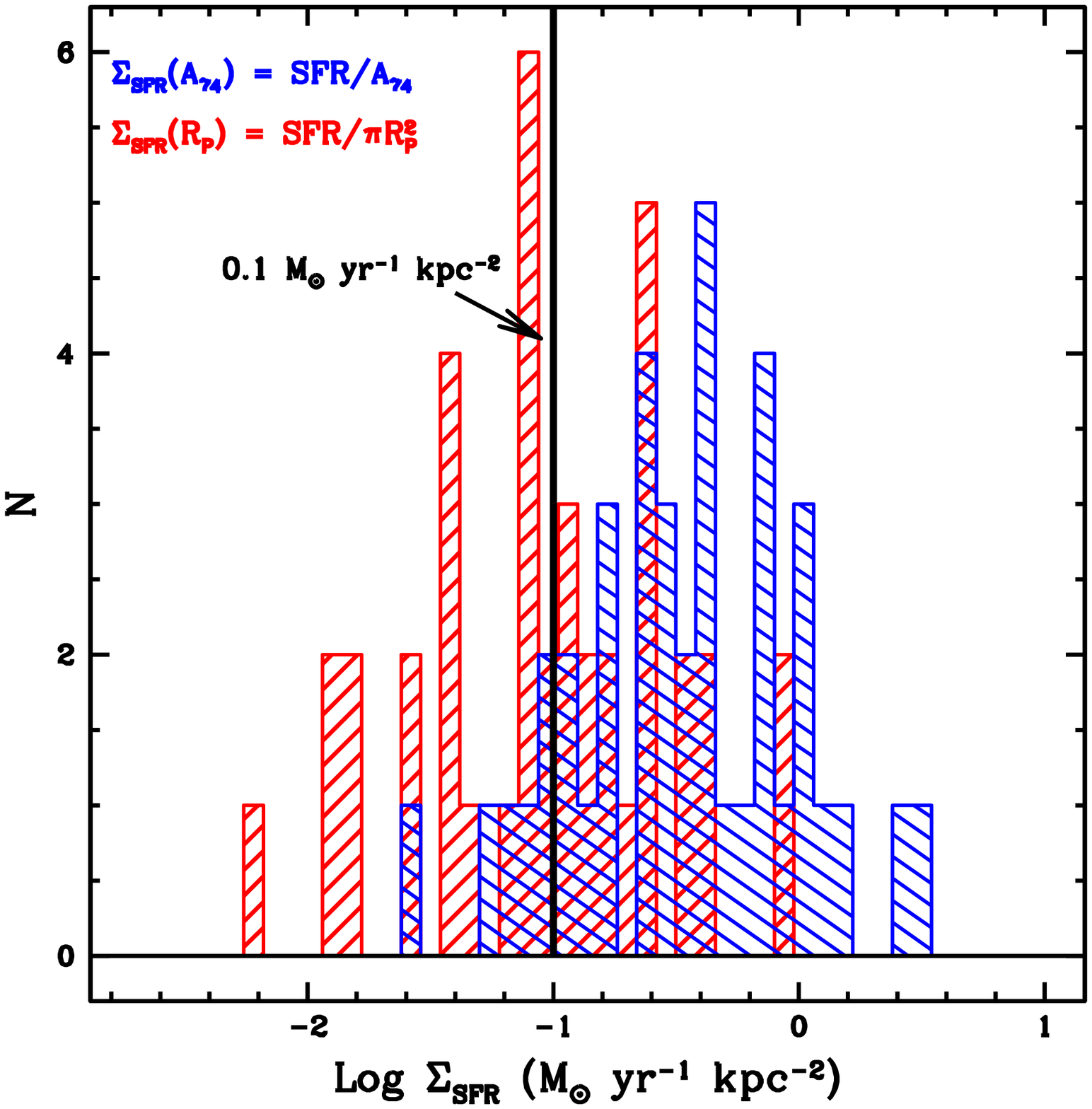}
\caption{Distributions of star-formation rate surface densities $\Sigma_{\rm SFR}$(A$_{74}$) and $\Sigma_{\rm SFR}$(R$_{\rm P}$), where these quantities are calculated using SFR and either A$_{74}$ or $\pi$R$_{\rm P}^2$. Greater than 70\% of objects have $\Sigma_{\rm SFR}$(R$_{\rm P}$) $>$ 0.1 M$_{\odot}$ yr$^{-1}$ kpc$^{-2}$, the local threshold for driving an outflow \citep{heckman2002}. The two objects with the highest $\Sigma_{\rm SFR}$(A$_{74}$) and $\Sigma_{\rm SFR}$(R$_{\rm P}$) -- 12015914 and 12100420 -- are both extremely compact galaxies with A$_{74}$ ($\pi$R$_{\rm P}^2$) areas of $\sim$ 3 (10) kpc$^2$.}
\label{sfrsd_histograms}
\end{figure}

Figure \ref{sfrsd_histograms} shows the distributions of $\Sigma_{\rm SFR}$(A$_{74}$) and $\Sigma_{\rm SFR}$(R$_{\rm P}$). Irrespective of the area measurement assumed, most objects have star-formation rate surface densities above the local threshold of 0.1 M$_{\odot}$ yr$^{-1}$ kpc$^{-2}$ thought to be necessary for driving an outflow \citep{heckman2002}. While this is by construction for $\Sigma_{\rm SFR}$(A$_{74}$), the fact that greater than 70\% of objects also have $\Sigma_{\rm SFR}$(R$_{\rm P}$) $>$ 0.1 M$_{\odot}$ yr$^{-1}$ kpc$^{-2}$ leads to the prediction that if a threshold $\Sigma_{\rm SFR}$ is the sole requirement for driving an outflow, the majority of objects ought to exhibit blueshifted Fe~II features. Of course, other factors such as local environment, halo mass, galaxy escape speed, and viewing angle, as well as the S/N of the data, may strongly influence the observed fraction of galaxies hosting winds. As discussed above, we do not correct A$_{74}$ areas for either dust or inclination effects. We find kinematic evidence for 1$\sigma$ significant outflows based on Fe~II absorption lines in 40\% of the sample (Section \ref{fitting}), although we note that the Mg~II features often exhibit strong blueshifts of their centroids and blue wings even when the Fe~II lines do not (Section \ref{sec: discussion}). 

\section{Modeling Absorption Lines} \label{sec: modeling}

\subsection{Systemic Redshift} \label{sec: sys_z}

Spectral analysis of outflowing gas requires an accurate determination of systemic redshift, $z_{\rm sys}$. Nebular emission lines such as [O~II] $\lambda \lambda$3726/3729, [O~III] $\lambda \lambda$4959/5007, and the Balmer series trace star forming regions and are consequently assumed to be at rest with respect to a galaxy's center of mass. While both the DEIMOS and LRIS spectral datasets are inclusive of nebular features, it is preferable to calculate $z_{\rm sys}$ from the LRIS spectra in order to ensure constancy in both slit position angle and spatial sampling for the $z_{\rm sys}$ and outflow measurements. We constructed a linear, S/N-weighted sum of three template spectra -- post-starburst, old stellar population absorption line, and emission line -- and used the DEEP2 {\tt IDL} pipeline to compare the template spectrum with each science spectrum. In the estimation of systemic redshifts, only lines with rest wavelengths longer than 3000 \AA\ were fit in order to exclude bluer features tracing winds. The $\chi^2$ of the fit was minimized according to redshift, and a best-fit $z_{\rm sys}$ was extracted. For 63/72 objects ($\sim$ 90\%), it was possible to extract $z_{\rm sys}$ from the LRIS spectra. For the nine objects in the sample without [O~II] spectral coverage in the LRIS data, $z_{\rm sys}$ was calculated from the DEIMOS spectra using the same technique. In comparing measurements of $z_{\rm sys}$ with those given for the DEIMOS data in the DEEP2 catalogs, we find a mean discrepancy and standard deviation of $\langle$$\delta z$$\rangle$ = 9 $\times$ 10$^{-5}$ (16 km s$^{-1}$) and $\sigma$$_{\delta z}$ = 2.6 $\times$ 10$^{-4}$ (42 km s$^{-1}$), respectively.  These results are consistent with the values obtained when considering the entire LRIS sample \citep{martin2012}. %See lris_z.all.dat from Alison's 4/23/10 email; these values are consistent with the values for the whole LRIS sample; see my 9/19/11 notes.

\subsection{Fitting Lines -- Fe~II Centroids} \label{fitting}

\begin{figure*}
%\begin{center}$
%\begin{array}{c}
%\includegraphics[width=6.5in]{Mstar_SFR.eps} \\ %SFR_galex_0609.pro
%\includegraphics[width=6.5in]{Mstar_SSFR.eps} \\
%\end{array}$
%\end{center}
\centering
\includegraphics[width=6.5in]{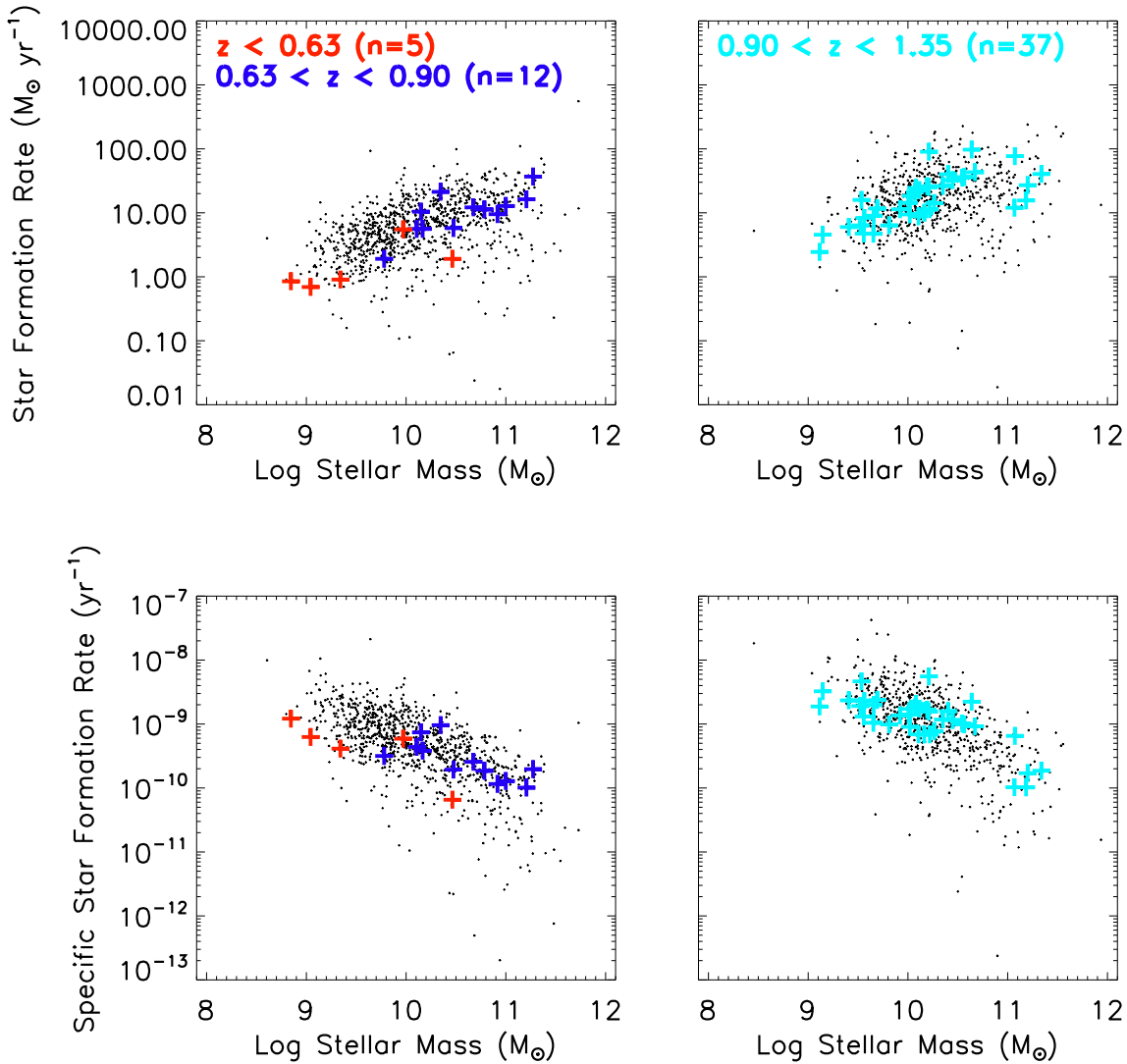}
\caption{\emph{Top left: }SFR versus stellar mass, where the small background points indicate EGS galaxies at 0.63 $<$ \emph{z} $<$ 0.90 without LRIS follow-up and colored crosses mark galaxies observed with LRIS, divided into two redshift bins. \emph{Top right:} Same as the plot on the left, for the redshift interval 0.90 $<$ $z$ $<$ 1.35. We note that the data presented in this paper have stellar masses and SFRs consistent with the parent DEEP2 sample. \emph{Bottom: }Stellar mass versus sSFR, where symbols are as above.}
\label{Mstar_SFR}
\end{figure*}

Simulations of galactic winds suggest that multiple fronts of outflowing material are present along any one sightline \citep[e.g.,][]{fujita2009}. While fitting several outflow components simultaneously would in principle provide the best diagnostic of these superimposed winds, the finite spectral resolution and S/N of observations typically limit fitting to a single wind component that is assumed to trace the integrated bulk motion of the outflow. The measurements we present here, which derive from and are discussed more fully in \citet{martin2012}, make the approximation of a single outflow velocity. Furthermore, we do not decompose the observed absorption profile into both a wind component and an absorption trough at the systemic velocity, where the latter may arise from stationary material present in stellar atmospheres or in the surrounding interstellar medium. Several recent outflow studies have modeled absorption lines as arising from the sum of wind and systemic components \citep{weiner2009,chen2010,coil2011}, although the implementation of such a technique is strongly dependent on the quality of the spectroscopic data. \citet{steidel2010} note that the limited resolution and finite S/N of typical spectroscopic observations at \emph{z} = 1--3 often preclude fitting a more complex model than a single absorption profile. In light of the S/N of our data, we do not apply a correction for material at the systemic velocity\footnote{As the majority of the sample is composed of actively star-forming galaxies, stellar absorption in the Fe~II and Mg~II lines is assumed to be minimal \citep[e.g.,][]{bruzual2003}. While objects with \emph{U}--\emph{B} colors between the red sequence and the blue cloud (i.e., ``Green Valley" galaxies) may experience substantial stellar Mg~II absorption, we have verified that our results are qualitatively unchanged when the ten Green Valley galaxies in our sample are omitted from analyses.}. The outflow velocities presented here are accordingly conservative lower limits \citep{martin2012}.

In \citet{martin2012}, we fit a single-component model simultaneously to five resonance Fe~II lines at 2249.88, 2260.78, 2344.21, 2374.46, and 2586.65 \AA\ in the LRIS spectra. These lines trace cool (T $<$ 10$^4$ K) gas. Upon absorbing a resonance photon at one of these five wavelengths, the Fe~II ion can decay by either emitting a photon of equal wavelength as the one it just absorbed (scattering) or emitting a photon to intermediate fine-structure levels (fluorescence). The presence of these fine-structure levels ensures that the resonance absorption lines are not simply filled in with scattered re-emitted resonance photons \citep[``emission filling";][]{prochaska2011}. The Fe~II features at 2382.76 and 2600.17 \AA\ are more susceptible to emission filling due to their dearth of alternate decay paths \citep[][Martin et al. 2012b, in prep.]{prochaska2011} and we purposefully omit these lines from fitting as emission filling can shift the measured centroid of absorption lines to bluer wavelengths \citep[e.g.,][]{prochaska2011}. The Mg~II features at $\lambda \lambda$2796/2803 are particularly affected by emission filling as these transitions have no fine-structure levels; in the absence of dust extinction, all absorbed resonance photons are re-emitted to the ground state. We accordingly do not measure the centroids of Mg~II absorption but instead develop a method to parameterize the blue wing of the absorption profile (Section \ref{vmax}). 

\begin{figure}
\centering
\includegraphics[width=3.5in]{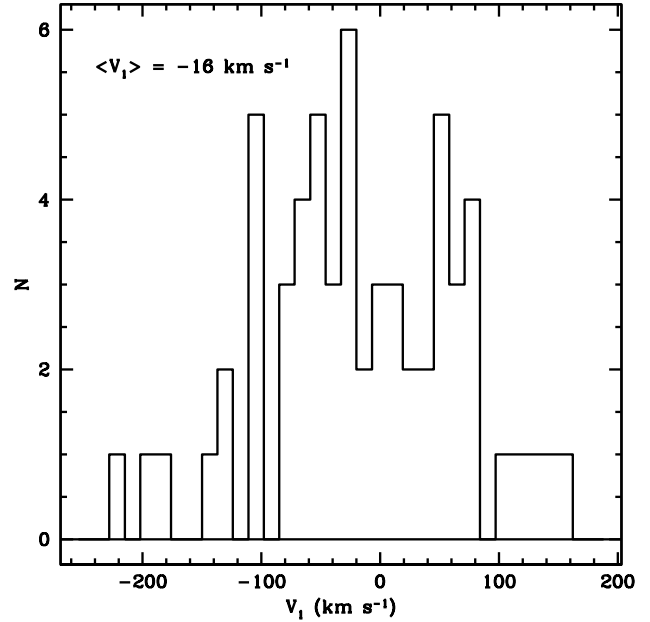}
\caption{Histogram of outflow velocities measured from fitting Fe~II interstellar absorption lines. Outflows (negative $V_1$ values) are seen at the 1$\sigma$ level in $\sim$ 40\% of the sample.  A non-negligible fraction of objects ($\sim$ 25\%) show positive Fe~II velocities at the 1$\sigma$ level. The sample as a whole is characterized by $\langle$$V_1$$\rangle$ $\sim$ --16 \kms, with a 1$\sigma$ scatter of 82 \kms.}
\label{match_info_histograms_vel_EGS}
\end{figure}

\begin{figure}
\centering
\includegraphics[trim=5in 0in 0in 0 in,clip,width=3.5in]{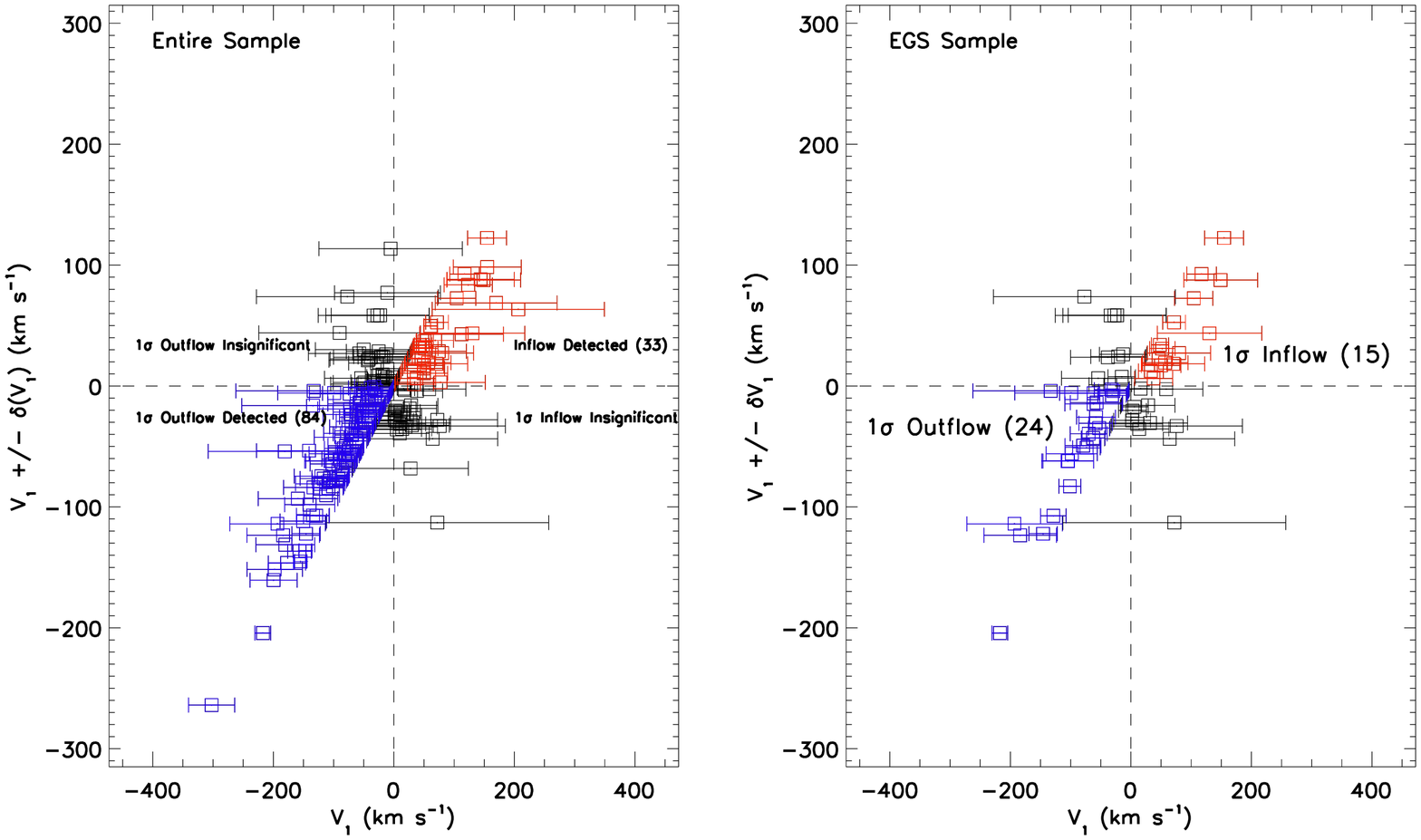}
\caption{A graphical representation of outflows and inflows, for the 61 objects with measured velocities. The y-axis is the measured Fe~II outflow velocity modulated by its 1$\sigma$ uncertainty, where the uncertainty was added to the velocity for objects showing outflows ($V_1$ $<$ 0 \kms) and subtracted from objects exhibiting inflows ($V_1$ $>$ 0 
\kms). With the Fe~II outflow velocity as the x-axis, the plot divides into quadrants: objects with significantly detected outflows cluster in the lower left (blue points) and galaxies with significantly detected inflows appear in the upper right (red points).}
\label{v_minv}
\end{figure}

The model fit to the Fe~II lines has four free parameters: Doppler shift, column density, Doppler width ($b$, where $b$ = $\sqrt{2}$$\sigma$ = FWHM/2$\sqrt{\rm ln2}$), and covering fraction. Due to the low spectral resolution and finite S/N of the observations, the Doppler shift fit to the five Fe~II lines is the primary quantity of interest; the other three parameters of the model will not be discussed in this paper. We measured velocities for 61/72 objects, where 11 objects had no significant absorption lines and therefore could not be modeled, and find velocities ranging from --217 \kms~to +155 \kms\ with a mean of --16 \kms\ and a 1$\sigma$ dispersion of 82 \kms ~(Figure \ref{match_info_histograms_vel_EGS}). This range and mean are similar to the values measured for the entire LRIS sample \citep{martin2012}. We define here the convention of employing ``$V_1$" to refer to the measured velocity shift of the deepest part of the Fe~II absorption line fit, relative to a systemic reference frame defined by [O~II] and Balmer emission lines. Negative $V_1$ values refer to blueshifts (``outflows") while positive $V_1$ values correspond to redshifts (``inflows"). Fe~II velocity shifts significant at the 1$\sigma$ (3$\sigma$) level are observed in $\sim$ 40\% (10\%) of the sample (Figure \ref{v_minv}). Within the sample of objects with 1$\sigma$ (3$\sigma$) outflows, $\langle$$V_1$$\rangle$ = --93 \kms~(--130 \kms). We can also parameterize the sample by stating what percentage of objects exhibit outflows at a certain threshold velocity \citep{martin2012}. The 1$\sigma$ error associated with the each object's velocity can be taken as the standard deviation of a Gaussian probability 
distribution for the outflow velocity of the object. Using these probability distributions, we calculated the fraction of objects in the sample having blueshifts of at least --40 \kms\ (i.e., the systemic redshift uncertainty; Section \ref{sec: sys_z}). We estimated errors on the outflow fraction via bootstrap resampling. We find that 40 $\pm$ 5\% of objects show outflows with blueshifts of at least --40 \kms. The outflow characteristics of the EGS objects presented in this paper are consistent with the properties of the parent sample discussed in \citet{martin2012}.

While the blueshifted signature of outflowing gas has been observed at a range of look-back times and in a variety of galaxy types \citep[e.g.,][]{steidel1996,franx1997,pettini2000,pettini2001,shapley2003,martin2005,veilleux2005,rupke2005,tremonti2007,weiner2009,steidel2010,coil2011}, detections of redshifted interstellar absorption lines -- i.e., gas inflows -- are much more elusive \citep{coil2011,giavalisco2011,rubin2011}. We defer a discussion of individual objects in our sample showing redshifted absorption lines to our accompanying demographics paper \citep{martin2012}.

In objects with blueshifted Fe~II profiles, we still observe significant absorption at the systemic velocity of the galaxy. These absorption profiles are similar to those presented in \citet{weiner2009} and \citet{rubin2010a} and are distinctly different from those appearing in \citet{tremonti2007}. The massive post-starburst galaxies at \z0.6 in the \citet{tremonti2007} sample exhibit Mg~II velocities ranging from 500-2000 \kms, where the profiles of Mg~II are blueshifted enough as to nearly eliminate any residual absorption at $z_{\rm sys}$. It is evident that the systems presented in this work represent a more typical star-forming population than the galaxies discussed in \citet{tremonti2007}. 

\subsection{Maximal Outflow Velocity} \label{vmax}

While outflow velocities can be measured from the centroids of Fe~II absorption lines, as discussed above, another complementary technique widely used for parameterizing outflow velocities utilizes the Mg~II doublet features at 2796.35 and 2803.53 \AA\ \citep{weiner2009,giavalisco2011,rubin2011}. As the Mg~II transitions are resonantly trapped -- such that no decay paths exist other than the re-emission of resonance photons -- these absorption lines are highly susceptible to filling from resonance emission lines. In the presence of a net outflow of gas, this filling shifts the centroid of Mg~II absorption to bluer wavelengths due to the sum of blueshifted and redshifted emission preferentially filling in the absorption around the systemic velocity. The most meaningful velocity measurements of Mg~II are therefore those made largely independently of the Mg~II centroid. We quantify shifts in Mg~II using a technique reliant on the profile of the blue side of the 2796 \AA\ feature; we do not use the 2803 \AA\ feature as it may be contaminated by emission from the 2796 \AA\ line. We adopt a methodology similar to the V$_{\rm 10\%}$ measurement of \citet{weiner2009}, as described in \citet{martin2012}. Firstly, we isolate the minimum of the 2796 \AA\ trough and proceed to shorter wavelengths until the sum of one pixel and its uncertainty, $\sigma$, is greater than a threshold value defined by the data's continuum S/N: 1.0 - $\frac{1.0}{\rm S/N}$. We record the wavelength value at which this threshold is first met, perturb the science spectrum by a value drawn from a Gaussian distribution of width $\sigma$, and repeat the same procedure a total of 1000 times. We iteratively compute the average of these 1000 wavelengths excluding outliers and adopt this value as the maximal outflow velocity of Mg~II, $V_{max}$(Mg~II). We measure an analogous maximal outflow velocity for the Fe~II 2374 \AA\ feature using the same technique, $V_{max}$(Fe~II). We choose to utilize the 2374 \AA\ feature over other Fe~II lines given that the blue wings of other Fe~II features are often contaminated by neighboring absorption lines;  the only feature close to the blue side of the 2374 \AA\ line is a fine-structure Fe~II$^*$ emission line at 2365 \AA\ (--1124 \kms). 

We report maximal outflow velocities only for composite spectra, in light of the following limitations of the maximal outflow velocity technique. Firstly, only resolved absorption lines will yield meaningful maximal outflow velocities, given that unresolved absorption lines simply reflect the instrumental profile as opposed to the intrinsic distribution of gas kinematics. Secondly, as discussed above, maximal outflow velocities are strongly dependent on the S/N of the spectroscopic data, with higher S/N data showing more blueshifted values. The uniformly high S/N of the composite spectra (27--39 pixel$^{-1}$) ensure that the effects of differential S/N are largely mitigated. We measure $V_{max}$(Mg~II) ranging from [--605, --855] \kms\ while $V_{max}$(Fe~II) varies from [--444, --614] \kms. The typical FWHM of the Mg~II absorption lines is 665 \kms, significantly larger than the instrumental resolution (FWHM = 435 \kms). Therefore, the Mg~II lines are resolved and yield meaningful $V_{max}$(Mg~II) measurements.

\section{Results} \label{sec: results}

With measurements of the star-forming and structural properties of the sample, we now turn to examining the correlations between these parameters and outflow velocity. We focus here on galaxy properties inferred from the ancillary data of the EGS (morphology, inclination, SFR, sSFR, $\Sigma_{\rm SFR}$, etc.). A discussion of how outflows relate to galaxy parameters such as stellar mass, color, redshift, and luminosity measured for all objects in the parent DEEP2 sample appears in \citet{martin2012}. A comparison of the kinematics of lines tracing high- and low-ionization interstellar features will also be presented in an upcoming paper (Shapley et al., in prep.). 

We adopt two complementary techniques when presenting the data: 1) direct comparison of individual outflow velocities with galaxy properties and 2) the construction of composite spectra based on subsamples of objects exhibiting similar star-forming or structural characteristics. These methods are both useful in that the former enables ones to draw conclusions about specific objects while the latter produces high S/N spectra from which global, averaged trends can be inferred across the sample as a function of a single property. In Table \ref{sampletable}, we present a summary of the data including redshift, luminosity, color, stellar mass, SFR, dust attenuation, areal measurements, and outflow velocity. Summaries of outflow velocities and correlation coefficients appear in Tables \ref{correlationtable} and {\ref{compositetable}}. 

We focus on results from the resonance Fe~II and Mg~II absorption features, using outflow velocities measured from both the centroids and blue wings of these lines. At the end of this section, we present basic analyses of the strength of Mg~II absorption. Forthcoming papers will include a more thorough discussion of both Mg~II and fine-structure Fe~II$^*$ emission (Kornei et al. 2012b, in prep. and Martin et al. 2012b, in prep.).

\subsection{Star Formation and Outflows}

The relationship between SFR and outflow velocity is one of the most commonly investigated correlations in the study of outflows \citep[e.g.,][]{martin2005,sato2009,weiner2009,rubin2010a}. \citet{martin2005} compiled data from a variety of local ULIRGs, LIRGs, and starburst dwarfs spanning four decades in SFR and showed that outflow velocity is related to SFR as V $\propto$ SFR$^{0.35}$. While this trend is consistent with the canonical picture of galactic-scale outflows arising from the combined effects of stellar and supernovae winds \citep{leitherer1995,veilleux2005}, recent studies probing more limited ranges in SFR have often failed to reproduce the \citet{martin2005} correlation \citep{rupke2005,steidel2010,law2012}. As the V $\propto$ SFR$^{0.35}$ relation flattens for galaxies with SFRs $\gtrsim$ 10--100 $M_{\odot}$ yr$^{-1}$ \citep{martin2005} -- the very interval probed by most investigations -- \citet{rupke2005} propose that the bulk of studies simply do not span a large enough SFR interval to unambiguously detect a trend. The inclusion of low-SFR dwarf galaxies with low outflow velocities may be necessary for detecting a correlation between outflow velocity and SFR \citep{heckman2011}. \citet{weiner2009} examined the Mg~II doublet in a sample of DEEP2 galaxies at \emph{z} = 1.4 and found a dependence of outflow velocity on SFR consistent with the result from \citet{martin2005}. This result is surprising given that the \citet{weiner2009} sample spans a much more limited SFR interval ($\sim$ 7--180 $M_{\odot}$ yr$^{-1}$) than the 0.1--1000 $M_{\odot}$ yr$^{-1}$ range of \citet{martin2005}. \citet{weiner2009} employed two different techniques to parametrize the velocity of outflowing Mg~II: a ``median velocity" containing 50\% of the absorption and a ``maximal velocity" where the outflow crosses 10\% or 25\% absorption. The median velocity method is most akin to the centroid technique used in this paper and in other recent studies of outflows \citep{rubin2010a,steidel2010,krug2010,law2012} while the maximal velocity is similar to our $V_{max}$(Mg~II) measurement. Relying on composite spectra, \citet{weiner2009} found that SFR was correlated with the 10\% maximal velocity of Mg~II such that V $\propto$ SFR$^{0.38}$. These authors did not find a significant trend between SFR and the Mg~II median velocity. In the analyses below, we employ the Spearman $\rho$ correlation test to examine how well variables are correlated. We report the Spearman rank-order correlation coefficient, $r_S$, and the number of standard deviations from the null hypothesis that the quantities are uncorrelated. 

\subsubsection{Individual Measurements}

\begin{figure*}
\begin{center}$
\begin{array}{c}
\includegraphics[trim=1.9in 0in 0.5in 0in,clip,width=3.5in]{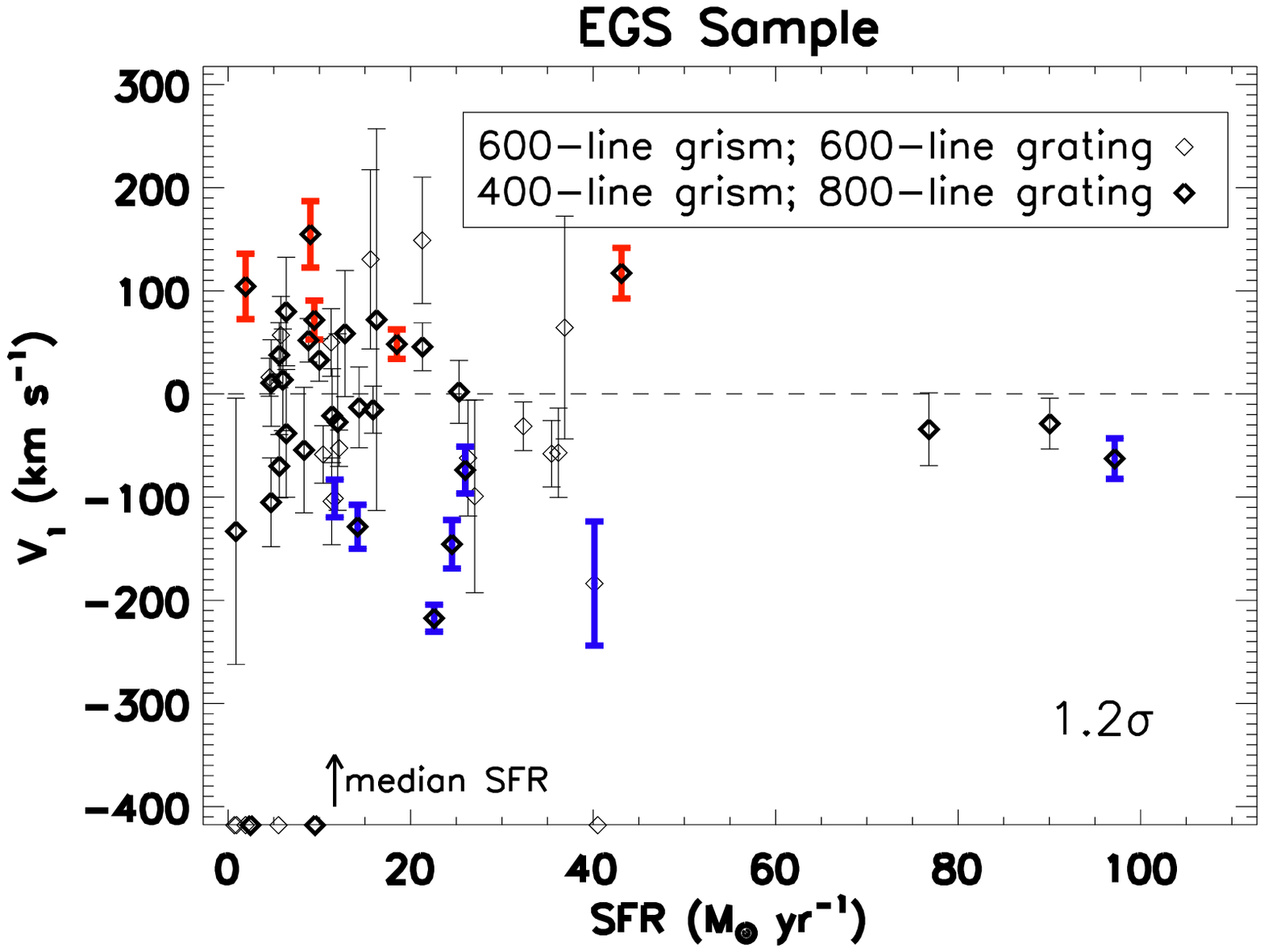}
\includegraphics[trim=1.9in 0in 0.5in 0in,clip,width=3.5in]{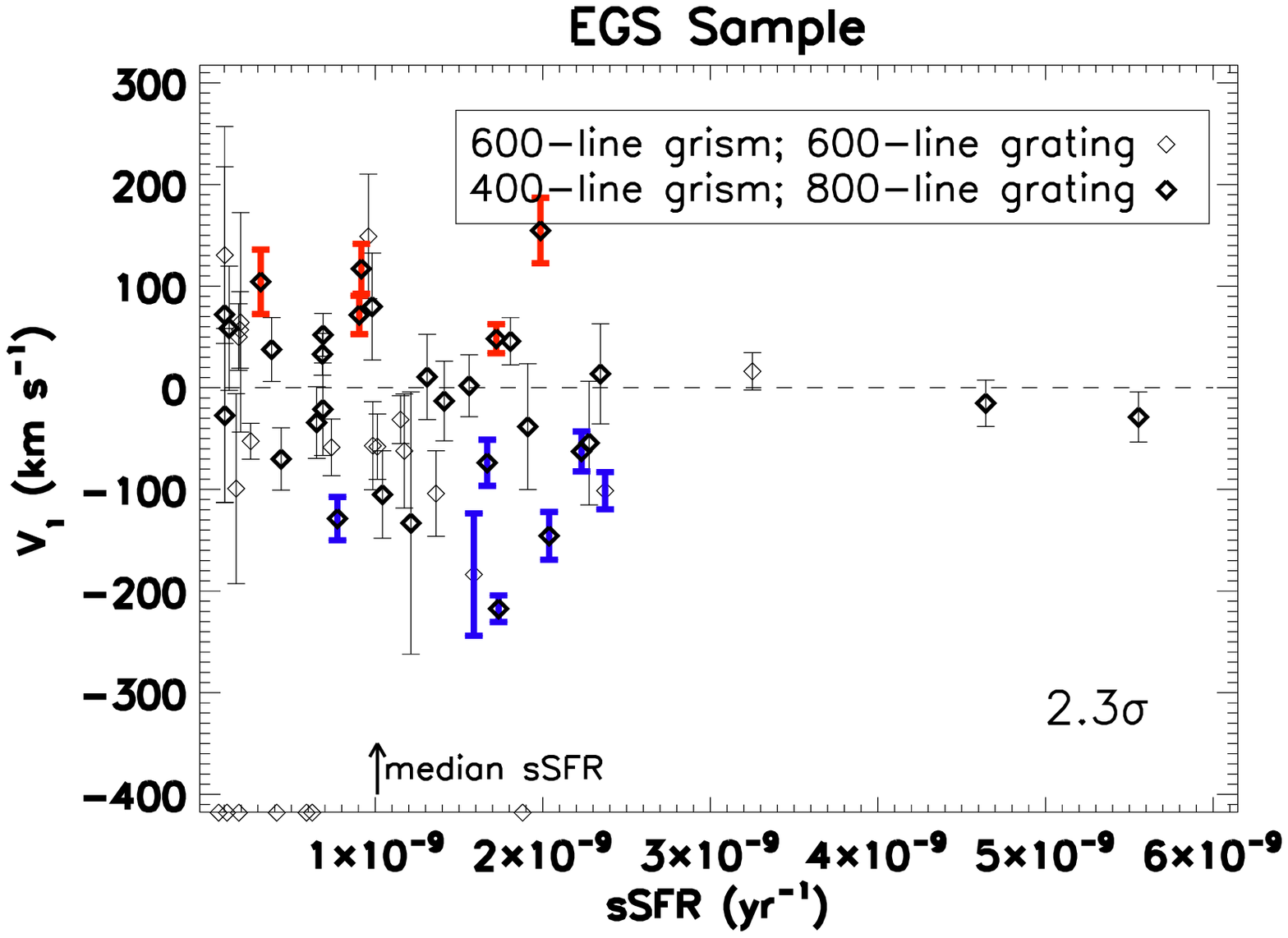}
\end{array}$
\end{center}
\caption{\emph{Left: }$V_1$ outflow velocity versus SFR. Objects with 3$\sigma$ detected outflows and inflows are indicated in blue and red, respectively. The grism and grating pair used in the LRIS observations is noted and the SFRs of objects without measured velocities (due to poor spectral signal-to-noise) appear along the x-axis. In the lower right, we indicate the significance level of the correlation. There is not a strong trend between SFR and outflow velocity in our sample, possibly due to the limited range in SFR probed by the data \citep[$\sim$ 1--100 M$_{\odot}$ yr$^{-1}$;][]{rupke2005,steidel2010,law2012}. \emph{Right: }$V_1$ outflow velocity versus sSFR, where a correlation is observed between $V_1$ and sSFR at the 2.3$\sigma$ ($r_S$ = --0.33) level. Symbols are the same as in the $V_1$ versus SFR plot.}
\label{SFR_uvcor_vel}
\end{figure*}

In the left panel of Figure \ref{SFR_uvcor_vel}, we plot $V_1$ versus SFR. We do not find a significant correlation ($r_S$ = --0.17; 1.2$\sigma$ from the null hypothesis) between SFR and $V_1$ across the sample, although this is not unexpected given the limited range in SFR probed by the data \citep[$\sim$ 1--97 $M_{\odot}$ yr$^{-1}$;][]{rupke2005,steidel2010,law2012}, the small sample size, the S/N of the data, and the fitting method used for measuring outflow velocities (centroid fitting of Fe~II). The three objects with the highest SFRs ($>$ 75 M$_{\odot}$ yr$^{-1}$) show small Fe~II blueshifts ($V_1$ $>$ --65 \kms); we propose that the sample's small dynamic range in SFR, the limited number of objects considered, the lack of high-resolution spectroscopy, and the methodology of employing centroid measurements of Fe~II may all contribute to this result. 

We observe a weak correlation between sSFR and $V_1$ outflow velocity (Figure \ref{SFR_uvcor_vel}, \emph{right}): $r_S$ = --0.33; 2.3$\sigma$ from the null hypothesis. As sSFR is a ratio between SFR (where higher values should produce stronger winds) and stellar mass (where larger values should hinder winds, assuming that the gravitational potential energy of a system increases with increasing stellar mass), one would expect that sSFR is a tracer of both the energetics and gravitational potential of a galaxy and should therefore correlate with the observed velocity of winds. While \citet{chen2010} do not find a trend between sSFR and either outflow velocity or the equivalent width of the outflow component in a large sample of Sloan Digital Sky Survey \citep[SDSS;][]{york2000} galaxies at \z0.1, these authors hypothesize that the lack of correlation with the strength of the outflow component may be due to the effects of dust shielding on the particular atomic species used to trace winds at low redshift. \citet{rubin2010a} also fail to find a trend between sSFR and outflow velocity at \z1.4, although these authors draw their conclusions solely based on co-added spectra\footnote{These stacked spectra have typical S/N values of 6.0 per pixel, comparable to our individual spectra (Figure \ref{composite_blue_EGS_smooth}).}. While \citet{prochaska2011} caution that stacked spectra show only a smeared version of the original emission and absorption features present in the component spectra, we find that the composite spectra discussed below reflect the trends of the individual measurements. Therefore, the fact that \citet{rubin2010a} do not observe a trend between sSFR and outflow velocity is more likely a result of the low S/N of their data rather than a direct consequence of relying on co-added spectra. 

\begin{figure*}
\begin{center}$
\begin{array}{c} %match_info_plots.pro
\includegraphics[trim=1.9in 0in 0.5in 0in,clip,width=3.5in]{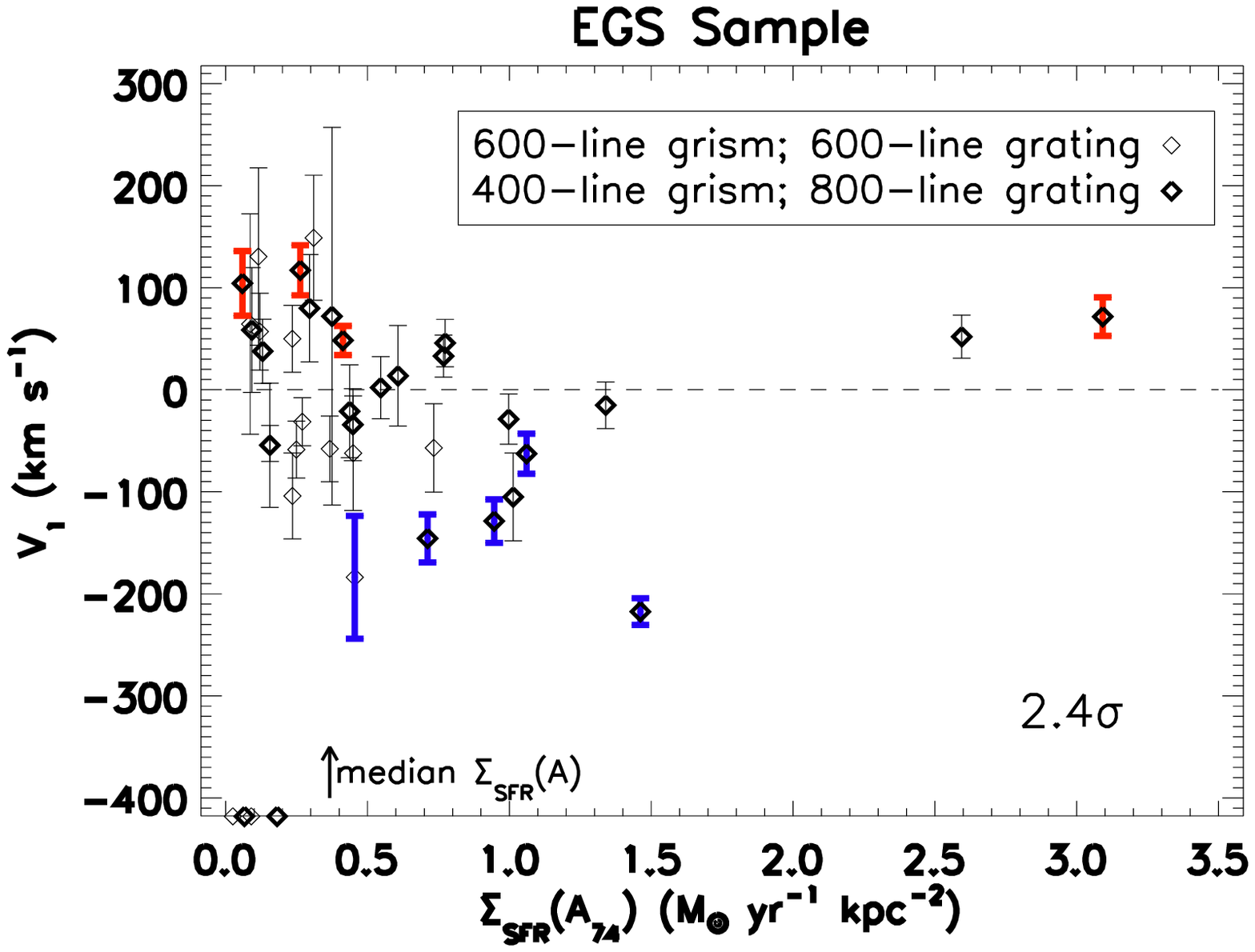}
\includegraphics[trim=1.9in 0in 0.5in 0in,clip,width=3.5in]{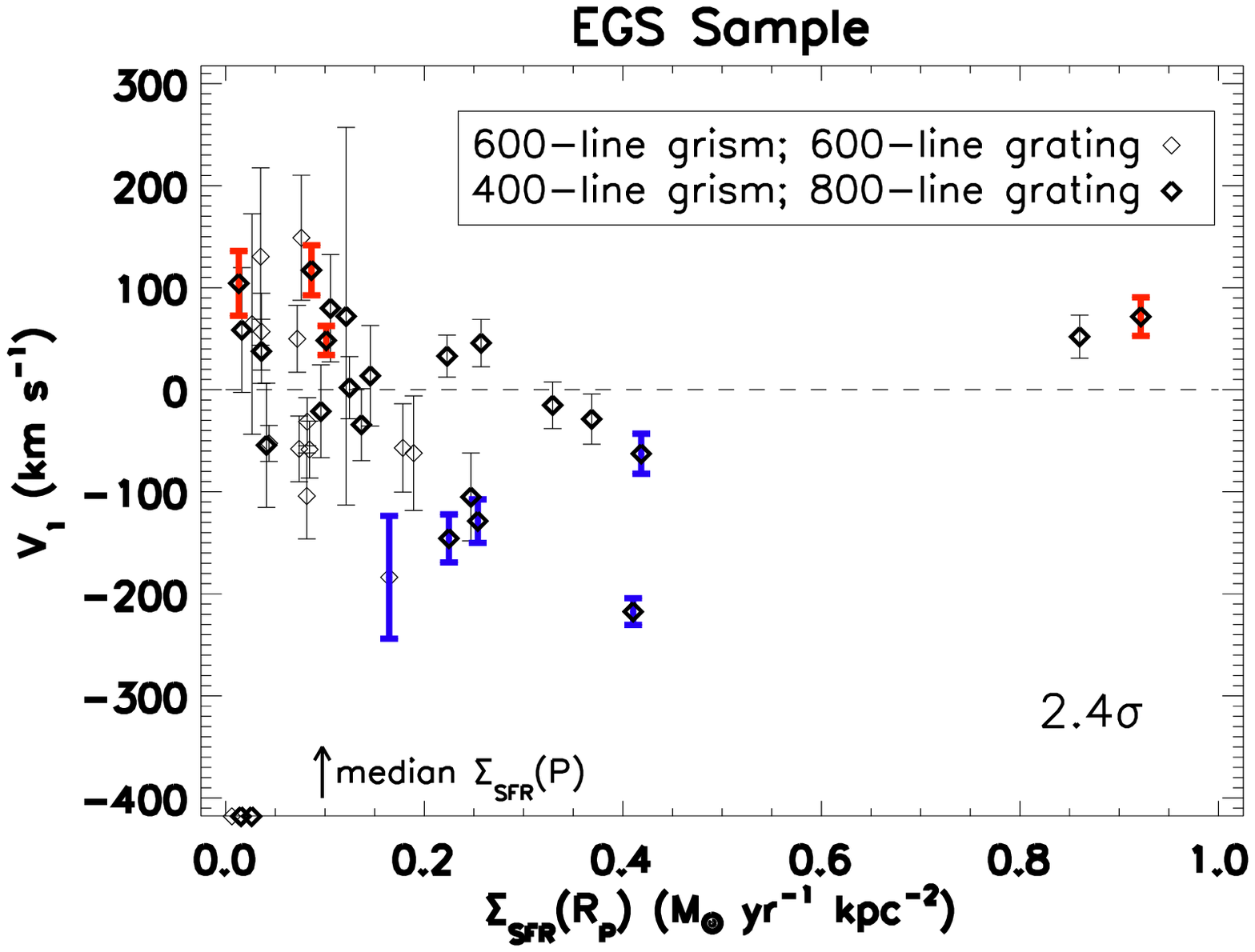}
\end{array}$
\end{center}
\caption{\emph{Left: }$V_1$ outflow velocity versus $\Sigma_{\rm SFR}$(A$_{74}$), where symbols are as in Figure \ref{SFR_uvcor_vel}. A correlation at the 2.4$\sigma$ level ($r_S$ = --0.40)  is observed between $V_1$ and $\Sigma_{\rm SFR}$(A$_{74}$), where objects with larger $\Sigma_{\rm SFR}$ show larger blueshifts in their Fe~II lines. \emph{Right: }$V_1$ outflow velocity versus $\Sigma_{\rm SFR}$(R$_{\rm P}$). All objects with 3$\sigma$ detections of outflows have $\Sigma_{\rm SFR}$(R$_{\rm P}$) $>$ 0.1 M$_{\odot}$ yr$^{-1}$ kpc$^{-2}$, in agreement with the $\Sigma_{\rm SFR}$ threshold proposed by \citet{heckman2002}. A 2.4$\sigma$ correlation ($r_S$ = --0.40) is observed between $V_1$ and $\Sigma_{\rm SFR}$(R$_{\rm P}$). When the two outliers at the highest $\Sigma_{\rm SFR}$ are removed from the sample -- objects 12015914 and 12100420 with extremely compact morphologies -- trends at the 3.1$\sigma$ level ($r_S$ = --0.54) are observed for both $\Sigma_{\rm SFR}$ estimates. \citet{law2012} observed a 2.1$\sigma$ correlation between outflow velocity and $\Sigma_{\rm SFR}$ in a sample of 35 star-forming galaxies at \z2, although other authors have noted weaker correlations \citep{rubin2010a,steidel2010}.}
\label{SFRSD_vel}
\end{figure*}

We find the strongest trend with $V_1$ outflow velocity when SFR is normalized by a representative galaxy area. Star-formation rate surface densities -- both $\Sigma_{\rm SFR}$(A$_{74}$) and $\Sigma_{\rm SFR}$(R$_{\rm P}$) -- exhibit pronounced trends with $V_1$ outflow velocity (Figure \ref{SFRSD_vel}). Both $\Sigma_{\rm SFR}$(A$_{74}$) and $V_1$ and $\Sigma_{\rm SFR}$(R$_{\rm P}$) and $V_1$ are correlated at the 2.4$\sigma$ level with $r_S$ =  --0.40. Omitting two extremely compact galaxies with uncertain area measurements yields 3.1$\sigma$ measurements from the null hypothesis and correlation coefficients of $r_S$ = --0.54 for both $\Sigma_{\rm SFR}$(A$_{74}$) and $\Sigma_{\rm SFR}$(R$_{\rm P}$). Such strong trends have not yet been reported in the literature at high redshift; \citet{rubin2010a} noted only a 1$\sigma$ correlation in a sample of 468 galaxies at 0.7 $<$ \emph{z} $<$ 1.5, \citet{steidel2010} also found a 1$\sigma$ correlation for 81 galaxies at \emph{z} $\simeq$ 2, and \citet{law2012} reported a 2.1$\sigma$ correlation for 35 optically-selected star-forming galaxies at \emph{z} = 1.5--3.6. However, the galaxy areas employed in the $\Sigma_{\rm SFR}$ calculations of \citet{rubin2010a}, \citet{steidel2010}, and \citet{law2012} correspond to half-light radii at rest-frame $\sim$ 2200 \AA, the extent of H$\alpha$ emission, and circularized effective radii, respectively. Given the clumpy distribution of star formation in distant galaxies \citep{lotz2008,law2012}, these areas may be overestimates of the regions corresponding to star-formation \citep{rubin2010a}\footnote{Simply overestimating galaxy areas is not sufficient to negate a trend with outflow velocity, however, since we find equally strong correlations with outflow velocity utilizing both $\Sigma_{\rm SFR}$(A$_{74}$) and $\Sigma_{\rm SFR}$(R$_{\rm P}$).}. Furthermore, the \citet{steidel2010} measurements are from ground-based data which are seeing-limited. The \citet{law2012} observations, while probing the bulk of the stellar mass with \emph{H}-band (rest-frame optical) \emph{HST} imaging, approximated the areas of clumpy galaxies by adopting the circularized effective radius of the brightest clump as representative of the size of the entire galaxy. This technique results in overestimates of $\Sigma_{\rm SFR}$, given that the areas of fainter clumps are neglected. In a more local sample of SDSS galaxies at \z0.1, \citet{chen2010} found no correlation between outflow velocity and $\Sigma_{\rm SFR}$. These authors note that the small dynamic range of their data (outflow speed = [120--160] \kms) may obscure trends. When we limit our sample to only objects showing outflows ($V_1$ $<$ 0 \kms), we find that star-formation rate surface density and $V_1$ are correlated at the $\sim$ 1$\sigma$ level (for both area measurements). We propose that inflows and outflows represent a continuum of gas cycling and therefore correlations with galaxy properties should be investigated inclusive of both redshifted and blueshifted gas. While SFR and $\Sigma_{\rm SFR}$(A$_{74}$) are weakly correlated, at the $\sim$ 2.3$\sigma$ level ($r_S$ = 0.37), we suggest that the scatter in the $V_1$ versus $\Sigma_{\rm SFR}$(A$_{74}$) relation prevents a similar correlation from being seen in the plot of $V_1$ versus SFR. All objects with 3$\sigma$ detections of outflows have $\Sigma_{\rm SFR}$(R$_{\rm P}$) $>$ 0.1 M$_{\odot}$ yr$^{-1}$ kpc$^{-2}$, in agreement with the $\Sigma_{\rm SFR}$ threshold proposed by \citet{heckman2002}. These objects with secure detections of outflows furthermore all have SFR $>$ 10 M$_{\odot}$ yr$^{-1}$, although we hypothesize that the $\Sigma_{\rm SFR}$ threshold is more fundamental given the stronger correlation of outflow velocity with $\Sigma_{\rm SFR}$ than with SFR. The trend between outflow velocity and star-formation rate surface density supports the theory that the shifts of absorption lines are due to gas flows as opposed to galactic rotation. The observation that the kinematic shear of [O~II] emission is systematically smaller than the shifts observed in the resonance Fe~II lines lends further credence to the assertion that gas flows are primarily responsible for the observed absorption line shifts. 

Using the outflow fraction methodology discussed in Section \ref{fitting}, we find that 26 $\pm$ 8\% of the low-$\Sigma_{\rm SFR}$(A$_{74}$) sample shows outflows with blueshifts of at least --40 \kms. In contrast, 48 $\pm$ 9\% of the high-$\Sigma_{\rm SFR}$(A$_{74}$) shows outflows at the same level. When the two galaxies with uncertain area measurements are removed from the high-$\Sigma_{\rm SFR}$(A$_{74}$) sample, the outflow fraction increases to 54 $\pm$ 9\%. The significant difference in outflow fraction between the low- and high-$\Sigma_{\rm SFR}$(A$_{74}$) samples is consistent both with the measurements from individual objects presented above and our findings based on composite spectra, discussed below. 

\subsubsection{Composite Spectra}

We constructed composite spectra from the binary division of data sorted according to SFR, sSFR, $\Sigma_{\rm SFR}$(A$_{74}$), $\Sigma_{\rm SFR}$(R$_{\rm P}$), A$_{\rm UV}$, and \emph{i}. We measured the shifts of the Fe~II interstellar absorption lines in the composites using the same techniques employed for the individual spectra (Section \ref{fitting}). We also measured maximal outflow velocities in both Mg~II and Fe~II for these composites. While the robustness of the velocity measurements benefit from the increased S/N of the stacked composites, it is important to remember that relying on composite spectra only reveals the global, averaged trends across the sample. 

Below, we discuss three measurements for each composite spectrum: $V_1$, $V_{max}$(Mg~II), and $V_{max}$(Fe~II). While the magnitudes of the $V_1$ shifts are not large, we emphasize that it is the \emph{difference} in $V_1$ between composite spectra that is primary interest; the apparent detections of inflows and outflows in individual composites may be spurious given the velocity uncertainties of the LRIS data. The results implied by $V_1$ and $V_{max}$(Fe~II) are corroborated by $V_{max}$(Mg~II), where shifts in the Mg~II lines are often visually obvious in the data.

\begin{figure*}
\begin{center}$
\begin{array}{c} %EGS_composite.pro
\includegraphics[trim = 0in 3in 0in 0in,clip,width=7in]{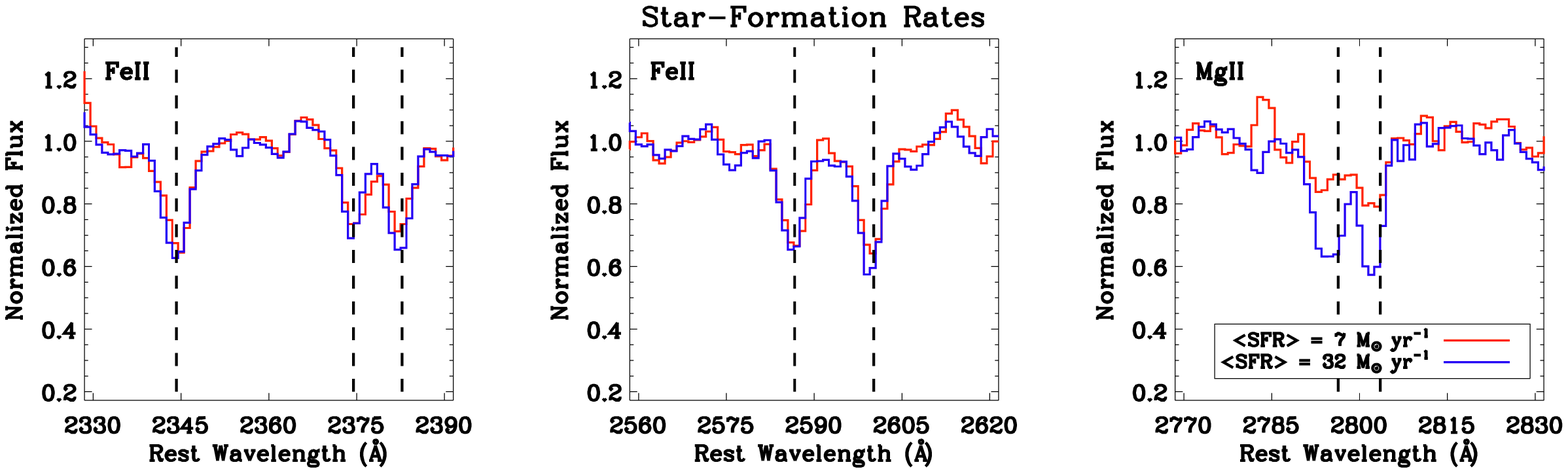} \\
\includegraphics[trim = 0in 3in 0in 0in,clip,width=7in]{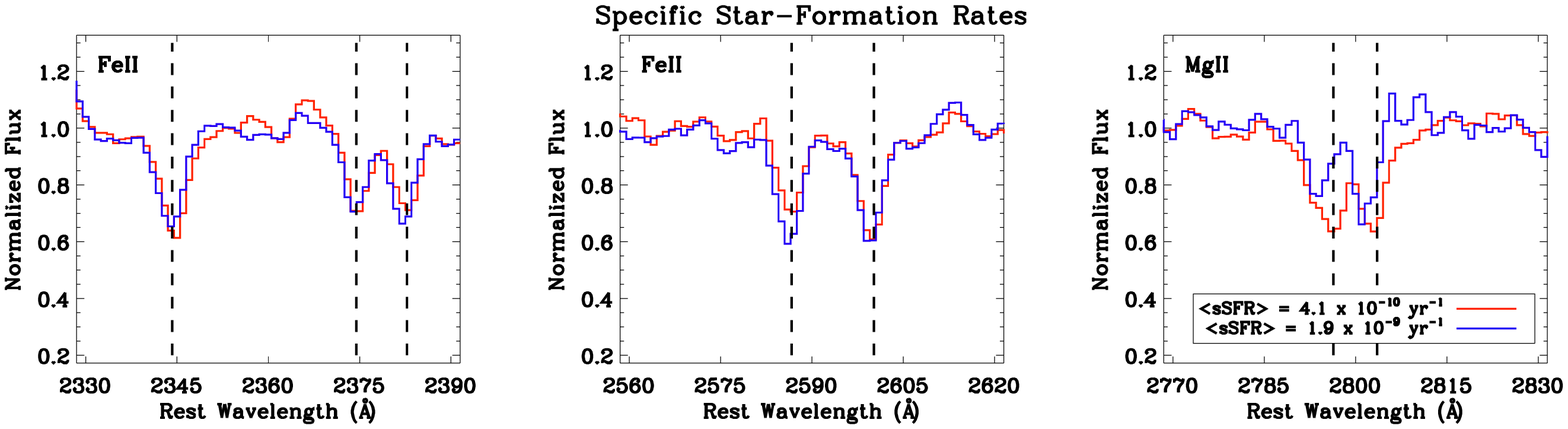}
\end{array}$
\end{center}
\caption{\emph{Top: }Composites of high and low SFR objects, where the dashed vertical lines indicate the rest-frame wavelengths of resonance Fe~II and Mg~II features. We measure $V_1$ values of --24 $\pm$ 9 \kms~for the high-SFR composite (blue line) and 29 $\pm$ 11 \kms~for the low-SFR composite (red line). The most pronounced spectral differences between the composites are observed around the Mg~II feature at $\sim$ 2800 \AA, where higher SFR objects exhibit deeper Mg~II absorption troughs on average. \emph{Bottom: } Same as above, for composite spectra assembled from high and low sSFR objects. We measure $V_1$ values of --34 $\pm$ 9 \kms~ for the high sSFR composite (blue line) and 27 $\pm$ 11 \kms~ for the low sSFR composite (red line). As with the SFR composites, the largest spectral differences are seen around Mg~II. Objects with high sSFR have bluer Mg~II centroids and less Mg~II absorption than low sSFR objects. A full discussion of Mg~II will appear in forthcoming papers (Kornei et al. 2012b, in prep. and Martin et al. 2012b, in prep.).}
\label{SFR_sSFR_smooth}
\end{figure*}

In the top row of Figure \ref{SFR_sSFR_smooth}, we compare high- and low-SFR composites and focus on the wavelength intervals around Fe~II and Mg~II (2370--2400, 2570--2650, and 2760--2840 \AA). The strengths of the Fe~II features are similar in both composites, although we find that the high-SFR composite ($\langle$SFR$\rangle$ = 32 M$_{\odot}$ yr$^{-1}$; extrema of [12, 97] M$_{\odot}$ yr$^{-1}$) exhibits a $V_1$ blueshift of --24 $\pm$ 9 \kms\ while the low-SFR composite ($\langle$SFR$\rangle$ = 7 M$_{\odot}$ yr$^{-1}$; extrema of [1,11] M$_{\odot}$ yr$^{-1}$) is best-fit with a model of redshifted gas with positive velocity 29 $\pm$ 11 \kms. These results are in the same sense as the trend of vigorously star-forming objects showing more pronounced outflows than lower SFR systems \citep{martin2005}. The SFR composites differ significantly in Mg~II strength, with larger SFR objects showing deeper absorption on average. The centroids of the Mg~II features are blueshifted in both the high- and low-SFR composites, so systemic absorption is not likely responsible for the difference in Mg~II strength. Furthermore, the red wings of the 2803 \AA\ features are remarkably similar between the two composites, so it is not immediately apparent that emission filling is affecting one spectrum more than the other. Since SFR and stellar mass are correlated (Figure \ref{Mstar_SFR}), Mg~II strength may depend on stellar mass. We investigate the correlation between Mg~II and stellar mass in Section \ref{sec: MgII}. Higher metallicities  are expected for more vigorously star-forming objects \citep{tremonti2004,noeske2007,elbaz2007}, although we note that because the Fe~II and Mg~II lines are saturated in our study (based on the differences between the observed lines ratios and those expected according to atomic theory) it is difficult to determine if metallicity is the driving factor behind absorption line strength. The kinematics of Mg~II are similar in both composites; the high (low) SFR composite has $V_{max}$(Mg~II) = --748 $\pm$ 94 (--614 $\pm$ 83) \kms. The composites have $V_{max}$(Fe~II) measurements of --514 $\pm$ 84 (--453 $\pm$ 104) \kms.

We compare sSFR composite spectra in the bottom row of Figure \ref{SFR_sSFR_smooth}. As with the SFR composites, there is minimal variation in the strength of the Fe~II resonance features between the high- and low-sSFR composites. Likewise, we find differences in the Fe~II kinematics between the two composites that are similar to the shifts seen in the SFR composites: objects with high sSFR ($\langle$sSFR$\rangle$ = 1.9 $\times$ 10$^{-9}$ yr$^{-1}$; extrema of [1.0 $\times$ 10$^{-9}$, 5.6 $\times$ 10$^{-9}$] yr$^{-1}$) exhibit an average blueshift of --34 $\pm$ 9 \kms~ while low sSFR objects ($\langle$sSFR$\rangle$ = 4.1 $\times$ 10$^{-10}$ yr$^{-1}$; extrema of [6.5 $\times$ 10$^{-11}$, 9.6 $\times$ 10$^{-10}$] yr$^{-1}$) are globally characterized by a Fe~II shift of 27 $\pm$ 11 \kms. The larger blueshifts seen in higher sSFR objects supports the correlation observed in individual galaxies (Figure \ref{SFR_uvcor_vel}). The sSFR composites are comparable in Mg~II strength but not in kinematics: the high (low) sSFR composite has $V_{max}$(Mg~II) = --605 $\pm$ 80 (--834 $\pm$ 148) \kms. The composites are characterized by $V_{max}$(Fe~II) measurements of --556 $\pm$ 94 (--456 $\pm$ 86) \kms. While the sense of the $V_{max}$(Mg~II) measurements is contrary to both the $V_1$ and $V_{max}$(Fe~II) results -- i.e., objects with larger sSFRs show smaller $V_{max}$(Mg~II) blueshifts than objects with lower sSFRs -- we note that the $V_{max}$(Mg~II) measurements of the high- and low-sSFR composites are consistent at roughly the 1$\sigma$ level. The large blueshift of $V_{max}$(Mg~II) for the low-sSFR composite may be explained by high-mass objects preferentially populating the low-sSFR composite. High-mass objects exhibit deeper and broader Mg~II absorption profiles than low-mass objects (Martin et al. 2012b, in prep.). Stellar Mg~II absorption and emission filling may also affect the observed profiles; the large differences in the red wings of the 2803 \AA\ features and the blueshift seen in the Mg~II centroids of the high-sSFR composite argue that emission filling may be present in the high-sSFR composite. 

\begin{figure*}
\centering
\includegraphics[width=7in]{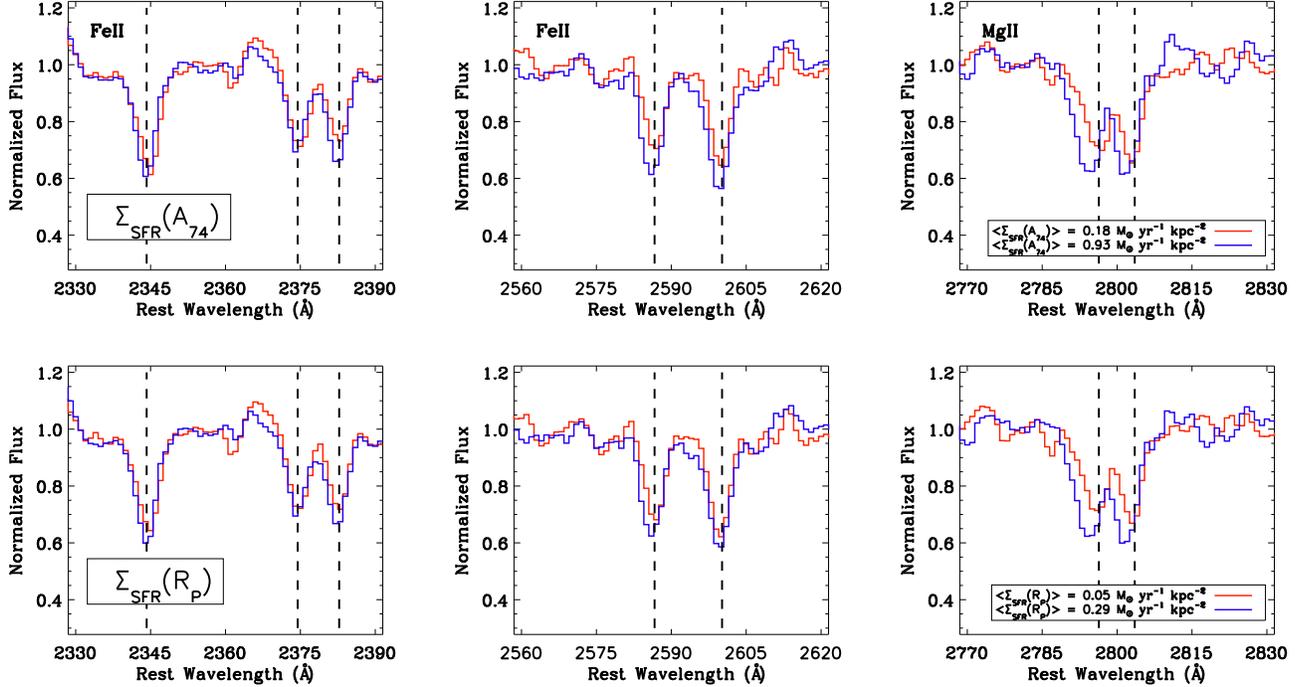}
\caption{\emph{Top: }Composite spectra assembled on the basis of $\Sigma_{\rm SFR}$(A$_{74}$). Kinematic shifts between the two composites in the Fe~II resonance lines are evident visually; we find a $V_1$ outflow velocity of --31 $\pm$ 7 \kms~ for the high-$\Sigma_{\rm SFR}$(A$_{74}$) composite (blue line) and 44 $\pm$ 15 \kms~for the low-$\Sigma_{\rm SFR}$(A$_{74}$) composite (red line). These pronounced differences are consistent with the strong trend observed between $\Sigma_{\rm SFR}$(A$_{74}$) and $V_1$ on a per-object basis (Figure \ref{SFRSD_vel}). The Mg~II features show a strong kinematic shift between the composites, with objects characterized by larger $\Sigma_{\rm SFR}$ exhibiting more blueshifted Mg~II absorption. \emph{Bottom: } Same as above, for $\Sigma_{\rm SFR}$(R$_{\rm P}$) composites. The high-$\Sigma_{\rm SFR}$(R$_{\rm P}$) composite is best-fit with a $V_1$ outflow velocity of --25 $\pm$ 6 \kms\ while the low-$\Sigma_{\rm SFR}$(R$_{\rm P}$) composite yields an outflow velocity of 33 $\pm$ 13 \kms. Objects with higher $\Sigma_{\rm SFR}$(R$_{\rm P}$) also show more strongly blueshifted Mg~II features. These results are in the same sense as the trends seen in the $\Sigma_{\rm SFR}$(A$_{74}$) composites, unsurprising given the multiplicative offset between the A$_{74}$ and $\pi$R$_{\rm P}^2$ areas (Figure \ref{Contour1}).}
\label{SFRSD_smooth}
\end{figure*}

In Figure \ref{SFRSD_smooth}, we plot composite spectra assembled according to $\Sigma_{\rm SFR}$. Kinematic shifts between the high- and low-$\Sigma_{\rm SFR}$ composites are visually evident and persist irrespective of which galaxy area measurement is employed. Quantitatively, we find $V_1$ shifts of --31 $\pm$ 7 and --25 $\pm$ 6 \kms\ for the high $\Sigma_{\rm SFR}$(A$_{74}$) and $\Sigma_{\rm SFR}$(R$_{\rm P}$) composites, respectively. In contrast, the low-$\Sigma_{\rm SFR}$ composites are best-fit with models having shifts of 44 $\pm$ 15 and 33 $\pm$ 13 \kms. These results, consistent with the $\sim$ 3$\sigma$ trends observed between outflow velocity and [$\Sigma_{\rm SFR}$(A$_{74}$), $\Sigma_{\rm SFR}$(R$_{\rm P}$)] on a per-object basis (Figure \ref{SFRSD_vel}), are in agreement with theoretical predictions that a higher density of star formation should lead to stronger outflows \citep{leitherer1995,veilleux2005,murray2011}. However, the most striking difference between the $\Sigma_{\rm SFR}$ composites are the blue wings of the Mg~II 2796 \AA\ features. We find visually-obvious shifts in the Mg~II doublet between the high- and low-$\Sigma_{\rm SFR}$ composites, where objects with larger star-formation rate surface densities show more blueshifted Mg~II absorption. Furthermore, the entire blue wing of the 2796 \AA\ feature is offset in the composites, suggestive of actual kinematic differences as opposed to centroid shifting from emission filling. The composites of high-$\Sigma_{\rm SFR}$ objects are characterized by $V_{max}$(Mg~II) = --855 $\pm$ 66 \kms~ and --862 $\pm$ 69 \kms, respectively. Conversely, the low-$\Sigma_{\rm SFR}$ composites have measured $V_{max}$(Mg~II) values of --640 $\pm$ 117 \kms~ and --668 $\pm$ 105 \kms, respectively. The corresponding $V_{max}$(Fe~II) measurements for the high-$\Sigma_{\rm SFR}$ composites are --611 $\pm$ 103 \kms\ and --614 $\pm$ 97 \kms, respectively. The low-$\Sigma_{\rm SFR}$ composites have $V_{max}$(Fe~II) values of --451 $\pm$ 86 \kms\ and --444 $\pm$ 87 \kms, respectively. The kinematic differences between the high- and low-$\Sigma_{\rm SFR}$ composites are among the largest observed for any pair of composites ($\Delta$$V_{max}$(Mg~II) = 215 $\pm$ 134 \kms\ and 194 $\pm$ 126 \kms; $\Delta$$V_{max}$(Fe~II) = 160 $\pm$ 134  \kms~ and 170 $\pm$ 130 \kms).

\begin{figure}
\centering
\includegraphics[trim=0in 4.8in 8.4in 1.77in,clip,width=3.5in]{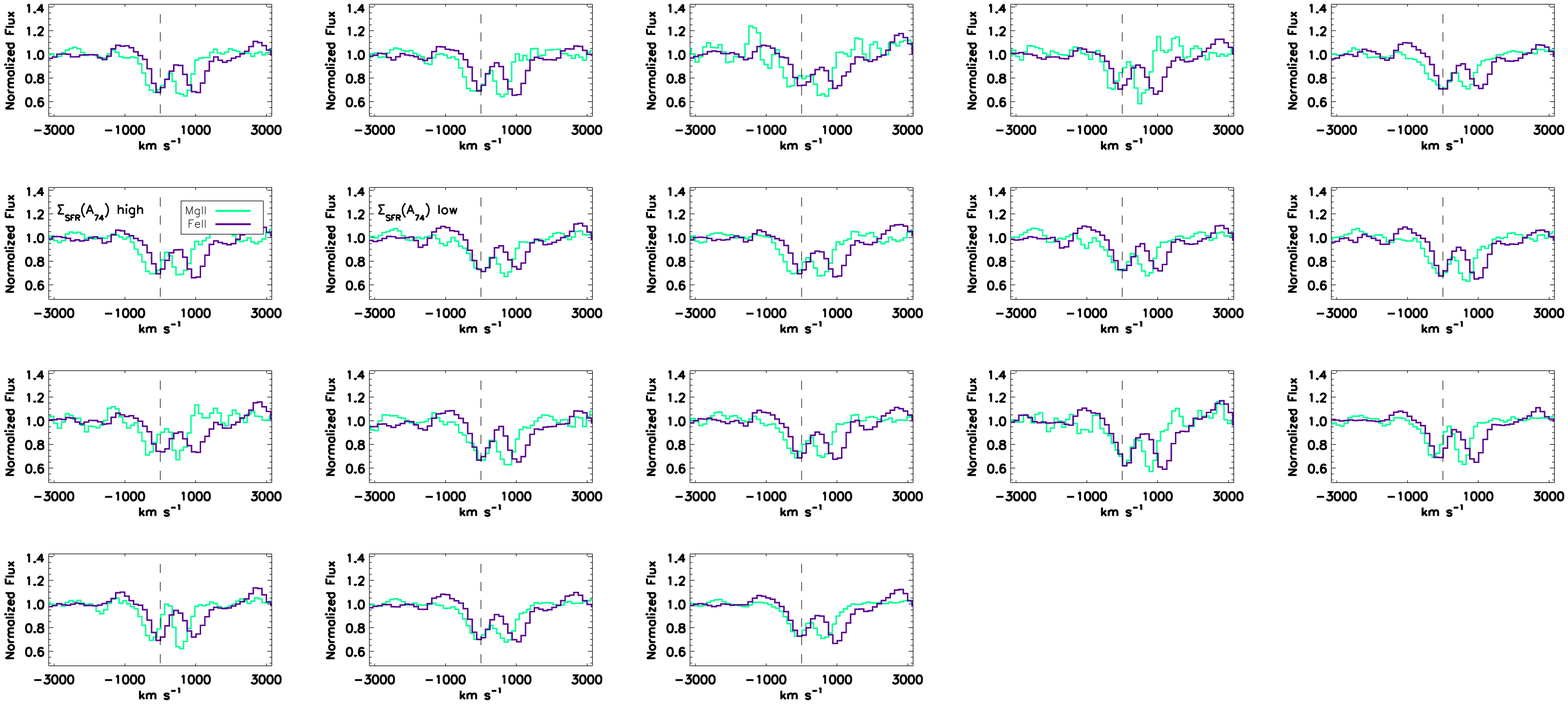}
\caption{Comparison of Fe~II 2374 and Mg~II 2796 profiles in velocity space, for the high and low $\Sigma_{\rm SFR}$(A$_{74}$) composite spectra. Note the pronounced variation between the blue wings of the features in the high $\Sigma_{\rm SFR}$(A$_{74}$) composite; $V_{max}$(Mg~II) is --855 $\pm$ 66 \kms\ while $V_{max}$(Fe~II) is --611 $\pm$ 103 \kms. High$\Sigma_{\rm SFR}$(A$_{74}$) systems show gas at high velocity, and this gas is better traced by Mg~II than by Fe~II (Section \ref{sec: discussion}). The absorption features at $\sim$ +770 and $\sim$ +1000 \kms\ are Mg~II 2803 and Fe~II 2382, respectively.}
\label{velocity_space_2796_2374}
\end{figure}

\begin{figure*}
\centering
\includegraphics[width=7in]{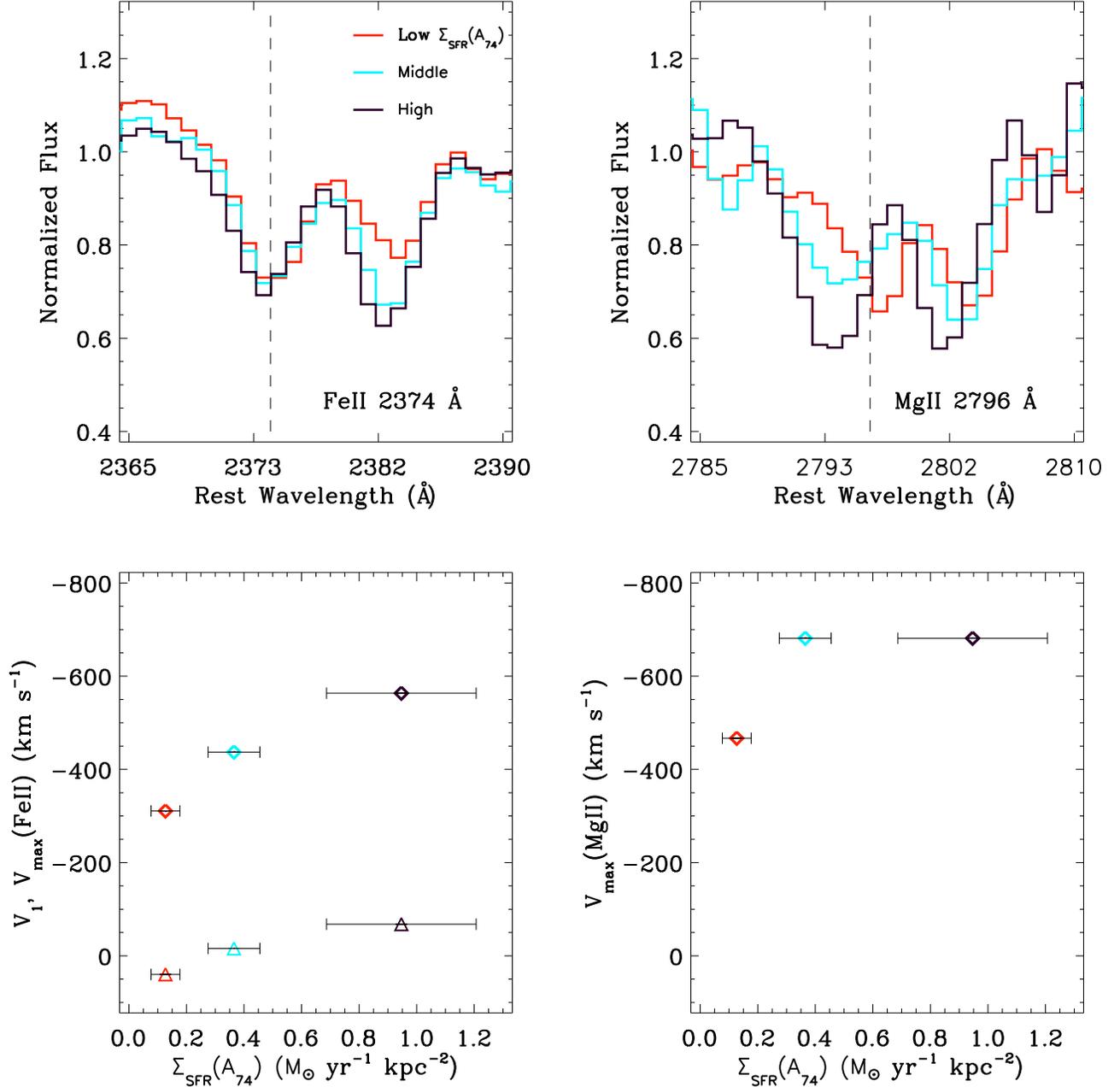}
\caption{\emph{Top: }The regions around Fe~II 2374 \AA\ and Mg~II 2796 \AA\ are shown for three composite spectra assembled from data ordered by $\Sigma_{\rm SFR}$(A$_{74}$). With increasing $\Sigma_{\rm SFR}$, the blue wings of both the Fe~II and Mg~II features extend to larger velocities. This is evidence of the increasing prevalence of high velocity gas with larger $\Sigma_{\rm SFR}$. \emph{Bottom: }We quantify the blue wings of Fe~II and Mg~II with $V_{max}$(Fe~II) and $V_{max}$(Mg~II) measurements. The left panel shows $V_{max}$(Fe~II)(diamonds) and $V_1$ (triangles) versus $\Sigma_{\rm SFR}$(A$_{74}$), where the errors bars on $\Sigma_{\rm SFR}$(A$_{74}$) indicate the 1$\sigma$ dispersion in each bin. The right panel shows the relationship between $V_{max}$(Mg~II) and $\Sigma_{\rm SFR}$(A$_{74}$). In all cases, more extended wing profiles tracing higher velocity gas are seen in systems with larger star-formation rate surface densities.}
\label{vmax_SFRSDtertiles}
\end{figure*}

The kinematic differences between Fe~II and Mg~II in the $\Sigma_{\rm SFR}$(A$_{74}$) composites are highlighted in Figure \ref{velocity_space_2796_2374}. On the left, the Fe~II 2374 \AA\ and Mg~II 2796 \AA\ lines of the high-$\Sigma_{\rm SFR}$(A$_{74}$) composite are overplotted in velocity space. The profiles of Fe~II and Mg~II absorption are clearly disparate, with the extended blue wing of Mg~II indicative of high-velocity gas not traced by Fe~II. On the right side of Figure \ref{velocity_space_2796_2374}, the same Fe~II and Mg~II transitions are shown for the low-$\Sigma_{\rm SFR}$(A$_{74}$) composite. In the regime of low-intensity star formation, the profiles of Fe~II and Mg~II are much more similar and Mg~II does not exhibit an extended blue wing. These results support the association of high star-formation rate surface densities with high-velocity gas, and Mg~II is observed to be better tracer of high-velocity gas than Fe~II. In order to better study the trend between outflow velocity and $\Sigma_{\rm SFR}$, we constructed composite spectra from three bins of $\Sigma_{\rm SFR}$(A$_{74}$). We find that the $V_1$, $V_{max}$(Fe~II), and $V_{max}$(Mg~II) measurements are all correlated with $\Sigma_{\rm SFR}$(A$_{74}$), where objects characterized by higher intensities of star formation have larger centroid and maximal outflow velocities (Figure \ref{vmax_SFRSDtertiles}). We observe that $V_{max}$(Fe~II) is a smooth function of $\Sigma_{\rm SFR}$(A$_{74}$), while $V_{max}$(Mg~II) appears to reach a maximum at $\Sigma_{\rm SFR}$(A$_{74}$) = 0.4 M$_{\odot}$ yr$^{-1}$ kpc$^{-2}$. The relatively coarse pixel sampling of the composite spectra -- $\sim$ 100 \kms\ -- is likely responsible for the plateau of $V_{max}$(Mg~II) values. From the composite spectra, it is evident that the equivalent width of blueshifted Mg~II absorption steadily increases with increasing $\Sigma_{\rm SFR}$(A$_{74}$). 

\begin{figure*}
\centering
\includegraphics[trim = 0in 3in 0in 0in,clip,width=7in]{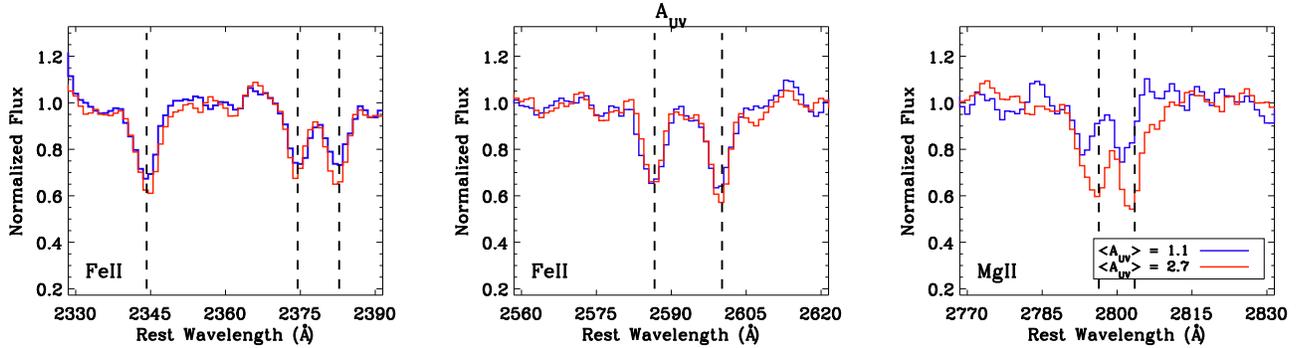}
\caption{Composite spectra assembled on the basis of A$_{\rm UV}$. Fitting the Fe~II resonance absorption lines yields a $V_1$ outflow velocity of --3 $\pm$ 10 \kms~ for the high-A$_{\rm UV}$ composite (red line) and --10 $\pm$ 9 \kms~ for the low-A$_{\rm UV}$ composite (blue line). Mg~II strength and kinematics are clearly correlated with A$_{\rm UV}$, where less attenuated objects show shallower Mg~II absorption troughs and more blueshifted features. In dusty objects, the lack of emission filling in the Mg~II features likely contributes to the observed absorption depth. These results are consistent with the trends observed for individual objects (Figure \ref{ID_MgII_SFR}).}
\label{AUV_smooth}
\end{figure*}

As previous observations have shown stronger interstellar absorption lines with increasing dust attenuation \citep{shapley2003}, indicative of an association between absorbing outflow material and dust, we investigated the relationship between A$_{\rm UV}$ and outflow velocity. In composite spectra assembled on the basis of A$_{\rm UV}$ (Figure \ref{AUV_smooth}), we find that objects with higher dust attenuation exhibit marginally stronger Fe~II absorption lines (especially in the 2344 \AA\ feature) and significantly stronger Mg~II absorption. These results support the correlation found by \citet{shapley2003}, where these authors observed that more attenuated Lyman Break Galaxies (LBGs) at \z3 showed deeper interstellar absorption lines. We find that objects with a variety of attenuation levels exhibit similar Fe~II kinematics (Figure \ref{AUV_smooth}); the centroids of the Fe~II lines in the high-A$_{\rm UV}$ composite ($\langle$A$_{\rm UV}$$\rangle$ = 2.7; extrema = [1.7, 4.2]) are best-fit with a velocity shift of --3 $\pm$ 10 \kms~ while the low-A$_{\rm UV}$ composite ($\langle$A$_{\rm UV}$$\rangle$ = 1.1; extrema = [0.0, 1.7]) shows a shift of --10 $\pm$ 9 \kms. These measurements are both consistent with no velocity shift. The $V_{max}$(Fe~II) measurements of the composites are consistent with each other within their 1$\sigma$ errors (--497 $\pm$ 88 \kms\ for the high-A$_{\rm UV}$ composite and --522 $\pm$ 91 \kms\ for the low-A$_{\rm UV}$ composite). \citet{law2012} find only a 0.7$\sigma$ correlation between outflow velocity and E(B--V) in a sample of 35 star-forming galaxies at \z2. Within a sample of local SDSS galaxies, \citet{chen2010} also find no evidence that attenuation level is linked to outflow velocity. 

While the Fe~II features of the two A$_{\rm UV}$ composites show only minimal variation in strength and kinematics, the Mg~II features are strikingly different between the high- and low-A$_{\rm UV}$ composites. We find that more attenuated objects have deeper Mg~II absorption troughs (consistent with a relative lack of emission filling or stronger absorption in dusty galaxies), Mg~II centroids roughly consistent with the systemic velocity, and $V_{max}$(Mg~II) = --732 $\pm$ 71 \kms\ while objects with lower A$_{\rm UV}$ have weaker absorption profiles, more strongly blueshifted centroids, and $V_{max}$(Mg~II) = --652 $\pm$ 89 \kms. \citet{prochaska2011} proposed outflow models in which dust has a pronounced effect on observed wind profiles: resonantly trapped photons (i.e., those arising from the Mg~II transitions at 2796 and 2803 \AA) have an increased probability of being absorbed by dust given their longer path lengths. Our observations support this hypothesis that Mg~II should show more variation with changing A$_{\rm UV}$ levels than the Fe~II transitions that are not resonantly trapped. However, \citet{prochaska2011} note that extreme levels of extinction ($\tau_{\rm dust}$ $\sim$ 10) are necessary in order to measure qualitative changes in line profiles, although we see this effect at significantly lower attenuation levels ($\tau_{\rm dust}$ $\sim$ 2). 

\subsection{Trends with Inclination and Morphology} \label{inclination}

Observations of local starbursts have shown extraplanar outflows emanating perpendicular to galaxy disks \citep[e.g.,][]{heckman1990,lehnert1996,cecil2001,strickland2009}. If this geometry persists at higher redshifts, one would expect the measured outflow velocities in the sample to depend strongly on inclination as the wind becomes increasingly oriented toward the observer. Studies to date have primarily focused on testing this hypothesis in local samples, given the advantages of apparent magnitude and spatial resolution. \citet{chen2010} used a sample of $\sim$ 150,000 SDSS galaxies and assembled stacks of the Na D absorption lines on the basis of galaxy inclination measured from two dimensional de Vaucouleurs \citeyearpar{devaucouleurs1948} fits to  \emph{ugriz} data. These authors found a striking correlation of increasing outflow velocity with decreasing inclination, consistent with the physical picture of winds escaping perpendicular to galaxy disks. Similarly, \citet{heckman2000} observed 18 local starbursts in the Na D lines and measured an increase in the fraction of objects exhibiting outflows when the sample was restricted to objects with \emph{i} $\leq$ 60$^{\circ}$. In the high-redshift regime, \citet{weiner2009} examined 118 galaxies at \emph{z} = 1.4 with \emph{HST} \emph{I}-band imaging and extracted minor-to-major axis ratios (\emph{b/a}) using SExtractor \citep{bertin1996}. These authors did not find a correlation between axis ratio and wind strength or outflow velocity; \citet{weiner2009} propose that the uncertainties on the axis ratios of small, irregular galaxies imaged in the rest-frame \emph{U}-band may be substantial. \citet{law2012} looked at 35 optically-selected star-forming galaxies at \emph{z} = 1.5--3.6 with \emph{b/a} estimated from GALFIT \citep{peng2002,peng2010} modeling of \emph{HST} F160W imaging and also found no correlation between outflow velocity and \emph{b/a}. It is important to note that these results at high redshift are much more uncertain than the \citet{chen2010} and \citet{heckman2000} findings, given the very different spatial resolutions achievable at \emph{z} = 0 and \z2. Furthermore, while \emph{b/a} and inclination are often presumed to be related by \emph{i} = cos$^{-1}$[\emph{b}/\emph{a}] \citep{tully1977}, \citet{law2012} caution that this conversion relies on the thin-disk approximation which may not be valid for star-forming galaxies with significant vertical velocity dispersions \citep{law2009,forsterschreiber2009}. However, in the absence of a significant sample of high-redshift galaxies for which spatially-resolved velocity dispersions have been measured, this simple conversion continues to be employed. As some of the objects in our sample furthermore resemble local star-forming disks more closely than the clumpy, irregular galaxies at \z2 with large vertical velocity dispersions, the thin-disk approximation may not be a gross assumption here. 

\begin{figure}
\centering
\includegraphics[width=3.5in]{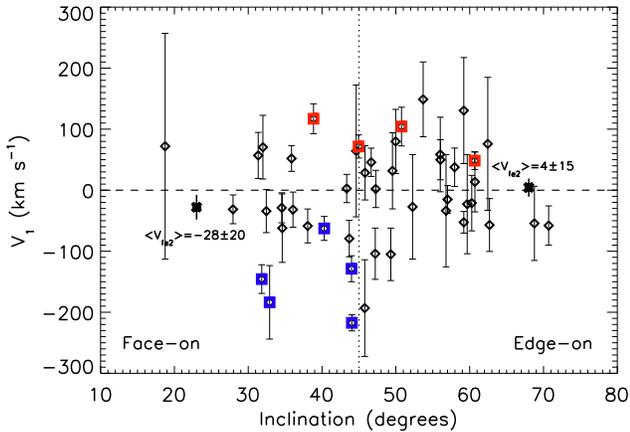}
\caption{$V_1$ outflow velocity versus galaxy inclination, where inclination was estimated from axial ratios: \emph{i} = cos$^{-1}$[\emph{b}/\emph{a}] \citep{tully1977}. Significant detections of outflows or inflows at the 3$\sigma$ level are highlighted in blue and red, respectively. While individual inclination measurements are likely uncertain by $\sim$ 10$^{\circ}$ due to measurement error in the rest-frame UV, we note that a simple division of the sample at \emph{i} = 45$^{\circ}$ yields a subgroup of more face-on galaxies with $\langle$$V_1$$\rangle$ = --28 $\pm$ 20 \kms\ and a set of more inclined galaxies with $\langle$$V_1$$\rangle$ = 4 $\pm$ 15 \kms. These findings are in the same sense as the results of \citet{chen2010}, who observed a correlation between outflow velocity and inclination in a sample of $\sim$ 150,000 local SDSS \citep{york2000} galaxies such that more face-on systems showed a stronger outflow component.} 
\label{ID_BA_dV}
\end{figure}

We investigate the relationship between galaxy inclination and outflow velocity using inclinations estimated from SExtractor \citep{bertin1996} axial measurements of summed \emph{HST} \emph{V}+\emph{I} detection images. In Figure \ref{ID_BA_dV}, we plot outflow velocity versus inclination. While inclination estimates for individual objects are likely uncertain by $\sim$ 10$^{\circ}$ due to measurement error in the rest-frame UV, highlighting the objects showing $\ge$ 3$\sigma$ $V_1$ outflows reveals that all such five objects have \emph{i} $<$ 45$^{\circ}$. When we make a simple division of the data at \emph{i} = 45$^{\circ}$, we additionally find that the sample of more edge-on systems has $\langle$$V_1$$\rangle$ = 4 $\pm$ 15 \kms\ while the sample of predominantly face-on systems is characterized by $\langle$$V_1$$\rangle$ = --28 $\pm$ 20 \kms. These results are consistent with the physical picture of a biconical wind profile where outflowing material preferentially emanates perpendicular to the galactic disk. \citet{weiner2009} investigated the relationship between outflow velocity and galaxy inclination at \z1.4 using \emph{HST} imaging, although these authors relied on observations in a single band (\emph{I}). The \citet{weiner2009} imaging data are accordingly shallower than ours by a factor $\sim$ $\sqrt{2}$, although we propose that the main reason we find a correlation between inclination and outflow velocity while \citet{weiner2009} do not lies in the relative S/N ratios of our respective spectroscopic data. The individual spectra used by \citet{weiner2009} have a S/N of $\sim$ 1 per resolution element near 2820 \AA. Conversely, the spectra presented in this paper have a S/N of $\sim$ 11 per resolution element in the vicinity of the Fe~II lines. The co-added spectra employed by \citet{weiner2009} for the majority of their analyses (although not inclination) have S/N of 16--28 per resolution element, while the spectral stacks used in this paper have S/N of 49--68 per resolution element. 

\begin{figure*}
\begin{center}$
\begin{array}{c} %inclination.pro
\includegraphics[width=5in]{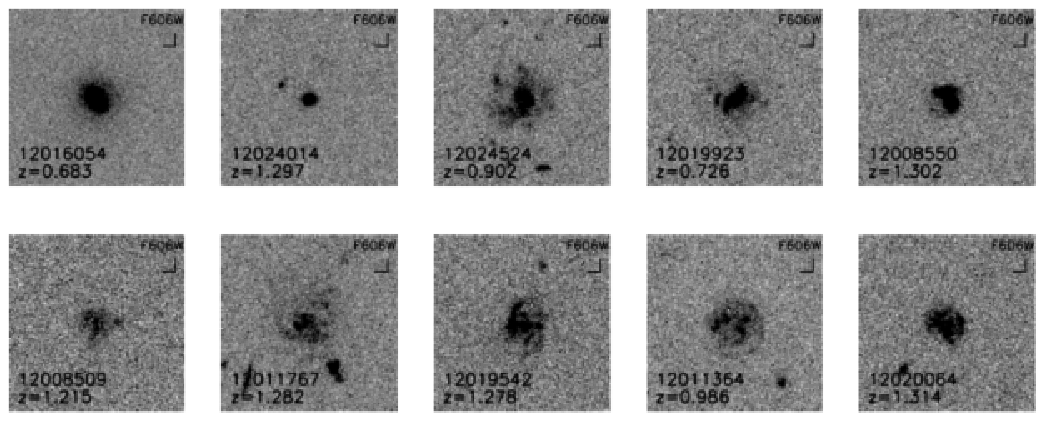} \\
\includegraphics[trim = 0in 2.8in 0in 0in,clip,width=6.5in]{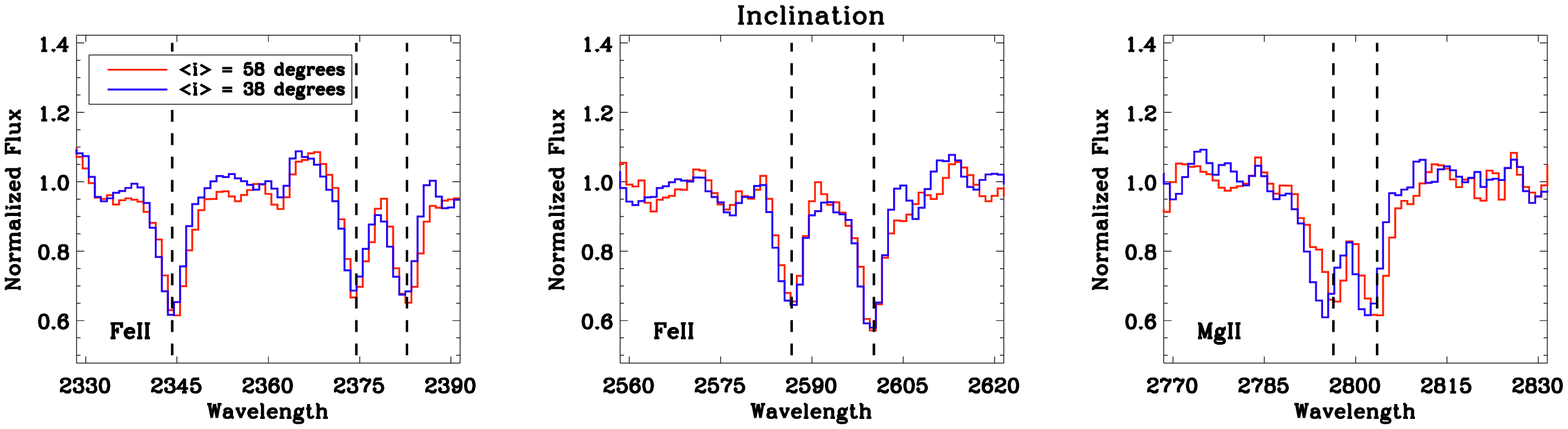} \\
\includegraphics[width=5in]{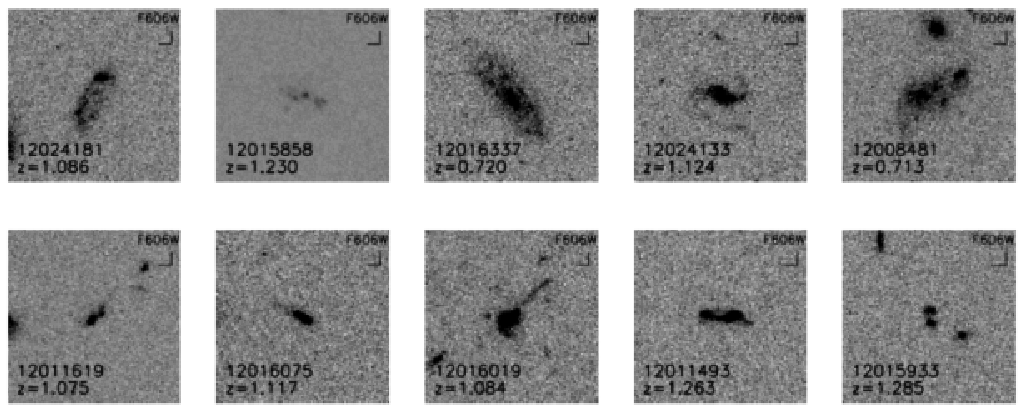}
\end{array}$
\end{center}
\caption{Composite spectra assembled from objects with high and low inclinations, respectively, where inclinations were estimated from \emph{HST} \emph{V}+\emph{I} band imaging assuming intrinsic circular disks: \emph{i} = cos$^{-1}$[\emph{b}/\emph{a}] \citep{tully1977}. We find that the composite spectrum composed of high inclination objects (red line) is best-fit with $V_1$ = 28 $\pm$ 11 \kms~while the composite spectrum assembled from low inclination galaxies (blue line) has a $V_1$ outflow velocity of --19 $\pm$ 9 \kms. Even though the overall shifts of each composite from 0 \kms~are small in magnitude, the kinematic difference between the two composites is significant. These results are consistent with the trend of stronger blueshifts with decreasing inclination, as noted by \citet{chen2010} for a local sample of SDSS galaxies. \emph{V}-band thumbnails of the ten lowest and highest inclination objects are shown above and below the composite spectra, respectively.}
\label{inclination_all}
\end{figure*}

We also utilized the statistical power of our sample in assembling composite spectra of high- and low-inclination objects, where $\langle i \rangle$ = 58$^{\circ}$ and 38$^{\circ}$, respectively. In Figure \ref{inclination_all}, we compare these composite spectra and show thumbnails of the ten highest (lowest) inclination objects below (above) the composite spectra for visual reference. While the magnitude of the shifts with respect to systemic velocity are small in both cases, we do find that galaxies with lower inclinations exhibit bluer Fe~II absorption troughs ($V_1$ = --19 $\pm$ 9 \kms) than more inclined galaxies ($V_1$ = 28  $\pm$ 11 \kms), qualitatively consistent with the results of \citet{chen2010}. Similarly, objects with lower inclinations show more blueshifted Mg~II lines ($V_{max}$(Mg~II) = --811 $\pm$ 140 \kms) than more edge-on systems (--692 $\pm$ 95 \kms). Likewise, the composite of lower (higher) inclination systems are characterized by $V_{max}$(Fe~II) = --557 $\pm$ 101  (--464 $\pm$ 77) \kms. 

The morphology of star-forming galaxies hosting outflows has only recently been investigated \citep{martin2005,sato2009,weiner2009,rubin2010a,law2012}. In the local universe, there is an implicit bias toward concluding that outflows are preferentially hosted by mergers. As systems with large SFRs are more likely to exhibit outflows \citep{martin2005,rupke2005} and local galaxies with vigorous star formation are often characterized by disrupted morphologies indicative of either major or minor mergers \citep[e.g.,][]{sanders1996}, an association between mergers and outflows naturally develops. In a study by \citet{martin2005} of 18 ULIRGs at 0.04 $<$ \emph{z} $<$ 0.16 characterized by merger signatures, these authors found evidence for neutral gas outflows in 15 systems ($>$ 80\%), consistent with \citet{rupke2005}. At larger redshifts, where vigorous star formation and merger morphology appear to be decoupled \citep{law2007,swinbank2010}, \citet{sato2009} still found that outflows traced by Na D were preferentially hosted by galaxies undergoing mergers. 

\begin{figure}
\centering
\includegraphics[width=3.5in]{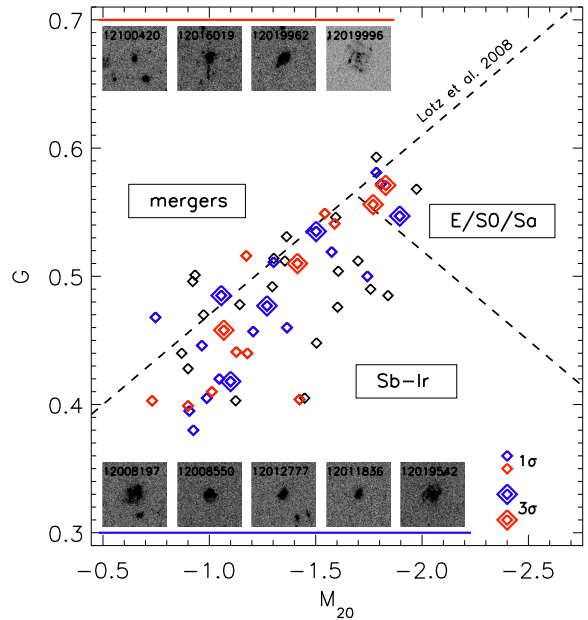}
\caption{Outflow demographics as a function of the quantitative morphological parameters $G$ and M$_{20}$, where parameter space can be partitioned into regions encompassing the classical morphological groups -- mergers, ellipticals, and spirals/irregulars \citep{lotz2008}. We highlight in blue (red) objects showing blueshifted (redshifted) Fe~II resonance absorption lines, reserving small diamonds for 1$\sigma$ significant $V_1$ velocities and larger diamonds for 3$\sigma$ significant $V_1$ measurements. $HST$ $V$-band thumbnails are shown for the nine objects with 3$\sigma$ velocity dectections. Both inflows and outflows occur in galaxies spanning a variety of morphologies; merging systems do not preferentially host winds. These results are consistent with the findings of \citet{law2012} at higher redshift, although \citet{sato2009} noted that outflows were preferentially seen in merger candidates at 0.11 $<$ \emph{z} $<$ 0.54.}
\label{G_M20}
\end{figure}

We use our sample of \z1 objects to study the prevalence of outflows in systems with large star-formation rates but a wide range of morphologies. We use quantitative morphological parameters from \citet{lotz2004} to investigate the light distributions of galaxies hosting outflows. We employ the Gini coefficient ($G$) and the second order moment of the brightest 20\% of each galaxy's flux, $M_{20}$. Figure \ref{G_M20} shows how objects in ($G$, $M_{20}$) space divide into classical early-type (E/S0/Sa), late-type (Sb--Irr), and merger candidates \citep{lotz2008}. We find that objects exhibiting outflows and inflows span ($G$, $M_{20}$) morphological parameter space; systems classified as mergers do not appear to drive winds any more frequently than galaxies classified as E/S0/Sa or Sb--Irr. Objects with $G$, $M_{20}$ values indicative of mergers, in fact, qualitatively host a relative dearth of outflows and inflows compared with other morphological classes. Both \citet{weiner2009} and \citet{rubin2010a} also found no trend between morphology and outflows in populations of star-forming galaxies. 

%\citet{law2011} also present data that galaxies identified as mergers in (G, $M_{20}$) space or by pair classification are statistically equivalent to non-mergers in terms of both physical parameters (stellar mass, SFR, color, etc.) and gas-phase kinematics. The one robust difference between mergers and non-mergers found by \citet{law2011} is galaxy size, where mergers having significantly smaller radii and correspondingly higher star-formation rate surface densities than non-mergers. We do not reproduce the same trend in our data, however; we compare the 11 objects classified as mergers in (G, $M_{20}$) space with the non-mergers and find a Kolmogorov-Smirnov probability of 93\% (72\%) that the two samples are drawn from the same parent distribution of $\pi$$R_{\rm P}^2$ (A$_{74}$) areas.

\begin{figure*}
\centering
\includegraphics[width=6in]{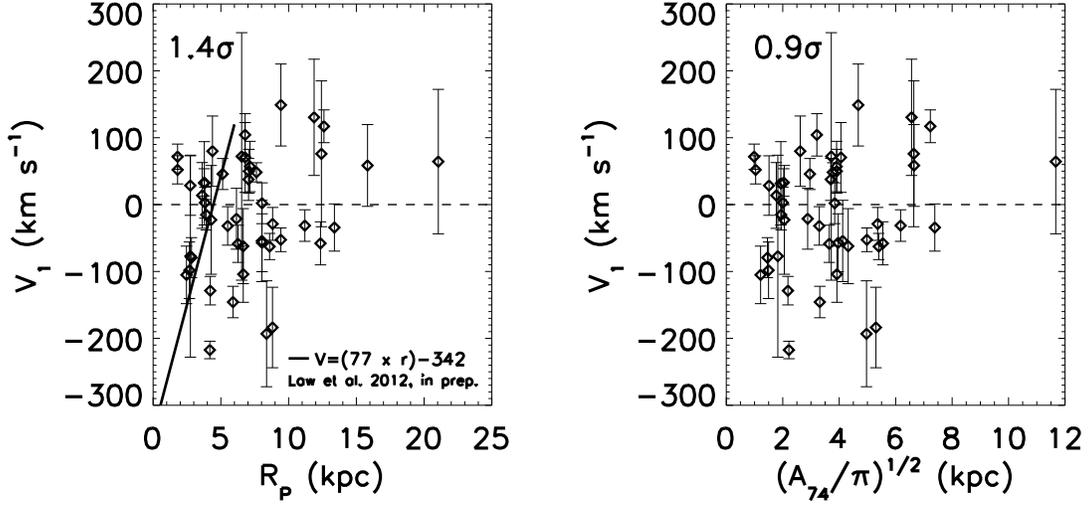}
\caption{\emph{Left:} Outflow velocity versus Petrosian radius. We overplot the relation from \citet{law2012}, where these authors found a strong correlation between outflow velocity and size in a sample of galaxies at \emph{z} = 1.5--3.6. The \citet{law2012} objects have semi-major axes \emph{r} $<$ 6 kpc, while our sample is characterized by $\langle$R$_{\rm P}$$\rangle$ =  7.1 kpc. We find no statistical evidence in our sample that Petrosian radius and outflow velocity are correlated. \emph{Right:} Outflow velocity versus the ``radius" of the A$_{74}$ area: (A$_{74}$/$\pi$)$^{1/2}$. As with the Petrosian radius, we find no evidence indicating a relation between outflow velocity and (A$_{74}$/$\pi$)$^{1/2}$.}
\label{size_relation}
\end{figure*}

\citet{law2012} studied the prevalence of outflows with galaxy size. These authors parameterized size using both the half-light radius along the semi-major axis (\emph{r}) and the circularized effective radius ($r_e$ = \emph{r$\sqrt{b/a}$}). \citet{law2012} found a $>$ 2$\sigma$ correlation with both size estimates, where the smallest galaxies exhibited the largest interstellar blueshifts with respect to the systemic frame defined by H$\alpha$. In Figure \ref{size_relation}, we use our data to plot outflow velocity versus two size estimates: Petrosian radius and (A$_{74}$/$\pi$)$^{1/2}$ (the ``radius" of the A$_{74}$ area). We show the best-fit line derived by \citet{law2012}, V = 77\emph{r} -- 342 \kms, for reference, but note that only $\sim$ 40\% of our sample has Petrosian radii falling in the same range as the \citet{law2012} data. We find no evidence for a correlation between outflow velocity and either Petrosian radius or (A$_{74}$/$\pi$)$^{1/2}$. 

\subsection{Mg~II Equivalent Width \& Kinematics} \label{sec: MgII}

\begin{figure*}
\centering
\includegraphics[trim = 0in 0in 0in 0.1in,clip,width=7in]{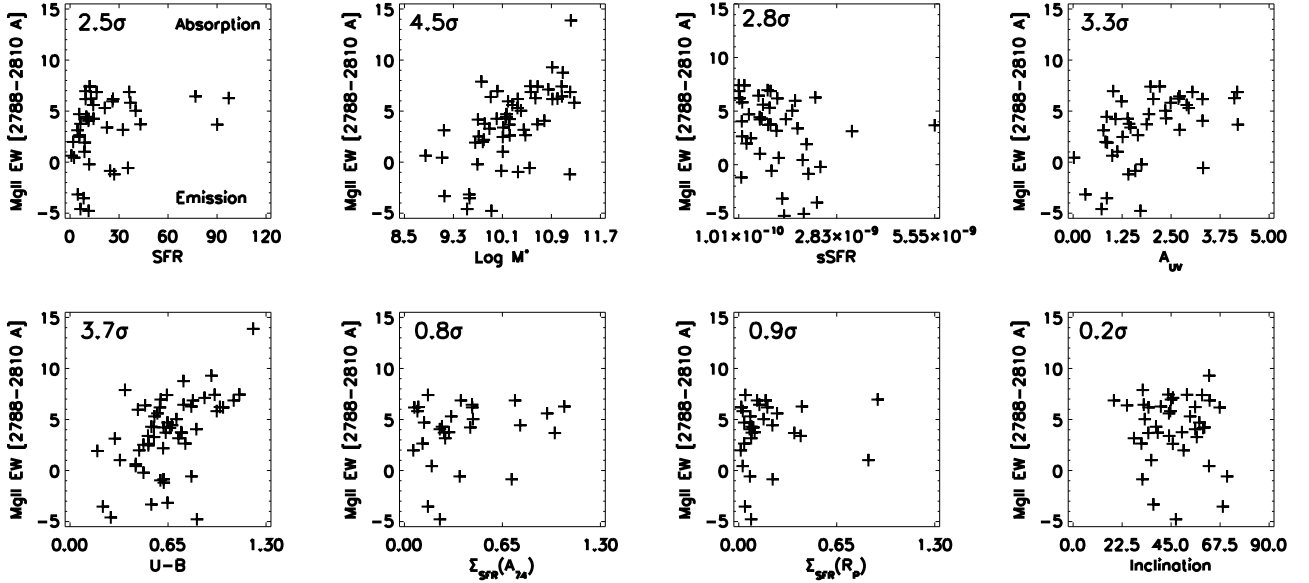}
\caption{Scatter plots of various galaxy properties versus Mg~II equivalent width, for galaxies in the EGS. The strength of Mg~II was estimated from a simple summation of flux over the interval 2788--2810 \AA\ encompassing both features of the doublet. We find that the strongest correlation, at 4.5$\sigma$ ($r_S$ = 0.61), is observed between Mg~II equivalent width and stellar mass. A thorough discussion of Mg~II will appear in Kornei et al. 2012b, in prep. and Martin et al. 2012b, in prep.}
\label{ID_MgII_SFR}
\end{figure*}

Motivated by the range of Mg~II absorption strengths observed in the sample, we investigated the primary driver of Mg~II equivalent width. We estimated the equivalent width of Mg~II by summing flux over the interval 2788--2810 \AA, thereby encompassing both features of the Mg~II doublet. We tested the correlation between Mg~II equivalent width and SFR, stellar mass, sSFR, A$_{\rm UV}$, \emph{U}--\emph{B} color, $\Sigma_{\rm SFR}$(A$_{74}$), $\Sigma_{\rm SFR}$(R$_{\rm P}$), and inclination, respectively (Figure \ref{ID_MgII_SFR}). Consistent with \citet{martin2012}, we find that stellar mass and Mg~II equivalent width are the most strongly correlated, at the 4.5$\sigma$ level ($r_S$ = 0.61). \citet{bordoloi2011} also note the correlation between Mg~II equivalent width and stellar mass in a sample of COSMOS galaxies at 0.5 $<$ \emph{z} $<$ 0.9. Stellar Mg~II absorption may be present in objects with large stellar masses; population modeling is necessary in order to definitively address the contribution of stellar absorption. 

Among the SFR, sSFR, and A$_{\rm UV}$ composite spectra, we find two families based on the strength and kinematics of Mg~II. The low-SFR, high-sSFR, and low-A$_{\rm UV}$ composite spectra exhibit weaker, blueshifted Mg~II features with little absorption in the red wings of the lines. On the other hand, the low-sSFR and high-A$_{\rm UV}$ composite spectra show stronger Mg~II absorption at roughly the systemic velocity, with substantial absorption in the lines' red wings. The high-SFR composite spectrum, with its strong, blueshifted Mg~II absorption, is anomalous in that appears to be included in both families. We note that the differences in the Mg~II profiles of these SFR, sSFR, and A$_{\rm UV}$ composite spectra are mainly limited to the red wings of the lines and are distinct from the shifting of the blue wings and centroids observed in the $\Sigma_{\rm SFR}$(A$_{74}$) and $\Sigma_{\rm SFR}$(R$_{\rm P}$) composite spectra. We propose that objects in the low-SFR, high-sSFR, and low-A$_{\rm UV}$ composite spectra are dominated by emission filling while the galaxies populating the low-sSFR and high-A$_{\rm UV}$ composite spectra experience significant systemic absorption. Furthermore, since stellar mass is correlated with SFR, sSFR, and A$_{\rm UV}$, the variation in Mg~II properties seen in the SFR, sSFR, and A$_{\rm UV}$ composite spectra may reflect the primary correlation between stellar mass and Mg~II equivalent width. Accordingly, the low-SFR, high-sSFR, and low-A$_{\rm UV}$ composite spectra are preferentially populated with low-mass objects with weaker Mg~II absorption \citep{martin2012}. Higher mass objects may have more absorption and also less emission filling. Lower levels of emission filling indicate attenuation by dust, which may be more prevalent in higher mass objects. Future work involving modeling of the stellar component of Mg~II absorption will help illuminate the connection between Mg~II and stellar populations. 

\section{Discussion}\label{sec: discussion}

In this section, we present a discussion and interpretation of our results. We observe the strongest trend between outflow velocity and $\Sigma_{\rm SFR}$, a finding that is supported both by theoretical work and also by observations at lower redshifts. We discuss the prevalence of outflows in the sample and consider the uncertainties of using different tracer species and measurement diagnostics to parameterize a single characteristic outflow velocity. 

\subsection{Outflow Velocity and the Star-Formation Rate Surface Density}

Our finding of a strong correlation between outflow velocity and the \emph{concentration} of star formation -- as opposed to the global SFR -- is consistent with previous observations. In a sample of star-forming galaxies at \emph{z} = 1.5--3.6, \citet{law2012} reported a 2.1$\sigma$ correlation between outflow velocity and $\Sigma_{\rm SFR}$. At a comparable redshift, \citet{genzel2011} used SINFONI observations of five galaxies to measure a higher incidence of broad H$\alpha$ profiles in star-forming clumps with larger H$\alpha$ surface brightnesses. Since broad wings are indicative of gas moving at high velocities and larger H$\alpha$ surface brightnesses can be directly linked to larger star-formation rate surface densities, these authors posit that their results may be evidence for galactic winds being modulated by the intensity of star formation. The lack of broad H$\alpha$ components in fainter interclump regions of the \citet{genzel2011} sample furthermore suggests that winds may require some threshold $\Sigma_{\rm SFR}$. Similarly, in a sample of $\sim$ 50 galaxies at \emph{z} = 1.2--2.6 with high H$\alpha$ surface brightnesses, \citet{letiran2011} found that systems with the highest star-formation rate surface densities were more likely to exhibit a broad H$\alpha$ component. In searching for a physical explanation for the incidence of broad H$\alpha$ kinematics, which are interpreted as arising in outflowing gas, \citet{letiran2011} used the ratio of [S~II] 6716 / [S~II] 6731 to infer electron densities \citep{osterbrock1989}. Based on conversions between electron density and thermal pressure, these authors found that the systems with the highest star-formation intensities were characterized by pressures similar to the values found in local starbursts exhibiting galactic winds \citep[e.g.,][]{lehnert1996}. \citet{letiran2011} accordingly proposed that thermal pressure may modulate the incidence of galactic winds. 

In addition to thermal pressure arising from the hot gas expelled by supernovae, radiation pressure on dust grains is also commonly invoked as a mechanism for driving galactic winds \citep[e.g.,][]{socrates2008,murray2011}. When dust grains experience a force from impinging photons, the cool gas that is coupled to the dust is also accelerated. Several authors have examined the relative contributions of thermal and radiation pressure in driving galactic winds. \citet{murray2011} propose that radiation pressure is the primary driver lofting gas clouds above the galactic disk in the first $\sim$ 4 Myr before supernovae explode. These authors hypothesize that a critical gas surface density is required to provide the initial radiation pressure to accelerate the gas cloud above the galactic disk. Once a bubble of gas is sufficiently separated from the galactic plane, ram pressure from the hot winds of later generations of supernovae accelerates the gas \citep[with a force comparable to the force from radiation pressure;][]{murray2005} above the galactic disk. The predicted scaling relation between outflow velocity and the SFR due to ram pressure alone is shallow:  V $\propto$ SFR$^{1/5}$ \citep{ferrara2006} or V $\propto$ SFR$^{1/4}$ \citep{heckman2000}. On the other hand, for a purely radiatively driven wind, the outflow velocity is expected to scale roughly linearly with the SFR \citep{sharma2011}. Using data from the \z3 LBG study of \citet{shapley2003}, \citet{ferrara2006} find a scaling relation between SFR and outflow velocity of the form V $\propto$ SFR$^{2/3}$. However, these authors make the assumption that outflow velocity can be calculated from the velocity difference between interstellar absorption lines and Ly$\alpha$ emission. As Ly$\alpha$ is resonantly scattered and often observed to be redshifted by several hundred km s$^{-1}$ relative to the systemic velocity frame defined by H$\alpha$ emission \citep{steidel2010}, the outflow velocities employed by \citet{ferrara2006} may very well be overestimates. Indeed, analysis of the \citet{shapley2003} data reveals that composite spectra differing in SFR have nearly identical gas kinematics as traced by interstellar absorption lines alone. The \citet{ferrara2006} conclusion that outflow velocity and SFR are related via the form V $\propto$ SFR$^{2/3}$ is primarily driven by kinematic changes in Ly$\alpha$ as opposed to true variation in the interstellar species tracing the wind. The \citet{shapley2003} data suggest that outflow velocity varies much more weakly, if at all, with SFR. Observations by \citet{martin2005} (V $\propto$ SFR$^{0.35}$) are not grossly inconsistent with the SFR$^{1/5}$ scaling predicted by \citet{ferrara2006} for acceleration by ram pressure alone, although various authors have suggested that galactic winds are likely accelerated by a combination of radiative and ram pressures \citep{sharma2011,murray2011}. We do not find a significant trend of outflow velocity with SFR, although we note that the relatively low resolution of our data paired with the limited range of SFRs probed may preclude a robust measurement of a trend, particularly if the scaling is shallow. 

As discussed in \citet{genzel2011}, the efficiency of radiation pressure in driving galactic winds can be linked to the surface density of star formation (i.e., $\Sigma_{\rm SFR}$) via simple heuristic arguments.  Environments with higher star-formation rate surface densities have higher gas surface densities \citep{kennicutt1998}. Gas surface density, in turn, is correlated with the dust surface density and, analogously, the dust opacity. The radiation force on dust grains, $\dot{p}_{\rm rad}$, is more efficient in dustier environments \citep[$\dot{p}_{\rm rad}$ $\propto$ $\tau L$/\emph{c}, where $\tau$ is the optical depth to radiation, \emph{L} is the bolometric luminosity, and \emph{c} is the speed of light;][]{veilleux2005}. As systems with higher star-formation rate surface densities will produce more radiation to begin with (having a larger concentration of star formation), it is expected that the combination of higher radiation levels and the efficient radiative coupling found in high $\Sigma_{\rm SFR}$ environments -- due to larger concentrations of dust -- will translate into effective driving of galactic winds. In the sample discussed here, however, we do not find that dustier systems show stronger outflows; the high- and low-A$_{\rm UV}$ composite spectra have $V_1$, $V_{max}$(Fe~II), and $V_{max}$(Mg~II) values consistent at the 1$\sigma$ level. This null result may indicate that radiation pressure does not dominate the acceleration of galactic winds, or that our dataset is simply too limited in size and S/N to unambiguously detect a trend between A$_{\rm UV}$ and outflow velocity. 

Several authors have estimated the form of the scaling relation between $\Sigma_{\rm SFR}$ and outflow velocity. In the context of outflows being driven from disrupted giant molecular clouds, \citet{murray2011} parameterize the ejection velocity of gas due to radiation pressure as proportional to the square of the gas surface density, $\Sigma_g$:

\begin{equation} v_{ej} \propto \frac{R_d^2 \Sigma_g^2}{v_c^2} \end{equation}

\noindent where $v_{ej}$ is the ejection velocity (assumed to be analogous to the outflow velocity), $R_d$ is the galaxy disk radius, and $v_c$ is the galaxy circular velocity. Assuming the Kennicutt-Schmidt \citeyearpar{kennicutt1998} star-formation law has the form $\Sigma_{\rm SFR}$ $\propto$ $\frac{\Sigma_g}{\tau_{\rm dyn}}$, where $\tau_{\rm dyn}$ is the dynamical timescale (${R_d}$/$v_c$), \citet{murray2011} alternatively express $v_{ej}$ as proportional to the square of the star-formation rate surface density:

\begin{equation} v_{ej} \propto \frac{R_d^4 \Sigma_{\rm SFR}^2}{v_c^4}.\end{equation}

The additional terms corresponding to $R_d$ and $v_c$ may translate into scatter in the $v_{ej}$ -- $\Sigma_{\rm SFR}$ relation; using our estimates of the Petrosian radius and the width of the [OII] emission feature as proxies for $R_d$ and $v_c$, respectively, we find that $\tau_{\rm dyn}$ spans a factor of $\sim$ 30 in our sample. \citet{murray2011} also derive a critical surface density of star formation required to launch a galactic wind by imposing the criterion that $v_{ej}$ $>$  $v_c$. These authors find that this critical density is dependent on both $R_d$ and $v_c$; a local $L_*$ galaxy has $\Sigma_{\rm SFR}^{\rm crit}$ $\approx$ 0.06 $M_{\odot}$ yr$^{-1}$ kpc$^{-2}$. This result is consistent with the observationally-motivated $\Sigma_{\rm SFR}$ threshold derived by \citet{heckman2002}: $\Sigma_{\rm SFR}^{\rm crit}$ $\approx$ 0.1 $M_{\odot}$ yr$^{-1}$ kpc$^{-2}$.

\citet{strickland2004} suggest an alternative scaling relation between outflow velocity and $\Sigma_{\rm SFR}$ assuming winds are driven by thermal pressure and that hot gas blows out of the galaxy three scale heights above the disk:

\begin{equation} v_{ej} \propto \left({\frac{\Sigma_{\rm SFR}}{\rho_0}}\right)^{1/3} \end{equation}

\noindent where $\rho_0$ is the gas density. Following \citet{chen2010}, the above relation can be equivalently expressed using the Kennicutt-Schmidt \citeyearpar{kennicutt1998} law and the approximation that the gas density is proportional to the gas surface density divided by the scale height of the disk, $H_z$:

\begin{equation} \label{eqn: strickland} v_{ej} \propto \Sigma_{\rm SFR}^{2/21} H_z^{1/3} \end{equation} 

\citet{strickland2004} adopt $\Sigma_{\rm SFR}$ $\propto$ $\Sigma_g^{1.4}$ as the form of the Kennicutt-Schmidt \citeyearpar{kennicutt1998} law. If one instead uses the same functional form of the Kennicutt-Schmidt \citeyearpar{kennicutt1998} law employed by \citet{murray2011}, one finds that $v_{ej}$ is independent of $\Sigma_{\rm SFR}$. Equation \ref{eqn: strickland} is consistent with the shallow scaling between outflow velocity and $\Sigma_{\rm SFR}$ recorded by \citet{chen2010}, although it is important to note that the small dynamic range of the \citet{chen2010} data (120 \kms\ $<$ $v_{ej}$ $<$ 160 \kms) may obscure trends. However, as mentioned above, \citet{chen2010} do find a significant correlation between $\Sigma_{\rm SFR}$ and the equivalent width of the Na D outflow component. The equivalent width of the outflow component is dependent on the velocity spread, covering fraction, and optical depth of the absorbing gas. 

The power-law relations discussed above can be generalized as being of the form $v_{ej}$ $\propto$ $\Sigma_{\rm SFR}^{\alpha}$, where $\alpha$ = 2.0 or 0.1. We use our $V_1$ and $\Sigma_{\rm SFR}$(A$_{74}$) data with the {\tt IDL} routine {\tt MPFITFUN} \citep{markwardt2009} to fit a power law of the form $V_1$ = A$\Sigma_{\rm SFR}$(A$_{74}$)$^{\alpha}$ + B, where A, $\alpha$, and B are free parameters. We choose to include only objects exhibiting outflows ($V_1$ $<$ 0 \kms) and we furthermore impose the prior that the additive term of the power law, B, is 0 (i.e., $V_1$ converges to 0 as $\Sigma_{\rm SFR}$(A$_{74}$) goes to 0). When we fit a power law with $\alpha$ =  2.0 or 0.1 to our data, we find a better fit with $\alpha$ = 2.0 (although the $\chi^2$ values per degree of freedom are large in both cases: 6.3, 7.5). Removing any priors on the slope, we recover a best-fit $\alpha$ of 0.8. However, we caution that the relatively large errors on our outflow velocities prevent a robust discrimination of $\alpha$. 

We can also bin the data in order to reduce the uncertainties on the velocity measurements. We divided the individual data points into three groups according to $\Sigma_{\rm SFR}$(A$_{74}$) and fit a power law of the same form as discussed above to the average $V_1$ and $\Sigma_{\rm SFR}$(A$_{74}$) values in each bin. We recover a best-fit slope of 0.3. When we bin all the data, including the objects showing apparent inflows but still excluding the two outliers at $\Sigma_{\rm SFR}$(A$_{74}$) $>$ 2.0, we find a best-fit slope of 1.6. As these various fits do not converge on a slope, it is clear that a larger sample of measurements is needed, ideally over a wider dynamic range in $\Sigma_{\rm SFR}$(A$_{74}$). Outflow velocities derived from higher resolution data will be furthermore beneficial for drawing meaningful conclusions about the slope of the power law relating outflow velocity and star-formation rate surface density. 

\subsection{Prevalence of Winds}

In the sample discussed here -- 72 objects in the EGS, representing a subset of the data discussed in \citet{martin2012}, -- we find 1$\sigma$ significant detections of blueshifted Fe~II absorption in $\sim$ 40\% of the sample. Likewise, 40 $\pm$ 5\% of objects show outflows with blueshifts of at least --40 \kms. This 40\% detection rate is a lower limit due to the incompleteness of measuring winds in low S/N spectra \citep{martin2012}. In stating that blueshifts are seen in $\sim$ 40\% of the sample, it is important to remember that the detection rate can be explained by degenerate assumptions about the Fe~II gas covering fraction and the percentage of galaxies hosting winds. If one assumes that the winds are isotropic with a covering fraction of 100\%, we can conclude that $\sim$ 40\% of galaxies host winds. On the other hand, 100\% of the sample could in fact host outflows if the Fe~II absorbing gas covers only $\sim$ 40\% of each galaxy. We favor the latter scenario -- in which winds are present in a majority of galaxies and yet observable only in a fraction of sources due to an average covering fraction less than unity -- as our work supports the collimation of winds: $\sim$ 70\% of the sample has star-formation rate surface densities above the threshold level predicted for driving outflows while only $\sim$ 40\% of objects show blueshifted Fe~II absorption lines. Other authors have also suggested that winds are collimated \citep{rupke2005,chen2010,bordoloi2011}. Imaging observations of winds in local starburst galaxies also reveal that outflows preferentially escape perpendicular to galactic disks \citep[e.g.,][]{heckman1990,deyoung1994,shopbell1998} and therefore do not cover the entirety of galaxies. We caution, however, that as high-redshift galaxies look less and less disk-dominated \citep{lotz2004,law2007,law2011}, the interpretation of outflows emanating particular to a presumed disk is complicated. Galactic winds appear to be ubiquitous in samples of clumpy, unresolved galaxies at \emph{z} = 2--3 \citep[e.g.,][]{shapley2003}. The galaxies in our \z1 sample, while generally not grand-design spirals, do not show the preponderance of disturbed morphologies characteristic of \emph{z} $\ge$ 2 samples \citep[e.g.,][]{law2007}. 

Our outflow detection frequency of 40\% is similar to the detection frequency reported for the parent sample of objects observed with LRIS \citep{martin2012}, indicative that the EGS and parent samples do not differ significantly in outflow properties. We also find that our outflow detection frequency is consistent with observations based on absorption lines in both local samples \citep{rupke2005} and at \z1.4 \citep{weiner2009}. Based on the absorption depth of Mg~II in a co-added spectrum of galaxies, these authors inferred that at least half of the galaxies in their sample exhibit outflows of cool, low-ionization gas. We note that the kinematics and absorption strength of the Mg~II lines in the \citet{weiner2009} composite spectrum are similar to those seen in our composite spectrum after accounting for the different spectral resolutions of our respective data \citep{martin2012}. This similarity is suggestive that the galactic winds probed by Mg~II are fundamentally alike in both our sample and the \citet{weiner2009} sample. 

While the kinematics of Mg~II in our composite spectrum are similar to those of the \citet{weiner2009} composite (i.e., offset from the systemic velocity by several hundred \kms), we find no blueshift in the Fe~II lines in our composite spectrum ($V_1$ = --5 $\pm$ 6 \kms). The spread of both positive and negative $V_1$ velocities observed for individual objects (Figure \ref{match_info_histograms_vel_EGS}) is consistent with a global average velocity close to 0 \kms, but the negligible blueshift of Fe~II lines in the composite spectrum may also be due to the smaller dynamic range of gas velocities probed by Fe~II than Mg~II \citep{rubin2010a,prochaska2011}. We discuss potential explanations for this decreased dynamic range below.

\subsection{Diagnostics of Winds}

Outflowing galactic winds can be measured using a variety of samples: background quasars probing galactic halos along the line-of-sight, galaxy-galaxy pairs, and isolated galaxies in which the stellar disk provides a continuum for absorption by foreground interstellar gas. Even considering only the last technique -- which we employ in this paper -- a range of outflow velocity measurements will be obtained for a galaxy depending on which interstellar tracer species are used and which parameterizations of absorption line kinematics are adopted. 

From the extensive literature of galactic winds published to date, it is evident that a range of tracer elements are used in tandem with a variety of measurement techniques to parametrize the velocity of outflowing gas. Before we discuss how the choices of tracer elements and line-fitting methods may strongly affect the derived outflow velocity and its scaling with galaxy properties, it is important to acknowledge the intrinsic uncertainty of extracting a single representative outflow velocity from a physical system characterized by many interacting clouds and shells of gas at a range of radii accelerated by a population of supernovae and massive stars injecting energy and momentum over a long timescale ($>$ 100 Myr). In the absence of extremely high resolution and high S/N spectroscopy, the kinematics extracted from the centroids of absorption lines must be thought of as tracing only the gross average of multiple fronts of outflowing gas. With this limitation in mind, we now turn to a discussion of how the choice of tracer species may impact the measured properties of the outflows. 

Absorption line studies are necessarily affected by the choice of absorption line probe. Authors often rely on relatively abundant species whose transitions are redshifted into the optical window of optimized detector sensitivity. Local studies typically employ Na D while investigations of higher redshift populations may use transitions of  Si~II, C~II, Al~II, Fe~II, Mg~II, or Mg~I in the rest-frame UV. As these species differ in abundance, ionization potential, susceptibility to emission filling, and (perhaps) spatial distribution, it is unsurprising that their associated outflow velocities differ as well. For instance, Na D stellar absorption is known to be stronger in older stellar populations; if the outflow and stellar components are not separated by means of modeling, the outflow velocity can be significantly underestimated and trends between galaxy properties and winds can become skewed or lost \citep[e.g.,][]{chen2010,coil2011}. The different ionization energies of Na~I, Fe~II, Mg~II, and Mg~I (5.1, 16.1, 15.0, and 7.6 eV, respectively) result in these species tracing different populations of gas: Na~I and Mg~I, with low ionization potentials, require dense, shielded gas in order to survive while Fe~II and Mg~II are less easily ionized and therefore survive in more widespread environments. Emission filling affects particular transitions of species to a differing degree depending on the relative availability and oscillator strengths of transitions decaying to excited ground states. The Mg~II doublet presents a particularly extreme case of emission filling. The doublet energy levels are such that these transitions are trapped; the absorption of a resonance photon necessarily results in the re-emission of another resonance photon. The Mg~II doublet is therefore strongly susceptible to emission filling and measurements of the centroids of these transitions may yield different kinematic information (i.e., larger blueshifts) relative to that derived from features less affected by emission filling \citep{martin2012}. The coincidental presence of additional spectral features adjacent to absorption lines of interest may also contaminate outflow signatures (e.g., He~I $\lambda$5876 $\sim$ 15 \AA\ blueward of Na D). 

\begin{figure}
\centering
\includegraphics[width=3.5in]{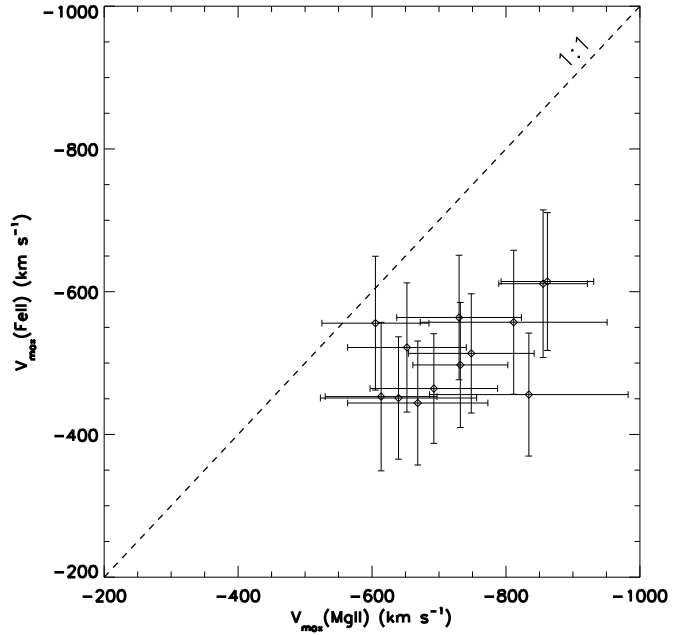}
\caption{$V_{max}$(Mg~II) versus $V_{max}$(Fe~II), for the 13 composite spectra discussed in this paper. $V_{max}$(Mg~II) is systematically more blueshifted than $V_{max}$(Fe~II), as discussed in Section \ref{sec: discussion}.}
\label{compare_vmax_2796_2374}
\end{figure}

Differences in oscillator strengths also likely impact derived outflow velocities. In the case of comparing velocities measured from Fe~II and Mg~II transitions, the larger oscillator strength of the Mg~II line at 2796 \AA\ ($f_{\rm 12}$ = 0.60) compared with the oscillator strengths of the Fe~II lines at 2344, 2374, and 2587 \AA\ used in our analyses ($f_{\rm 12}$ = 0.04--0.11) means that Mg~II is optically thick at lower column densities than Fe~II. Mg~II is a better probe of rarefied gas than Fe~II simply due to its larger cross-section for absorption. \citet{bordoloi2011} found that Mg~II equivalent widths decrease with increasing distance from galaxies. If equivalent width is a proxy for gas column density (neglecting the effects of saturation) then one can conclude that Mg~II gas column density decreases with increasing distance from galaxies. Given that Mg~II is an effective tracer of low-density gas, and presuming that low-density gas is found far from galaxies, the larger outflow velocities inferred for Mg~II compared with Fe~II \citep[Figure \ref{compare_vmax_2796_2374};][]{rubin2010a,prochaska2011} are suggestive that the speed of galactic winds increases with increasing galactocentric radius. This inference that galactic winds are accelerating is consistent with recent work by \citet{martin2009} and \citet{steidel2010}. \citet{martin2009} studied a sample of ULIRGs at \z0.25 and concluded that winds are accelerating based on the assumption of a spherical outflow and measurements of how gas covering fractions vary with velocity. These authors found that gas covering fractions decreased with increasing outflow velocity. Pairing this observation with the prediction that a spherical outflow will suffer geometrical dilution (i.e., reduction of its covering fraction) as it breaks up and expands to larger galactocentric radii, \citet{martin2009} concluded that outflowing gas is accelerating. \citet{steidel2010} showed that models of an accelerating wind provided the best match to observations at \emph{z} $\sim$ 2--3 of interstellar absorption lines arising from a range galaxy impact parameters (\emph{b} $\sim$ 0--200 kpc). Simulations also suggest that higher velocity gas may be located at larger galactocentric radii simply due to differences in travel time \citep[e.g.,][]{dallavecchia2012}. Observations of decreasing gas column densities with increasing outflow velocity seen in the gravitationally-lensed Lyman Break Galaxy cB58 \citep{pettini2002} provide further evidence of the connection between low-density gas, large galactocentric distances, and large outflow velocities. 

The effects discussed above are all dependent on the choice of tracer species. We now examine a more tangible aspect of the data analysis: how do different measurements of outflow velocity impact the derived velocities and subsequent scaling relations with galaxy properties? Two techniques feature prominently in the literature for quantifying the blueshift of an absorption line tracing an outflow. Some authors rely on fitting the centroid of an absorption line; i.e., measuring the wavelength at which the bulk of the absorption occurs and assuming that this velocity characterizes the outflowing gas \citep{shapley2003,martin2005,rubin2011,coil2011,law2012}. Other authors parameterize a maximal outflow velocity by measuring the blue shoulder of an absorption line at some fractional level of the continuum \citep[e.g., 90\%,][]{weiner2009,coil2011}. 

These two techniques, measuring different characteristics of the outflow, are complementary and both methods have associated uncertainties. In the case of the centroid measurement, one is susceptible to mistakenly attributing the entire absorption profile to an outflow when in fact the profile should be decomposed into both an outflow signature and a stellar absorption (or stationary ISM) component. Failing to correct for the gas not entrained in the outflow will yield an underestimate of the derived outflow velocity \citep{weiner2009,coil2011}, although such a correction is difficult and highly uncertain for low-resolution, low-S/N data \citep{steidel2010}. In our sample, the Fe~II absorption centroid that we measure is unlikely to be significantly contaminated by stellar absorption due to the young, star-forming nature of the majority of our galaxies; stellar Fe~II absorption is more prevalent in older populations \citep[e.g.,][]{bruzual2003}. The maximal outflow velocity method is dependent on resolved absorption lines; low spectral resolution data of unresolved absorption lines will merely reflect the instrumental profile and therefore measurements of a supposed wing profile will be meaningless. Furthermore, maximal outflow velocity measurements are strongly dependent on spectroscopic S/N (Section \ref{vmax}). We mitigate this effect in our present work by limiting our analyses of maximal outflow velocities to measurements from the composite spectra which have largely uniform S/N. 

%Mg~II is an $\alpha$-element produced in SNe II while Fe~II is made more slowly in SNe Ia; the time delay between SNe II and SNe Ia is at least 10$^8$ years \citep{mcwilliam1997}. In a chemically young environment, one would expect that Mg~II would be more abundant than Fe~II. \textbf{Do we have any estimates of age?} This difference in intrinsic abundance, paired with the increased dust depletion suffered by Fe~II compared with Mg~II \citep{savage1996}, creates an environment in which Mg~II is superabundant compared \textbf{to/with} Fe~II. The prevalence of Mg~II ensures that this species samples the kinematics of winds more fully than Fe~II, which is perhaps simply too tenuous to detect at the highest wind velocities. 

%As Mg is an $\alpha$-element, if the majority of Mg~II ions in galaxies result from the ejecta of SNe II one would expect these ions to be efficiently entrained in supernova winds. Since Fe~II is produced more slowly in SNe Ia, if present more in the surrounding ISM, may be less well entrained in the wind. \textbf{Can we say anything about the relative spatial distributions of SNe II and SNe Ia? Is one type more energetic than the other? Acknowledge Marcel Haas at STScI for this idea?}

\section{Summary and Conclusions} \label{sec: conclusions}

We utilized spectroscopic and imaging observations to investigate the properties and prevalence of outflowing galactic winds in a sample of 72 objects at 0.7 $<$ $z$ $<$ 1.3 in the Extended Groth Strip. These data are part of a larger study discussed in \citet{martin2012}. We used LRIS spectroscopy to study interstellar absorption lines in the rest-frame UV, including transitions of Fe~II and Mg~II. With \emph{GALEX}, \emph{HST}, and \emph{Spitzer} imaging from the AEGIS dataset, we employed two complementary analysis techniques to investigate how galaxy properties are correlated with the prevalence and strength of winds: 1) direct comparison of individual outflow velocities with galaxy properties and 2) construction of composite spectra based on subsamples of objects exhibiting similar star-forming or structural characteristics. We provide below a numbered list of our main conclusions in summary and expand on these points in the following paragraphs. 

1. Approximately 40\% of the sample exhibits $>$ 1$\sigma$ outflows in Fe~II.

2. We find a 3.1$\sigma$ trend between outflow velocity and $\Sigma_{\rm SFR}$.

3. Outflow velocity and SFR are only weakly correlated.

4. There is no apparent link between outflowing gas and host galaxy morphological type. 

5. Face-on galaxies exhibit more blueshifted Fe~II centroids than edge-on galaxies. 

6. The kinematics of Fe~II and Mg~II gas are often discrepant, with Mg~II preferentially tracing higher velocity gas. 

The sample includes objects exhibiting Fe~II Doppler shifts ranging from --217 \kms~to +155 \kms, where approximately 40\% of the sample exhibits $>$ 1$\sigma$ outflows in Fe~II. We find that 40 $\pm$ 5\% of the sample is characterized by Fe~II bllueshifts of at least --40 \kms; this outflow fraction is consistent with that of the parent sample discussed in \citet{martin2012}. We used SFRs inferred from \emph{GALEX} imaging in tandem with galaxy area estimates to measure $\Sigma_{\rm SFR}$. Given the clumpy morphologies of star-forming galaxies at \z1, we developed a new technique to estimate the area of only UV-bright clumps. Using these new ``clump areas" and also galaxy areas corresponding to Petrosian radii, we measured $\Sigma_{\rm SFR}$. The majority of objects have $\Sigma_{\rm SFR}$ $>$ 0.1 M$_{\odot}$ yr$^{-1}$ kpc$^{-2}$, yet we observe significant Fe~II blueshifts in only a minority of the sample. The lower fraction of objects showing outflows relative to that meeting the $\Sigma_{\rm SFR}$ threshold for driving a wind is evidence that winds are collimated and therefore observable only over a limited range of inclinations. We find a strong ($\sim$ 3$\sigma$) trend between outflow velocity and $\Sigma_{\rm SFR}$ such that objects with higher star-formation rate surface densities show more blueshifted Fe~II absorption. Composite spectra assembled on the basis of $\Sigma_{\rm SFR}$ also support this correlation, with $V_1$, $V_{max}$(Fe~II), and $V_{max}$(Mg~II) all varying toward more blueshifted values with increasing $\Sigma_{\rm SFR}$. At the same time, outflow veflocity and SFR are only weakly correlated. We propose that the data span too narrow a range in SFR (roughly two orders of magnitude) to show a trend. While the range of $\Sigma_{\rm SFR}$ is similarly limited (0.03 $\lesssim$ $\Sigma_{\rm SFR}$(A$_{74}$) $\lesssim$ 3.0), the trend between outflow velocity and $\Sigma_{\rm SFR}$ may be more fundamental and therefore recoverable over a smaller dynamic range. 

Based on quantitative morphological analyses of \emph{V}+\emph{I} \emph{HST} imaging, we do not find any link between outflowing gas and galaxy structure. Rather, galaxies experiencing winds appear to span the classical morphological regimes from disks to spheroidals to mergers. We estimated galaxy inclinations from axis ratios in the \emph{HST} imaging and found that face-on systems exhibit stronger Fe~II blueshifts than edge-on galaxies. This result lends credence to the canonical picture of outflowing winds emanating perpendicular to galactic disks. Motivated by recent results of faster outflows in smaller galaxies, we investigated the relationship between outflow velocity and size, but found no evidence for a trend. 

Quantifying the prevalence of outflows requires an understanding of the physical processes affecting the absorption lines tracing winds. Different elemental tracer species often yield discrepant outflow velocities, for reasons including emission filling, differences in oscillator strengths, and perhaps simply intrinsic variations in gas cloud kinematics. Furthermore, the variety of techniques available for quantifying the speed of galactic winds (centroid fitting, maximal velocity measurements from a blue wing, etc.) complicate the parameterization of a single outflow velocity. Future studies utilizing higher resolution spectroscopic data and complementary spatially-resolved imaging to determine galaxy morphologies and the spatial extent of winds will be crucial for better understanding the link between galactic outflows and their host galaxies at a range of redshifts. 

\begin{acknowledgements}
We appreciate the referee for his or her helpful comments that improved the quality of this paper. We thank Ben Weiner for providing measurements of [OII] widths and his composite spectrum of star-forming galaxies. K.A.K. is grateful for support from a Dissertation Year Fellowship at UCLA. A.E.S. acknowledges support from the David and Lucile Packard Foundation. A.L.C. has been supported by the Alfred P. Sloan Foundation and NSF CAREER award AST-1055081. This study was supported in part by the NSF under contract AST--0909182 (CLM). A portion of this work was completed at the Aspen Center for Physics (CLM). This study makes use of data from AEGIS, a multiwavelength sky survey conducted with the Chandra, GALEX, Hubble, Keck, CFHT, MMT, Subaru, Palomar, Spitzer, VLA, and other telescopes and supported in part by the NSF, NASA, and the STFC. We also wish to recognize and acknowledge the very significant cultural role and reverence that the summit of Mauna Kea has always had within the indigenous Hawaiian community. We are most fortunate to have the opportunity to conduct observations from this mountain.
\end{acknowledgements}

\clearpage

%\bibliography{/Users/katherinekornei/Documents/Paper2/Outflow_refs}

\clearpage

\clearpage
\LongTables

\begin{deluxetable}{lllllllllllr}
%\rotate
\tablecaption{Sample Parameters\tablenotemark{a}}
\tabletypesize{\small}
\tablewidth{0pt}
\tablehead{
\colhead{ID} & \colhead{RA\tablenotemark{b}} & \colhead{Dec\tablenotemark{c}} & \colhead{\emph{z$_{sys}$}} & \colhead{M$_{\rm B}$} & \colhead{\emph{U}--\emph{B}} & \colhead{$M_*$} & \colhead{SFR\tablenotemark{d}} & \colhead{A$_{\rm UV}$\tablenotemark{e}} & \colhead{A$_{74}$\tablenotemark{f}} & \colhead{$\pi$R$_{\rm P}^2$\tablenotemark{g}} & \colhead{$V_1$ $\pm$ $\delta V_1$\tablenotemark{h}} \\ \colhead{} & \colhead{(J2000)} & \colhead{(J2000)} & \colhead{} & \colhead{} & \colhead{} & \colhead{(log $M_{\odot}$)} & \colhead{(M$_{\odot}$ yr$^{-1}$)} & \colhead{} & \colhead{(kpc$^2$)} & \colhead{(kpc$^2$)} & \colhead{(km s$^{-1}$)} 
%\multicolumn{1}{c}{ID}
%& \multicolumn{1}{c}{RA\tablenotemark{b}}
%& \multicolumn{1}{c}{Dec\tablenotemark{c}}
%& \multicolumn{1}{c}{\emph{z$_{sys}$}}
%& \multicolumn{1}{c}{M$_{\rm B}$}
%& \multicolumn{1}{c}{\emph{U}--\emph{B}}
%& \multicolumn{1}{c}{$M_*$}
%& \multicolumn{1}{c}{SFR\tablenotemark{d}}
%& \multicolumn{1}{c}{A$_{\rm UV}$\tablenotemark{e}}
%& \multicolumn{1}{c}{A$_{74}$\tablenotemark{f}}
%& \multicolumn{1}{c}{$\pi$R$_{\rm P}^2$\tablenotemark{g}}
%& \multicolumn{1}{c}{V$_{\rm fe2}$\tablenotemark{h}} \\
%\colhead{}
%\hline
%\hline
%& \multicolumn{1}{c}{(J2000)}
%& \multicolumn{1}{c}{(J2000)}
%& \multicolumn{1}{c}{}
%& \multicolumn{1}{c}{}
%& \multicolumn{1}{c}{}
%& \multicolumn{1}{c}{(log $M_{\odot}$)}
%& \multicolumn{1}{c}{(M$_{\odot}$ yr$^{-1}$)}
%& \multicolumn{1}{c}{}
%& \multicolumn{1}{c}{(kpc$^2$)}
%& \multicolumn{1}{c}{(kpc$^2$)}
%& \multicolumn{1}{c}{(km s$^{-1}$)} \\
%\colhead{}
%\hline
%& \multicolumn{1}{l}{}
%& \multicolumn{1}{l}{}
}
\startdata
12008116 & 14 17 25.67 & 52 30 25.19 & 0.74664 & --19.69 & 0.64 & 10.1 & 5 & 1.9 & 43 & 155 & 37 $\pm$ 31 \\
12008166 & 14 17 21.59 & 52 28 03.62 & 1.28545 & --19.69 & 0.32 & 9.7 & \nodata & 0 & 11 & 44 & 31 $\pm$ 62 \\
12008197 & 14 17 21.84 & 52 29 19.62 & 0.98022 & --21.19 & 0.80 & 10.6 & 97 & 4.1 & 91 & 232 & --62 $\pm$ 19 \\
12008364\tablenotemark{i} & 14 17 09.44 & 52 29 08.95 & 0.77957 & --19.30 & 0.55 & 9.8 & \nodata & 0 & 0 & 0 & --33 $\pm$ 92 \\
12008441 & 14 17 16.76 & 52 28 39.59 & 0.83273 & --20.81 & 0.99 & 10.9 & 12 & 2.6 & 138 & 786 & 58 $\pm$ 61 \\
12008445 & 14 17 04.91 & 52 27 48.85 & 1.27739 & --20.69 & 0.51 & 10.2 & \nodata & 0 & 77 & 220 & --193 $\pm$ 79 \\
12008456 & 14 17 10.67 & 52 30 11.63 & 0.90230 & --19.93 & 0.73 & 9.8 & 6 & 1.4 & 21 & 60 & 79 $\pm$ 52 \\
12008481 & 14 17 10.04 & 52 28 39.45 & 0.71341 & --21.03 & 0.93 & 10.9 & \nodata & 0 & 138 & 484 & 75 $\pm$ 109 \\
12008509 & 14 17 03.23 & 52 30 33.13 & 1.21577 & --19.80 & 0.36 & 9.7 & \nodata & 0 & 51 & 144 & 70 $\pm$ 52 \\
12008550 & 14 16 56.74 & 52 29 52.48 & 1.30249 & --21.22 & 0.62 & 10.0 & 24 & 1.5 & 34 & 109 & --145 $\pm$ 23 \\
12008553 & 14 17 02.57 & 52 29 38.86 & 0.90156 & --21.56 & 1.12 & 10.5 & \nodata & 0 & 6 & 25 & --79 $\pm$ 29 \\
12008591 & 14 17 14.74 & 52 27 57.13 & 0.77192 & --18.37 & 0.53 & 9.1 & \nodata & 0 & 11 & 29 & \nodata \\
12008811 & 14 16 55.32 & 52 30 24.91 & 1.21556 & --20.79 & 0.67 & 10.1 & 9 & 0.8 & 13 & 44 & 33 $\pm$ 20 \\
12011364 & 14 18 28.46 & 52 33 16.22 & 0.98642 & --20.75 & 0.74 & 10.2 & 90 & 4.2 & 90 & 244 & --28 $\pm$ 24 \\
12011428 & 14 18 28.35 & 52 31 47.76 & 1.28408 & --19.87 & 0.18 & 9.6 & 9 & 0.8 & \nodata & \nodata & 154 $\pm$ 32 \\
12011493 & 14 18 27.94 & 52 33 39.53 & 1.26377 & --20.74 & 0.62 & 10.2 & 11 & 1 & 26 & 118 & --21 $\pm$ 45 \\
12011600 & 14 18 25.03 & 52 31 08.95 & 0.43665 & --17.67 & 0.43 & 8.8 & 0 & 0.9 & \nodata & \nodata & --133 $\pm$ 129 \\
12011619 & 14 18 24.68 & 52 32 48.66 & 1.07449 & --19.30 & 0.43 & 9.1 & 2 & 0 & 13 & 93 & \nodata \\
12011742 & 14 18 16.19 & 52 32 16.77 & 1.33578 & --20.81 & 0.71 & 10.2 & \nodata & 0 & 12 & 45 & 2 $\pm$ 23 \\
12011767 & 14 18 24.75 & 52 32 55.43 & 1.28170 & --22.18 & 0.75 & 11.0 & 76 & 2.7 & 171 & 562 & --34 $\pm$ 35 \\
12011836 & 14 18 16.05 & 52 31 48.06 & 0.92707 & --19.90 & 0.57 & 10.2 & 14 & 2.8 & 14 & 55 & --128 $\pm$ 21 \\
12012500 & 14 17 21.28 & 52 34 01.05 & 0.86494 & --20.28 & 0.89 & 10.6 & \nodata & 0 & 34 & 95 & --31 $\pm$ 28 \\
12012764 & 14 17 03.83 & 52 33 00.36 & 1.23532 & --20.26 & 0.61 & 9.7 & \nodata & 0 & \nodata & \nodata & --62 $\pm$ 47 \\
12012777 & 14 17 03.22 & 52 31 42.52 & 1.27426 & --21.06 & 0.51 & 10.1 & 22 & 1.4 & 15 & 55 & --217 $\pm$ 13 \\
12012817 & 14 16 59.78 & 52 31 09.48 & 1.21583 & --20.92 & 0.89 & 10.8 & \nodata & 0 & 7 & 23 & 28 $\pm$ 44 \\
12012842 & 14 16 56.45 & 52 33 13.77 & 1.31484 & --21.68 & 0.75 & 11.0 & \nodata & 0 & \nodata & \nodata & 5 $\pm$ 23 \\
12012871 & 14 17 00.37 & 52 33 38.53 & 1.34433 & --20.75 & 0.55 & 10.0 & 14 & 1.3 & \nodata & \nodata & --13 $\pm$ 39 \\
12013002 & 14 16 50.03 & 52 33 46.59 & 1.21841 & --20.22 & 0.64 & 9.5 & 4 & 0.3 & \nodata & \nodata & 10 $\pm$ 42 \\
12013145 & 14 16 55.46 & 52 32 15.81 & 1.34063 & --19.66 & 0.27 & 9.5 & 6 & 0.7 & \nodata & \nodata & --38 $\pm$ 61 \\
12013242 & 14 16 55.22 & 52 31 38.98 & 1.28679 & --21.31 & 0.45 & 10.1 & 25 & 1.2 & \nodata & \nodata & --73 $\pm$ 22 \\
12015177 & 14 18 55.51 & 52 37 18.84 & 0.98609 & --20.23 & 0.48 & 9.6 & 11 & 1.7 & \nodata & \nodata & --101 $\pm$ 18 \\
12015226 & 14 18 53.61 & 52 35 48.28 & 0.91600 & --19.30 & 0.29 & 9.1 & 4 & 0.7 & \nodata & \nodata & 16 $\pm$ 18 \\
12015295 & 14 18 55.92 & 52 37 07.05 & 0.49157 & --19.52 & 0.81 & 10.4 & 1 & 1.2 & \nodata & \nodata & \nodata \\
12015313 & 14 18 50.39 & 52 36 21.36 & 0.68009 & --20.72 & 1.21 & 11.2 & \nodata & 0 & \nodata & \nodata & \nodata \\
12015320 & 14 18 51.69 & 52 36 00.43 & 0.98573 & --22.09 & 1.05 & 11.3 & 40 & 2.6 & \nodata & \nodata & \nodata \\
12015563 & 14 18 34.29 & 52 36 49.30 & 1.28239 & --21.30 & 0.59 & 10.3 & \nodata & 0 & 7 & 22 & --98 $\pm$ 42 \\
12015643 & 14 18 49.35 & 52 36 08.38 & 0.24718 & --17.97 & 0.51 & 9.3 & 0 & 1.6 & \nodata & \nodata & \nodata \\
12015680 & 14 18 42.71 & 52 36 35.96 & 0.75867 & --20.47 & 0.55 & 10.3 & 21 & 2.9 & 68 & 279 & 148 $\pm$ 61 \\
12015682 & 14 18 49.05 & 52 36 29.65 & 1.28371 & --21.52 & 0.62 & 11.2 & 27 & 1.4 & \nodata & \nodata & --99 $\pm$ 93 \\
12015775 & 14 18 18.64 & 52 36 47.89 & 1.22428 & --19.36 & 0.22 & 9.6 & 4 & 0.7 & 4 & 19 & --105 $\pm$ 43 \\
12015792\tablenotemark{i} & 14 18 19.81 & 52 35 15.88 & 1.23068 & --21.23 & 0.95 & 11.0 & 12 & 2.2 & 0 & 0 & --27 $\pm$ 85 \\
12015858 & 14 18 26.49 & 52 36 08.41 & 1.23068 & --19.87 & 0.21 & 9.5 & 8 & 0.8 & 53 & 202 & --54 $\pm$ 60 \\
12015908 & 14 18 08.59 & 52 36 10.08 & 0.99924 & --20.25 & 0.64 & 10.0 & 21 & 3 & 27 & 82 & 45 $\pm$ 23 \\
12015914 & 14 18 22.10 & 52 35 27.04 & 1.10461 & --19.82 & 0.33 & 10.1 & 8 & 1.1 & 3 & 10 & 51 $\pm$ 21 \\
12015933 & 14 18 25.38 & 52 37 11.55 & 1.28514 & --20.34 & 0.66 & 9.7 & \nodata & 0 & 13 & 57 & --22 $\pm$ 81 \\
12016019 & 14 18 07.96 & 52 36 43.93 & 1.08467 & --20.71 & 0.62 & 10.0 & 18 & 1.9 & 44 & 182 & 48 $\pm$ 14 \\
12016050 & 14 18 08.67 & 52 35 13.82 & 0.97971 & --19.71 & 0.44 & 9.5 & 15 & 2.6 & 11 & 48 & --15 $\pm$ 22 \\
12016054 & 14 18 06.66 & 52 35 06.05 & 0.68377 & --20.73 & 1.08 & 11.2 & 16 & 4.1 & 43 & 134 & 72 $\pm$ 185 \\
12016075 & 14 18 05.98 & 52 34 08.42 & 1.11739 & --19.67 & 0.54 & 9.4 & 5 & 1.2 & 9 & 41 & 13 $\pm$ 49 \\
12016337 & 14 18 03.71 & 52 34 23.42 & 0.72026 & --20.27 & 1.01 & 10.9 & 9 & 3.3 & 144 & 617 & \nodata \\
12016903 & 14 17 12.78 & 52 34 28.40 & 1.15998 & --21.46 & 0.50 & 10.2 & 25 & 1 & 46 & 202 & 2 $\pm$ 30 \\
12017063 & 14 16 57.61 & 52 34 28.14 & 0.73912 & --19.79 & 0.51 & 10.1 & 5 & 1.2 & \nodata & \nodata & --70 $\pm$ 30 \\
12019542 & 14 18 49.94 & 52 40 22.15 & 1.27848 & --21.68 & 0.70 & 10.4 & 40 & 2.3 & 88 & 243 & --183 $\pm$ 60 \\
12019674 & 14 18 56.19 & 52 38 43.89 & 0.98491 & --21.48 & 0.96 & 11.1 & 15 & 2.2 & 135 & 443 & 130 $\pm$ 86 \\
12019697 & 14 18 49.69 & 52 37 42.18 & 0.26481 & --17.25 & 0.54 & 9.0 & 0 & 1.8 & 27 & 111 & \nodata \\
12019709 & 14 18 47.13 & 52 37 52.21 & 0.75792 & --19.90 & 0.48 & 9.7 & \nodata & 0 & 10 & 23 & --77 $\pm$ 151 \\
12019916 & 14 18 36.51 & 52 40 20.89 & 0.72849 & --20.26 & 0.83 & 10.7 & 11 & 3.3 & 48 & 156 & 49 $\pm$ 32 \\
12019923 & 14 18 35.21 & 52 39 42.89 & 0.72674 & --20.19 & 0.76 & 10.4 & 5 & 1.6 & 47 & 160 & 56 $\pm$ 37 \\
12019962 & 14 18 36.59 & 52 37 43.63 & 0.64444 & --19.22 & 0.45 & 9.7 & 1 & 0.8 & 32 & 144 & 104 $\pm$ 31 \\
12019973 & 14 18 28.61 & 52 40 53.43 & 0.81886 & --19.80 & 0.53 & 10.1 & 10 & 2.3 & 41 & 123 & --58 $\pm$ 27 \\
12019995 & 14 18 47.09 & 52 37 54.47 & 0.24888 & --19.06 & 0.52 & 9.8 & \nodata & 0 & 42 & 150 & \nodata \\
12019996 & 14 18 36.64 & 52 37 54.62 & 1.28128 & --21.71 & 0.63 & 10.6 & 43 & 1.8 & 164 & 499 & 117 $\pm$ 24 \\
12020031 & 14 18 41.23 & 52 39 34.32 & 0.82072 & --22.00 & 0.97 & 11.2 & 36 & 2.4 & 427 & 1391 & 64 $\pm$ 108 \\
12020064 & 14 18 46.56 & 52 38 13.78 & 1.31479 & --21.12 & 0.59 & 10.3 & 26 & 2 & 58 & 138 & --62 $\pm$ 56 \\
12020075 & 14 18 48.94 & 52 37 30.01 & 0.80829 & --20.52 & 0.64 & 10.6 & 12 & 1.9 & 77 & 279 & --52 $\pm$ 17 \\
12024014 & 14 18 50.28 & 52 42 05.78 & 1.29775 & --20.51 & 0.49 & 9.9 & \nodata & 0 & 6 & 17 & \nodata \\
12024045 & 14 18 51.65 & 52 40 56.08 & 0.24698 & --18.91 & 0.61 & 9.9 & 5 & 4.5 & 61 & 424 & \nodata \\
12024133 & 14 18 51.82 & 52 41 11.85 & 1.12415 & --21.14 & 0.81 & 10.5 & 36 & 3 & 49 & 202 & --57 $\pm$ 43 \\
12024181 & 14 18 49.60 & 52 42 36.79 & 1.08692 & --21.00 & 0.80 & 10.5 & 35 & 3.3 & 96 & 479 & --57 $\pm$ 32 \\
12024409 & 14 18 33.91 & 52 41 28.32 & 1.03579 & --20.59 & 0.84 & 9.9 & 11 & 1.7 & 48 & 138 & --104 $\pm$ 42 \\
12024524 & 14 18 33.28 & 52 41 10.89 & 0.90300 & --20.93 & 0.71 & 10.4 & 32 & 2.7 & 119 & 393 & --31 $\pm$ 23 \\
12100420 & 14 16 57.97 & 52 31 58.63 & 1.19952 & --20.63 & 0.60 & 10.0 & 9 & 1 & 3 & 10 & 71 $\pm$ 18 \\
%X & X & X & X & X & X & X & X & X & X & X & X \\
%X & X & X & X & X & X & X & X & X & X & X & X \\
%X & X & X & X & X & X & X & X & X & X & X & X \\
%X & X & X & X & X & X & X & X & X & X & X & X \\
%X & X & X & X & X & X & X & X & X & X & X & X \\
%X & X & X & X & X & X & X & X & X & X & X & X \\
%X & X & X & X & X & X & X & X & X & X & X & X \\

\enddata
\tablenotetext{a}{``\nodata" indicates no data.} 
\tablenotetext{b}{Units of right ascension are hours, minutes, and seconds.}
\tablenotetext{c}{Units of declination are degrees, arcminutes, and arcseconds.}
\tablenotetext{d}{Dust-corrected UV star-formation rate estimated from \emph{GALEX} observations.}
\tablenotetext{e}{Dust attenuation estimated from UV colors.}
\tablenotetext{f}{Galaxy ``clump area" (Section \ref{sec: clump}).}
\tablenotetext{g}{Galaxy Petrosian area.}
\tablenotetext{h}{$V_1$ outflow velocity measured from the simultaneous fit to five resonance Fe~II absorption lines (Section \ref{fitting}).}
\tablenotetext{i}{Objects with \emph{HST} imaging, although the S/N was too low to make robust area measurements.}
\label{sampletable}
\end{deluxetable}

\clearpage

\newpage

\begin{deluxetable}{lllll}
\centering
\tablecaption{Correlations Between $V_1$ Outflow Velocity and Galaxy Properties}
\tablewidth{0pt}
\tablehead{
\colhead{Property} & \colhead{Correlation with $V_1$\tablenotemark{a}} 
%\multicolumn{1}{c}{Property}
%& \multicolumn{1}{c}{Correlation with $V_1$\tablenotemark{a}}
}
\startdata
SFR & 1.2, --0.17 (47) \\  %from match_info_plots.pro
sSFR & 2.3, --0.33 (47) \\
$\Sigma_{\rm SFR}$(R$_{\rm P}$) & 2.4, --0.40 (36) \\ 
& 3.1, --0.54\tablenotemark{b} (34) \\
$\Sigma_{\rm SFR}$(A$_{74}$) & 2.4, --0.40 (36) \\
& 3.1, --0.54\tablenotemark{b} (34) \\
\emph{i} & 0.9, --0.14 (46) \\   %from ID_BA_dV.txt
R$_{\rm P}$ & --1.4, 0.20 (47) \\
(A$_{74}$/$\pi$)$^{1/2}$ & --0.9, 0.14 (47) \\
\enddata
\tablenotetext{a}{The first number indicates the number of standard deviations from the null hypothesis that the quantities are uncorrelated, according to the Spearman $\rho$ correlation test. A larger number of standard deviations means a higher likelihood that the data are correlated; negative values refer to inverse correlations. The second number is the correlation coefficient, $r_S$. The quantity in parentheses indicates the number of objects in the sample.}
\tablenotetext{b}{Correlations calculated omitting two outliers (12015914 and 12100420) whose extremely compact morphologies translated into large size uncertainties.}
\label{correlationtable}
\end{deluxetable}

\begin{deluxetable}{lllll}
\tablecaption{Composite Spectra}
\tablewidth{0pt}
\tablehead{
\multicolumn{1}{c}{Composite}
& \multicolumn{1}{c}{Average Value}
& \multicolumn{1}{c}{$V_1$\tablenotemark{a}} 
& \multicolumn{1}{c}{$V_{max}$(Mg~II)\tablenotemark{b}} 
& \multicolumn{1}{c}{$V_{max}$(Fe~II)\tablenotemark{c}} \\
\colhead{}
& \multicolumn{1}{c}{}
& \multicolumn{1}{c}{(\kms)}
& \multicolumn{1}{c}{(\kms)}
& \multicolumn{1}{c}{(\kms)}
}
\startdata
%more results.fitting.composites2.dat | egrep -v "#" | awk '{print $1" & "int($2)" $\\pm$ "int($3)" \& XXX \\\\"}' | sed 's/-/--/' | sed 's/\_//g' | egrep "lowi|highi|highUVcorrcom|lowUVcorrcom|highSFRSDpetr|lowSFRSDpetro|highSFRSDcomp|lowSFRSDcompo"
EGS & \nodata & --5 $\pm$ 6 & --730 $\pm$ 93 & --564 $\pm$ 87  \\
SFR -- \emph{high} & 32 $\pm$ 4 M$_{\odot}$ yr$^{-1}$ & --24 $\pm$ 9 & --748 $\pm$ 94 & --514 $\pm$ 84 \\
SFR -- \emph{low} & 7 $\pm$ 1 M$_{\odot}$ yr$^{-1}$ & 29 $\pm$ 11 & --614 $\pm$ 83 & --453 $\pm$ 104  \\
sSFR -- \emph{high} & 1.9 $\pm$ 0.2 $\times$ 10$^{-9}$ yr$^{-1}$ & --34 $\pm$ 9 & --605 $\pm$ 80 & --556 $\pm$ 94  \\
sSFR -- \emph{low} & 4.1 $\pm$ 0.6 $\times$ 10$^{-10}$ yr$^{-1}$ & 27 $\pm$ 11 & --834 $\pm$ 148 & --456 $\pm$ 86  \\
$\Sigma_{\rm SFR}$(R$_{\rm P}$) -- \emph{high} & 0.29 $\pm$ 0.05 M$_{\odot}$ yr$^{-1}$ kpc$^{-2}$ & --25 $\pm$ 6 & --862 $\pm$ 69 & --614 $\pm$ 97  \\
$\Sigma_{\rm SFR}$(R$_{\rm P}$) -- \emph{low} & 0.05 $\pm$ 0.007 M$_{\odot}$ yr$^{-1}$ kpc$^{-2}$ & 33 $\pm$ 13 & --668 $\pm$ 105 & --444  $\pm$ 87  \\
$\Sigma_{\rm SFR}$(A$_{74}$) -- \emph{high} & 0.93 $\pm$ 0.15 M$_{\odot}$ yr$^{-1}$ kpc$^{-2}$ & --31 $\pm$ 7 & --855 $\pm$ 66 & --611 $\pm$ 103 \\
$\Sigma_{\rm SFR}$(A$_{74}$) -- \emph{low} & 0.18 $\pm$ 0.02 M$_{\odot}$ yr$^{-1}$ kpc$^{-2}$ & 44 $\pm$ 15 & --640 $\pm$ 117 & --451 $\pm$ 86  \\
A$_{\rm UV}$ -- \emph{high} & 2.7 $\pm$ 0.13 & --3 $\pm$ 10 & --732 $\pm$ 71 & --497 $\pm$ 88  \\
A$_{\rm UV}$ -- \emph{low} & 1.1 $\pm$ 0.078 & --10 $\pm$ 9 & --652 $\pm$ 89 & --522 $\pm$ 91  \\
\emph{i -- high} & 58$^{\circ}$ $\pm$ 1$^{\circ}$ & 28 $\pm$ 11 & --692 $\pm$ 95 & --464 $\pm$ 77  \\
\emph{i -- low} & 38$^{\circ}$ $\pm$ 1$^{\circ}$ & --19 $\pm$ 9 & --811 $\pm$ 140 & --557 $\pm$ 101 \\
\enddata
\tablenotetext{a}{Outflow velocity measured from the simultaneous fit to five resonance Fe~II absorption lines, as described in Section \ref{fitting} and \citet{martin2012}.}
\tablenotetext{b}{Maximal outflow velocity measured from the 2796 \AA\ Mg~II line (Section \ref{sec: MgII}).}
\tablenotetext{c}{Maximal outflow velocity measured from the 2374 \AA\ Fe~II line (Section \ref{sec: MgII}).}
\label{compositetable}
\end{deluxetable}

\end{document}